\newcommand{\nc}{\newcommand}
\nc{\kms}{\,{km\,s$^{-1}$}}
\nc{\as}{\arcsec}
\nc{\HI}{H\,{\sc i}}
\nc{\HII}{H\,{\sc ii}}
\nc{\NII}{N\,{\sc ii}}
\nc{\FeII}{Fe\,{\sc ii}}
\nc{\OIII}{O\,{\sc iii}}
\nc{\CI}{C\,{\sc i}}
\nc{\hto}{H$_{2}$O}
\nc{\htmo}{H$_{2}^{16}$O}
\nc{\htio}{H$_{2}^{18}$O}
\nc{\am}{\arcmin}
\nc{\ciso}{C$^{18}$O}
\nc{\onetwoco}{$^{12}$CO}
\nc{\onethreeco}{$^{13}$CO}
\nc{\hctn}{HC$_{3}$N}
\nc{\htwo}{H$_{2}$}
\nc{\ot}{O$_{2}$}
\nc{\nht}{NH$_{3}$}
\nc{\htco}{H$_{2}$CO}
\nc{\htre}{H$_{3}$O${^+}$}
\nc{\msol}{{$\mathrm{M}_{\odot} $}}
\nc{\lsol}{{$\mathrm{L}_{\odot} $}}
\nc{\met}{CH$_{3}$OH}
\nc{\metiso}{$^{13}$CH$_{3}$OH}
\nc{\fas}{$\farcs$}
\nc{\chl}{H$_{2}$Cl$^{+}$}
\nc{\mum}{$\mu$m}
\nc{\htrep}{H$_3^+$}
\nc{\ohp}{OH$^+$}
\nc{\htop}{H$_2$O$^+$}
\nc{\htreop}{H$_3$O$^+$}
\nc{\CHp}{CH$^+$}
\nc{\hcop}{HCO$^+$}

\documentclass[bibyear]{aa}

\usepackage{graphicx}
\usepackage{txfonts}

\begin{document}

\title{{\it Herschel} and {\it Odin} observations of \hto, CO, CH,
  \CHp, and [\NII] in the barred spiral galaxy NGC\,1365
\thanks{{\it Herschel} is an ESA space observatory with science
  instruments provided by European-led Principal Investigator
  consortia and with important participation from NASA. {\it
      Herschel} was launched on May 14, 2009 and decommissioned on
    June 17, 2013.}
\thanks{{\it Odin} is a Swedish-led satellite project funded jointly
  by the Swedish National Space Board (SNSB), the Canadian Space
  Agency (CSA), the National Technology Agency of Finland (Tekes), the
  Centre National d'Etudes Spatiales (CNES), France, and the European
  Space Agency (ESA). The former Space division of the Swedish Space
  Corporation, today OHB Sweden, is the prime contractor also
  responsible for {\it Odin} operations. {\it Odin} was launched
    on February 20, 2001 and is still in active operation.}
}

\subtitle{Bar-induced activity in the outer and inner circumnuclear tori}

\author{Aa.\,Sandqvist\inst{1}
  \and \AA .\,Hjalmarson\inst{2} 
  \and B.\,Larsson\inst{1}
  \and U.\,Frisk\inst{3}
  \and S.\,Lundin\inst{4}
  \and G.\,Rydbeck\inst{2
  }
}

\institute{Stockholm Observatory, Stockholm University, AlbaNova
  University Center, SE-106 91 Stockholm, Sweden\\ \email{aage@astro.su.se}
\and Department of Space, Earth and Environment, Chalmers University of
Technology, Onsala Space Observatory, SE-439 92 Onsala, Sweden
\and Omnisys Instruments AB, Ringv\"agen 100E, SE-118 60 Stockholm,
Sweden
\and OHB Sweden, PO Box 1269, SE-164 29 Kista, Sweden
}

\offprints{\\ Aage Sandqvist, \email{aage@astro.su.se}}

\date{Received $<$8 July 2020$>$; accepted $<$22 December 2020$>$}

\abstract {The {\it Odin} satellite is now into its twentieth year of
  operation, much surpassing its design life of two years. One of its
  major astronomical pursuits was the search for and study of water
  vapor in diverse regions of the Solar System and the Milky Way
  galaxy. The {\it Herschel} space observatory was needed to detect
  water vapor in external galaxies.} {Our goal is to study
  the distribution and excitation of water vapor and other molecules
  in the barred spiral galaxy NGC\,1365.} {{\it Herschel} has observed
  the central region of NGC\,1365 in two positions, and both its SPIRE and PACS
  observations are available in the {\it Herschel} Science
  Archive.  {\it Herschel} PACS images have been produced of the 70
  and 160 $\mu$m infrared emission from the whole galaxy, and also of the
  cold dust distribution as obtained from the ratio of the 160 to 70
  $\mu$m images. The {\it Herschel} SPIRE observations have been
  used to produce simultaneously observed maps of the 557 GHz o-\hto,
  752 GHz p-\hto, 691 GHz CO($6-5$), 1037 GHz CO($9-8$), 537 GHz CH, 835
  GHz \CHp, and the 1461 GHz [\NII] lines (efficiently probing the warm
  ionized medium) in the inner bar and
  circumnuclear torus region; – however, these observations have
    no effective velocity resolution. For this reason {\it Odin} has recently 
  observed the 557 GHz ortho-\hto\ ground state line in the central
  region with high (5 \kms) spectral resolution.} { The emission and
  absorption of \hto\ at 557 GHz, with a velocity resolution of 5 \kms, has
  been marginally detected in NGC\,1365 with {\it Odin}. The water vapor is
  predominantly located in a shocked 15\as\ (1.3 kpc) region near some
  central compact radio sources and hot-spot \HII\ regions, close to the
  northeast component of the molecular torus surrounding the
  nucleus. An analysis of the \hto\ line intensities and velocities
  indicates that a shock-region is located here. This is corroborated
  by a statistical image deconvolution of our SEST CO($3-2$)
  observations, yielding 5\as\ resolution, and a study of our Very
  Large Array (VLA) \HI\
  absorption observations, as well as comparisons with published
  interferometric CO observations. Additionally, an enticing 20\as\ \HI\
  ridge is found to extend south-southeast from the nucleus,
  coinciding in position with the southern edge of an \OIII\ outflow
  cone, emanating from the nucleus. The molecular chemistry of the
  shocked central region of NGC\,1365 is analyzed with special emphasis
  on the CO, \hto\ and CH, \CHp\ results.} {The dominating activity 
  near the northeast (NE) torus component may have been triggered by the rapid
  bar-driven inflow into the circumnuclear torus causing cloud-cloud
  collisions and shocks, leading to the formation of stellar
  superclusters and, hence, also to more efficient PDR chemistry,
    which, here, may also benefit from cosmic ray focusing caused by
    the observed aligned magnetic field. The  
  very high activity near the NE torus component may
reflect the fact that the eastern bar-driven gas inflow into the NE
region is much more massive than the corresponding western gas inflow into the
southwest (SW) region. The \hto\ and \CHp\ emissions peak in the
NE torus region, but 
the CO and CH emissions are more evenly distributed across the whole
circumnuclear torus. The higher energy CO spectral line energy
distribution (SLED) is nicely modeled by a
low velocity (10 \kms) shock, which may as well explain the required CH
excitation and its high abundance in denser gas. The higher velocity
(40 \kms) shock required to model the \hto\ SLED in the NE torus region,
paired with the intense UV radiation from the observed massive young
stellar superclusters, may also explain the high abundance of \CHp\ in
this region. The nuclear \HI\ ridge may have been created by the action 
of outflow-driving X-ray photons colliding with ice-covered dust grains. A
precessing nuclear engine, as is suggested by the tilted massive
  inner gas torus, may be necessary to explain the various
nuclear outflows encountered.}


\titlerunning{{\it Herschel} and {\it Odin} observations of \hto, CO,
  CH, \CHp, and [\NII] in NGC\,1365}
\authorrunning{Aa. Sandqvist et al.}
\maketitle

\section{Introduction}
NGC\,1365 is a prominent barred supergiant spiral galaxy in the
Fornax cluster with a heliocentric velocity of +1632 \kms\ (the
velocity with respect to the Local Standard of Rest, $V_{\rm LSR} =
1613$ \kms, is used in this paper). In Fig. 1 we present a comparison
of a visible-light ``true-color'' image of NGC\,1365 from
  three exposures taken with the FORS1 camera on ESO’s VLT UT1, along
with an infrared $K$-band image ($0.9-2.5$ \mum) obtained with the
HAWK-I camera on the ESO VLT UT4 
telescope (ESO/P. Grosb\"ol). The infrared picture suggests a smaller
stellar bar, or disk, within the radius of the
Inner Lindblad Resonance (ILR) of 30\as\ (P.A.B. Lindblad et
al. \cite{lin96}) that is roughly perpendicular to the larger bar (as well
as elongated along the line of nodes). Alonso-Herrero et
al. (\cite{alo12}) performed a thorough study of the infrared
spectrum ($3.6 - 500$ \mum) of the central region of NGC\,1365 using the
{\it Herschel} Photodetector Array Camera and Spectrometer (PACS) and
Spectral Photometric Imaging REceiver (SPIRE), as well as {\it Gemini}
and {\it Spitzer}. From 
the {\it Herschel} Science Archive, we retrieved their 70 and 160 \mum\ PACS 
observations of NGC\,1365 and display 70 \mum\ and 160 \mum\ images in
Fig. 1, where we also indicate the high- and low-frequency
beams, respectively, of the SPIRE spectrometer observations presented
in Sect. 2.1. The galaxy displays a wide 
range of phenomena indicating activity – including a Seyfert 1.5 type
nucleus with strong, broad, and narrow H$\alpha$ lines. Ionized outflows from
the active galactic nucleus (AGN) with velocities up to a few hundred
\kms\ have been observed (e.g., Venturi et al. \cite{ven18} and
references therein). Kristen et al. (\cite{kri97}) used the Hubble Space 
Telescope to study the Seyfert nucleus and circumnuclear hot spots in
NGC\,1365 and found these hot spots to be resolved into a number of
bright compact condensations, which they interpreted as super star clusters
(SSC). The magnetic field in the central region has been mapped at the
NRAO VLA with a
resolution of 9\as\ by Beck et al. (\cite{bec05}). J\"ors\"ater \&\ van Moorsel
(\cite{jor95}) used the NRAO VLA to map the \HI\ emission
distribution in the arms and the bar. In this paper, we present the
\HI\ absorption results in the galaxy's central region, which were not discussed
by them. A rapidly spinning supermassive black hole has been
disclosed at the center of NGC\,1365 by Risaliti et
al. (\cite{ris13}). At an assumed distance of 18.6 Mpc (Madore et
al. \cite{mad98}), 1\as\ corresponds to 90 pc. For an extensive review
of NGC\,1365, see Lindblad (\cite{lin99}).

To set a solid background of observational knowledge, we now
  provide a rather detailed introductory review upon which the
  interpretation of our own observations can be based. In a flattened
  rotating proto-galaxy – very likely created in the potential well of
  a cosmological dark matter fluctuation (Persson et al. \cite{per10})
  – there exist intrinsic instability modes such as spiral and bar 
  instabilities, which may be triggered by encounters with nearby
  galaxies or by internal “noise” (e.g., P.O. Lindblad \cite{lin60};
  Lin \&\ Shu \cite{lin64}; Shu et al. \cite{shu73}; Sundelius et
  al. \cite{sun87}; Pfenniger \&\ Norman \cite{pfe90}; P.A.B. Lindblad
  et al. \cite{lin96}; P.O. Lindblad \cite{lin99}; Pinol-Ferrer et
  al. \cite{pin12}). Here the A. Toomre (\cite{too64}) (in)stability
  criterion is a useful constraint.  

Inside corotation, the bar pattern velocity is slower than the
rotational motion of the gas. According to modeling and observations (e.g.,
P.A.B. Lindblad et al. \cite{lin96}; P.O. Lindblad et
al. \cite{polin96}; P.O. Lindblad \cite{lin99};
Sakamoto et al. \cite{sak07}; Elmegreen et 
al. \cite{elm09}), the gas catches up with the bar, and when leaving it
on the front side of the bar experiences a strong shock with a drop of
velocity of up to a few hundred \kms. This leads to an inflow of gas
along the curved leading dust lanes of the bar (in the eastern bar 
estimated to be $\approx$22 \msol /year  
at an inflow velocity of $\approx$80 \kms), creating a rather massive
rotating torus inside the ILR. Here
the orbit crowding is expected to cause multiple cloud-cloud
collisions and shocks, leading to cloud coalescing into very massive
($\approx 10^{7}$ \msol) cloud complexes, as well as to the subsequent
birth of OB stars and massive (several $\times 10^{6}$ \msol) stellar
superclusters, as is observed in the outer NGC\,1365 
torus (Elmegreen et al. \cite{elm09}; Galliano et al. \cite{gal12};
Fazeli et al. \cite{faz19}). The 
aforementioned expectations were based upon the fact that
massive OB stars are born (mainly) in the spiral arms – where orbit
crowding and density wave streaming is causing cloud-cloud collisions and
shocks (cf. Roberts \&\ Stewart \cite{rob87}; Rydbeck et
al. \cite{ryd85}; Aalto et al. \cite{aal99}; Schinnerer et
al. \cite{sch10}) – as concluded from the Milky Way observational
result, that the number of giant \HII\ regions in a  
specified volume roughly scales as <$n$(\htwo)>$^2$, where <n(\htwo)> is
the mean \htwo\ density in the same volume (Scoville et al. \cite{sco86}). The
physical reason for the increased stellar formation activity is the
increased external pressure created by the spiral arm molecular cloud
collisions, triggering multiple massive gravitational collapses in
these clouds, already residing in their low internal pressure (low
temperature) virial equilibrium. This is a result of the efficient cooling
provided by spectral line radiation from the interstellar molecules
\onetwoco, \onethreeco, \hto, and other species – (astro)chemistry
(e.g., Irvine et al. \cite{irv87}; van Dishoeck \&\ Blake \cite{dis98};
van Dishoeck et al. \cite{dis13}) and (astro)physics in a necessary
collaboration (Goldsmith \&\ Langer \cite{gol78}; Takahashi et
al. \cite{tak85}; Neufeld et al. \cite{neu95}; see Hjalmarson \&\
Friberg \cite{hja88} for a discussion and additional references).

\begin{figure*}[ht]
\includegraphics[angle=0, width=.99\textwidth]{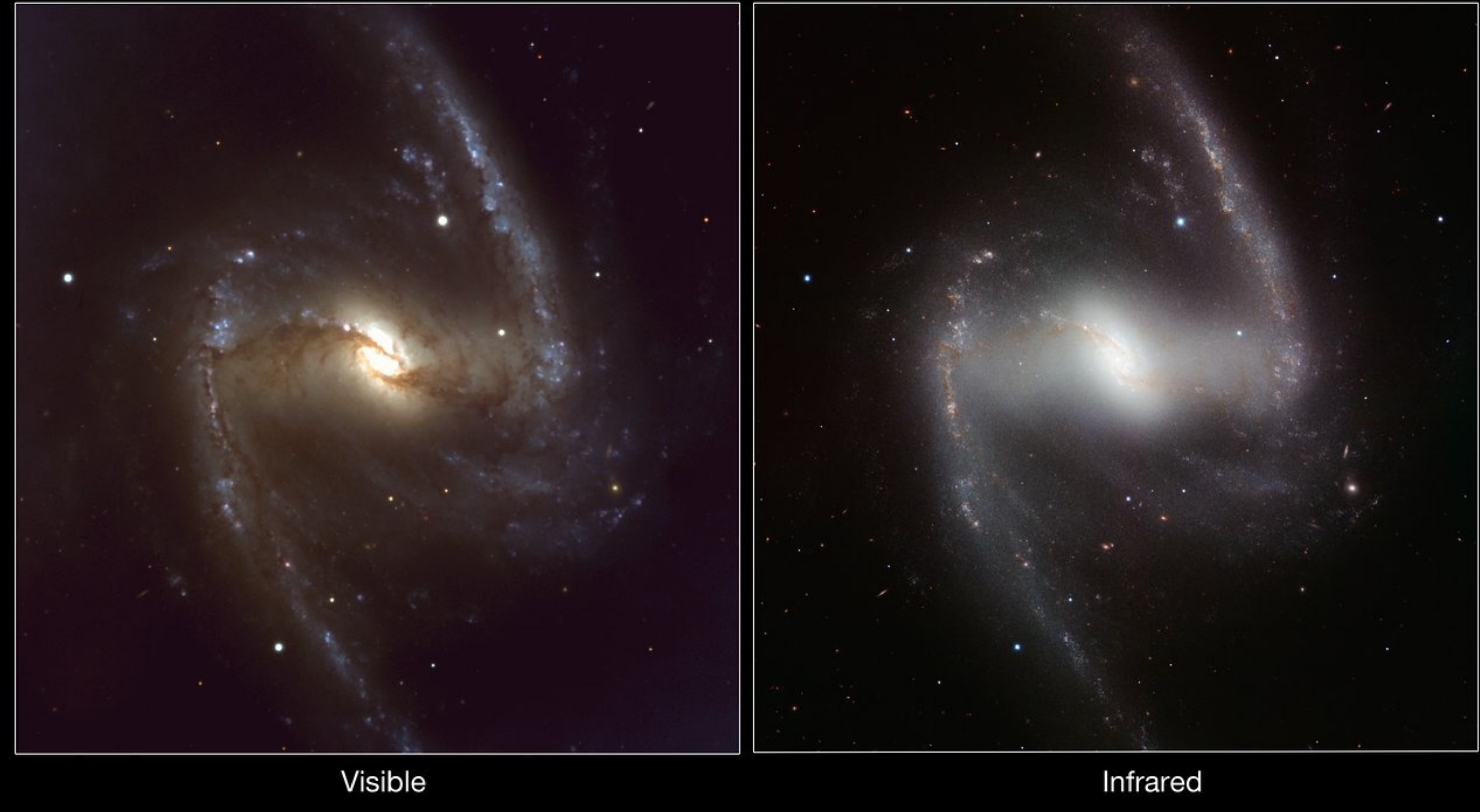}
\includegraphics[angle=0, width=.49\textwidth]{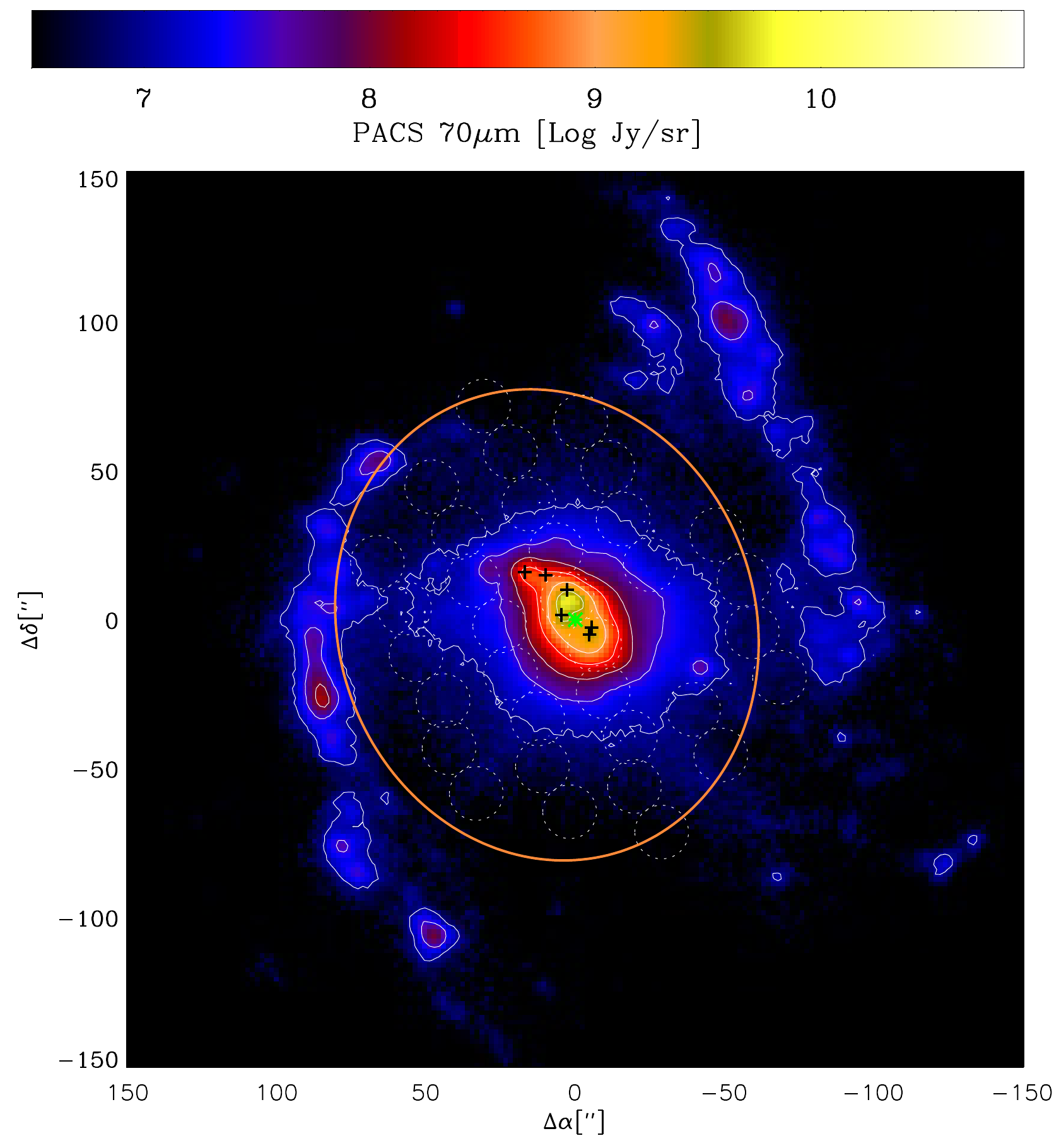}
\includegraphics[angle=0, width=.49\textwidth]{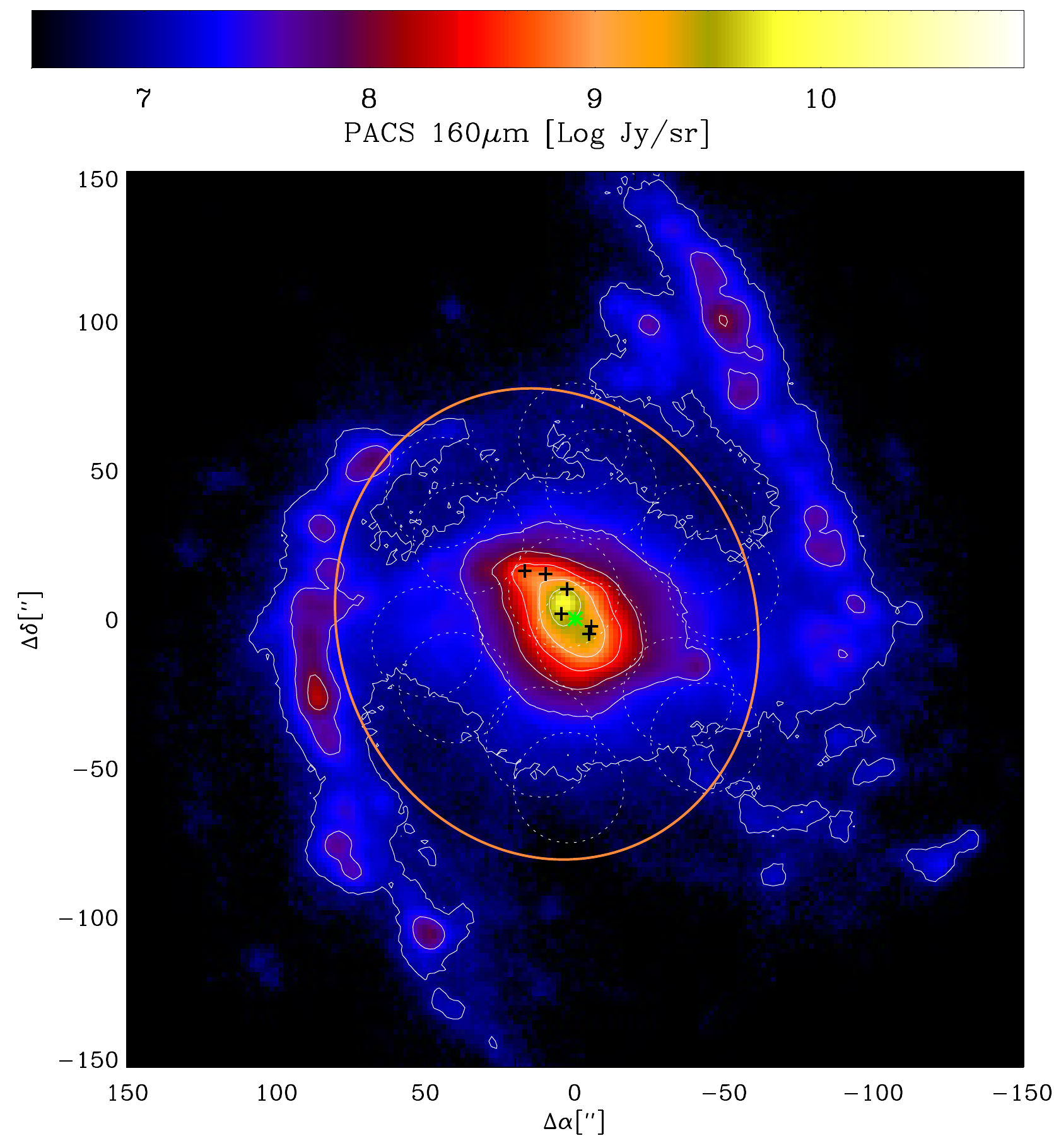} 
    \caption{{\it Top}: Visible-light ``true-color'' and infrared
   ($0.9-2.5$ \mum) images of NGC\,1365, obtained with ESO's VLT
    (ESO/P. Grosb\"ol). {\it Bottom}: The 70 \mum\ and 160 \mum\
    {\it Herschel} PACS images 
   of NGC\,1365 (beam sizes 5\farcs6 and 11\farcs3;
   Alonso-Herrero et al. \cite{alo12}) on which are superimposed the 
    high- and low-frequency beams (20\as\ and 40\as,
      respectively, dashed circles) of the SPIRE spectrometer for the
    two sets of observations (toward the NE and SW positions,
      see Sect. 2.1) and the resultant beam
    (140\as $\times$ 160\as, solid ellipse) of the {\it Odin}
    observations. The crosses, from top to bottom (in decreasing
    declination), represent the ``hot spot'' \HII\ regions L4, L11,
    L12, L1, L3, and L2 (Alloin et al. \cite{all81}). The equatorial
    offsets are with respect to the optical nucleus, which is marked
    with a green asterisk.}     
  \label{1}
\end{figure*}

The very active starburst region in the central parts of the NGC\,1365
bar has been studied in CO molecular lines at high resolutions: 
the $J=3-2$ CO line with 5\as\ effective resolution (Sandqvist
\cite{san99}; this paper), and the $J=2-1$ lines of $^{12}$CO,
$^{13}$CO, and C$^{18}$O with 2\as\ resolution (Sakamoto et
al. \cite{sak07}). Sandqvist et al. (\cite{san95}) performed VLA
aperture synthesis observations of the radio emission 
from ionized gas, revealing a lower level circumnuclear ring of radius
$\approx 9\as\ (\approx 800$ pc), interspersed with compact radio
sources, all of which have spectral indices indicating free-free
emission from hot ionized gas, except one, called ``F'', which has 
the spectral index characteristic of synchrotron radiation (see Table
2). They also mapped the distributions of the CO $J=1-0$ 
and $2-1$ lines using the SEST (Swedish ESO Submillimeter Telescope, now
”retired”).  In the nuclear region, the envisioned synchrotron jet is
enshrouded by a fan- or rather cone-shaped ionized gas outflow,
revealed by its visual \OIII\ emission (Fig. 7(4) and Hjelm \&\ Lindblad
\cite{hje96}, discussed in detail in the review paper by Lindblad
\cite{lin99}). More recently this kpc-size wide-angle outflow of
ionized gases was studied in detail by Lena et al. (\cite{len16})
and Venturi et al. (\cite{ven18}). Combes et 
al. (\cite{com19}) present Atacama Large Millimeter/submillimeter
Array (ALMA) observations of CO($3-2$) with a
resolution of $0\farcs1$ in a sample of seven Seyfert/LINER galaxies
which includes NGC\,1365, where they estimate a central AGN (black
hole) mass of $4 \times 10^6$ \msol\ by model fitting to the velocity
field observed inside a radius of $1\farcs4$. It appears that the
  rotation axis of the inner nuclear torus has a position angle (P.A.)
  and inclination (Incl.) – (P.A. $\approx 160\degr \pm 10\degr$;
  Incl. $ \approx 63\degr \pm 
10\degr$) – which markedly deviates from the common rotation axis of
the spiral arm disk, the bar, and the outer molecular gas torus of the
galaxy (P.A. $\approx  130\degr$; Incl. $ \approx 50\degr$). The
orientation of the (symmetry) axes of the nuclear outflows
(the wide-angle bi-conical ionized gas outflow, the synchrotron
radiation radio jet, as well as the narrow CO outflow outlined in
the present paper) all seem to be co-aligned with the galaxy rotation
axis. The tilted rotation axis of the inner nuclear torus, or accretion
disk, may suggest that we are witnessing the action of a precessing
nuclear engine, a matter which will be outlined in our forthcoming
discussion. The radius of the rotating inner torus, possibly a  black hole
accretion disk, was determined to be $\approx$0.3\as\ (26 pc) and its
estimated gas mass is $\approx 7 \times 10^6$ \msol. Their larger scale
mapping, in addition, confirms the outer rotating circumnuclear
molecular gas torus of radius 9\as\ (800 pc), already known from the
CO($3-2$) observations by Sandqvist (\cite{san99}) and the CO($2-1$)
SubMillimeter Array (SMA) mapping by Sakamoto et al. (\cite{sak07});
compare also Fig. 7 of this paper.  

The convincing signs of shocks in the NE outer torus region of
NGC\,1365 have indeed been observationally reported in terms of
vibration-rotation 2 $\mu$m \htwo\ emission lines, 
requiring shock excitation and also pronounced Br$\gamma$ line wings,
indicative of outflows and shocks emanating from three carefully studied
superclusters (Galliano et al. \cite{gal12}). Fazeli et
al. (\cite{faz19}), in their recent VLT SINFONI near-IR bservations of the
central 9\as\ $\times$\ 9\as\ region of NGC 1365, found strong
evidence of shocks – 
in terms of emission from several 2 $\mu$m vibration-rotation \htwo\ lines, as
well as from the “shock indicator line” $\lambda$1.644 $\mu$m “forbidden”
\FeII\ – all emissions being strong in the SW and eastern side of the
circumnuclear torus where the radio sources A, F, and H
are situated (See Table 2). Source F may be especially interesting since it
is the one being identified with the nuclear synchrotron radiation
jet and CO outflow discussed in Sects. 4.3.1 and 4.3.2.  

The outer much more massive torus is likely be an important reservoir of
matter, feeding the accretion disk of the supermassive black hole
(SMBH), but the inward matter transport process between the outer and
inner tori is still 
unclear. However, Lena et al. (\cite{len16}), in their study of the nuclear
outflow of ionized gas, also seem to “find kinematic
components that may trace gas which has lost angular momentum and is
slowly migrating toward the nucleus”. Here Fazeli et al. (\cite{faz19}) find
``a spiral-shaped molecular gas structure (at a scale <2\as) which
could indicate inward streaming motions''. Such kinematic features also
appear to be be visible in the ALMA data of Combes et al. and
hopefully will be discussed in their forthcoming study of gas
outflow. Here the observed existence of a narrow 
relativistic plasma jet outflow and an accompanying co-aligned
molecular gas outflow (jet) from the nuclear engine may be
important. The cold gas expulsion process may be an upscaled version
of the magnetohydrodynamics (MHD) generation of the bi-polar outflows
necessary to solve the angular 
momentum problem of star formation (e.g., Königl \&\ Pudritz
\cite{kon00}; Arce et al. \cite{arc07}), hence
supporting the accretion toward the central AGN. The launching
mechanism of the synchrotron jet as such may be a process closely
connected with the angular momentum loss required for mass accretion
onto and growth of the AGN (e.g., Blandford \&\ Begelman \cite{bla99};
Nayakshin \cite{nay14}; Garcia-Burillo et al. \cite{gar14}). For a
more initiated and detailed discussion, we here refer to Aalto et
al. (\cite{aal16}). 

The submillimeter line spectrum of the Seyfert galaxy
NGC\,1068 has been observed with {\it Herschel} SPIRE by Spinoglio et
al. (\cite{spi12}). While González-Alfonso et al. (\cite{gon10},
\cite{gon12}) find that their {\it Herschel} SPIRE and PACS observations of
multiple \hto\ lines in the ULIRGs (Ultra-Luminous Infrared Galaxies)
Mrk\,231, NGC\,4418, and Arp\,220 were best modeled
by FIR excitation (“radiation pumping”), the SPIRE observations of the
early-stage merger galaxy NGC\,6240 lead Meijerink et al. (\cite{mei13}) to
conclude that shock excitation must be the cause of the CO-SLED
(Spectral Line Energy Distribution). Submillimeter \hto\ lines have
been studied in detail in nine nuclei of actively star-forming
galaxies using the {\it  Herschel} HIFI, SPIRE, and PACS receivers (Liu
et al. \cite{liu17}). That sample did not include NGC\,1365 but there
are unpublished results which are available in the {\it Herschel}
SPIRE and PACS Archive, and they form partial basis of the present
paper. However, SPIRE does not yield spectral resolution capable of
resolving any velocity structure across the galaxy in the
signal. Since the 557 GHz (0.54 mm) \hto\ line is a sensitive
indicator of the existence of photodissociation regions (PDR) and
other physical and chemical processes, leading to enhanced \hto\
abundances such as outflows and shocks (e.g., Hjalmarson et
al. \cite{hja03}, \cite{hja05}; Gerin et al. \cite{ger16}), we have
used the {\it Odin} satellite to probe the central region of NGC\,1365
with high spectral resolution in this \hto\ line.

The outline of this paper reads as follows. After an Introduction,
containing a summary of the current knowledge of NGC\,1365 relevant
for the present communication, we present in Sect. 2 observations by
the {\it Herschel} and {\it Odin} space telescopes. In Sect. 3 our new results
from the {\it Herschel} PACS and SPIRE data and the complementary spectrally
resolved {\it Odin} data are reported together with support analyses of SEST
CO($3-2$) and VLA \HI\ data. In Sect. 4 our results concerning CO, \hto,
CH, \CHp, [\NII], and \HI\ are discussed in some detail. Our Summary and
conclusions are provided in Sect. 5. For the interested reader, we have
in Appendices B and C provided shortcut briefings on the interstellar
chemistry of \hto\ and CH, \CHp\ in prototype regions of our own Galaxy,
wherein the physical and chemical conditions and processes are more
accurately known, for comparison with observational results in
external galaxies.

\section{Observations}

\subsection{{\it Herschel} SPIRE}

The nuclear region of NGC\,1365 has been observed by {\it Herschel}
using the SPIRE Fourier-transform spectrometer in the
SpireSpectroPoint observing mode at two positions near the two maxima
of the central CO torus, NGC\,1365 NE: (J2000.0) 
$3^h33^m36\fs60, -36\degr08\arcmin20\farcs0$ ($+2\farcs8,+5\farcs4$),
and NGC\,1365 SW: (J2000.0) $3^h33^m35\fs90, -36\degr08\arcmin35\farcs0$
($-5\farcs7,-9\farcs6$) – the numbers in parentheses are equatorial
offsets from the optical nucleus. These observations, which are
unpublished, were made on 22 August 2010 with total integration times
of 3476 and 5640 seconds for the NE and SW positions,
respectively. The channel resolution of the spectrometer is 1.447 GHz
and the half-power beamwidth (HPBW) is roughly 35\as\ and 20\as\ for the 
low- and high-frequency parts of the spectra, respectively
(Swinyard et al. \cite{swi10}; cf. Fig. 1 of Spinoglio et
al. \cite{spi12}). We have retrieved these SPIRE observations from the
{\it Herschel} Science Archive.  

\subsection{{\it Odin}}
The observations of NGC\,1365 in the 557 GHz \hto\ line were performed
with the {\it Odin} space telescope in December 2016, March and December
2017, and February 2018. The satellite was commanded toward the
galaxy's optical nucleus located at $3^h33^m36\fs37,
-36\degr08\arcmin25\farcs4$ (J2000.0) (Lindblad
\cite{lin99}). Position-switching was performed in right ascension
with the OFF-position being displaced by $-600\arcsec$. The ON-source
integration time was 7.4, 51.4, 14.5, and 10.9 hours in December 2016,
March and December 2017, and February 2018, respectively. The backend
spectrometer was a 1050-channel AOS with a channel resolution of 1
MHz. The system temperature was about 3500 K. The HPBW beamwidth of
{\it  Odin} at this frequency is $2\farcm1$ and 
the beam efficency is 0.89 (Frisk et al. \cite{fri03}). The nominal pointing
accuracy of {\it Odin} is 10\as, but due to a systematic pointing
error, slightly different positions were actually observed in the
different periods. For the total averaged \hto\ profile, this error led
to an effective elliptical beam of 140\as $\times$ 160\as, centered at
an equatorial offset from the optical nucleus of (+10\as,
$-$2\as). This beam is also displayed in Fig. 1.

\section{Results}
\subsection{{\it Herschel} SPIRE and PACS results}

From the ratio of the PACS 160 \mum\ to 70 \mum\ images of NGC\,1365
presented in Fig. 1, we obtain a cold dust color temperature
distribution for the entire galaxy (see Fig. 2) in a manner similar to
Alonso-Herrero et al. (\cite{alo12}) who used the 100 \mum\ to 70 \mum\
PACS images for the central region.

\begin{figure}
  \resizebox{\hsize}{!}{\rotatebox{0}{\includegraphics{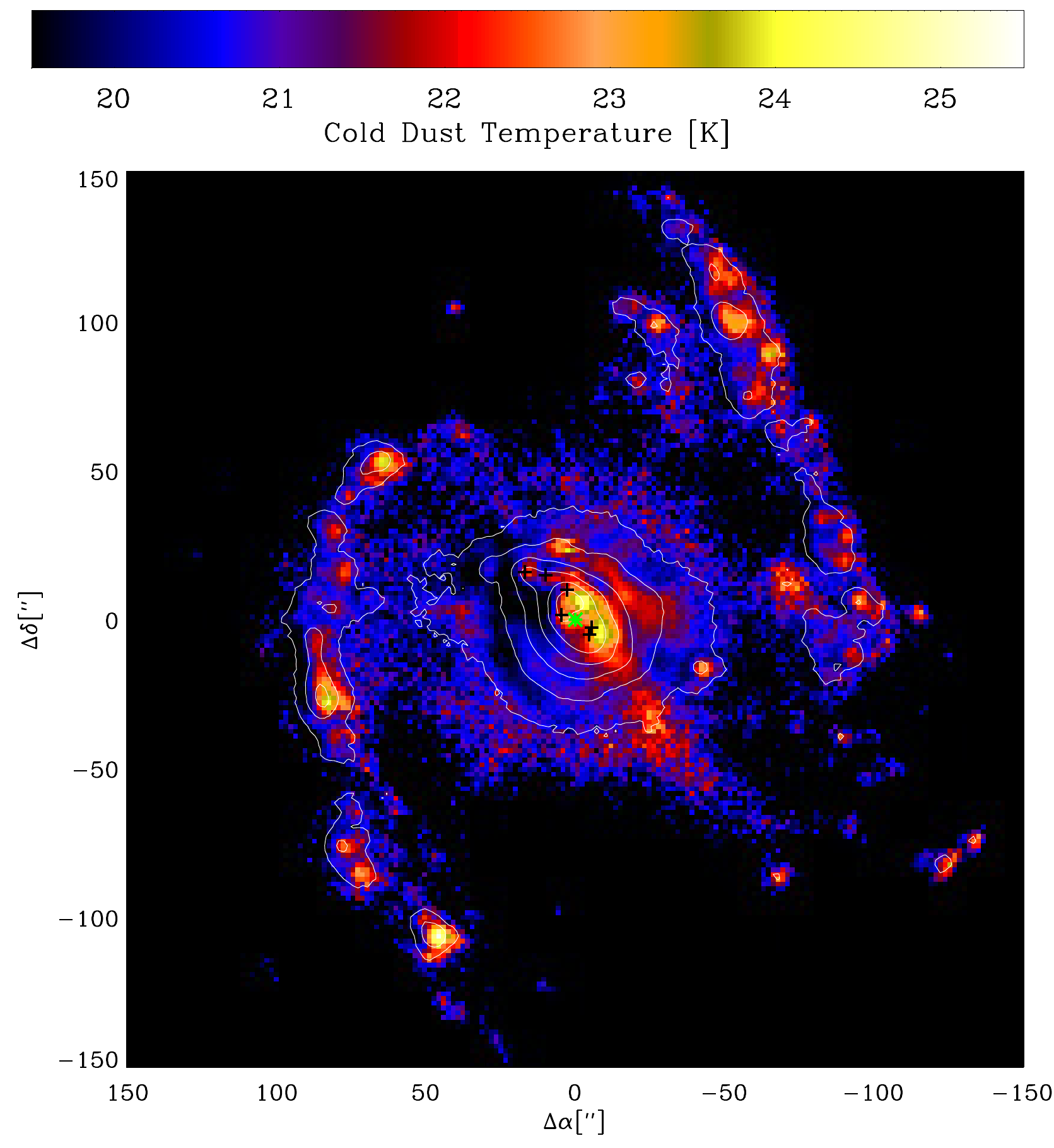}}} 
  \caption{Cold dust color temperature distribution in NGC\,1365,
    obtained from the ratio of the 160 \mum\ to 70 \mum\ {\it
      Herschel} PACS images in Fig. 1. The contours indicate the
    70 \mum\ intensities from Fig. 1. The crosses, from top to bottom (in
    decreasing declination), represent the ``hot spot'' \HII\ regions
    L4, L11, L12, L1, L3, and L2. The equatorial offsets are with
    respect to the optical nucleus, which is marked with a green
    asterisk.}          
  \label{2}
\end{figure}

The {\it Herschel} SPIRE apodized spectrum obtained toward the NGC\,1365 NE
torus position, corrected to a continuum source size of 14\as\ in a 40\as\ beam,
is shown in Fig. 3. A myriad of lines has been detected for the
following molecules: o-\hto\ and p-\hto, $^{12}$CO, CH, NH,
CH$^+$, and atoms: [C\,{\sc i}] and [N\,{\sc ii}]; their
properties are listed in Table 1. 

\begin{figure*}
  \resizebox{\hsize}{!}{\rotatebox{90}{\includegraphics{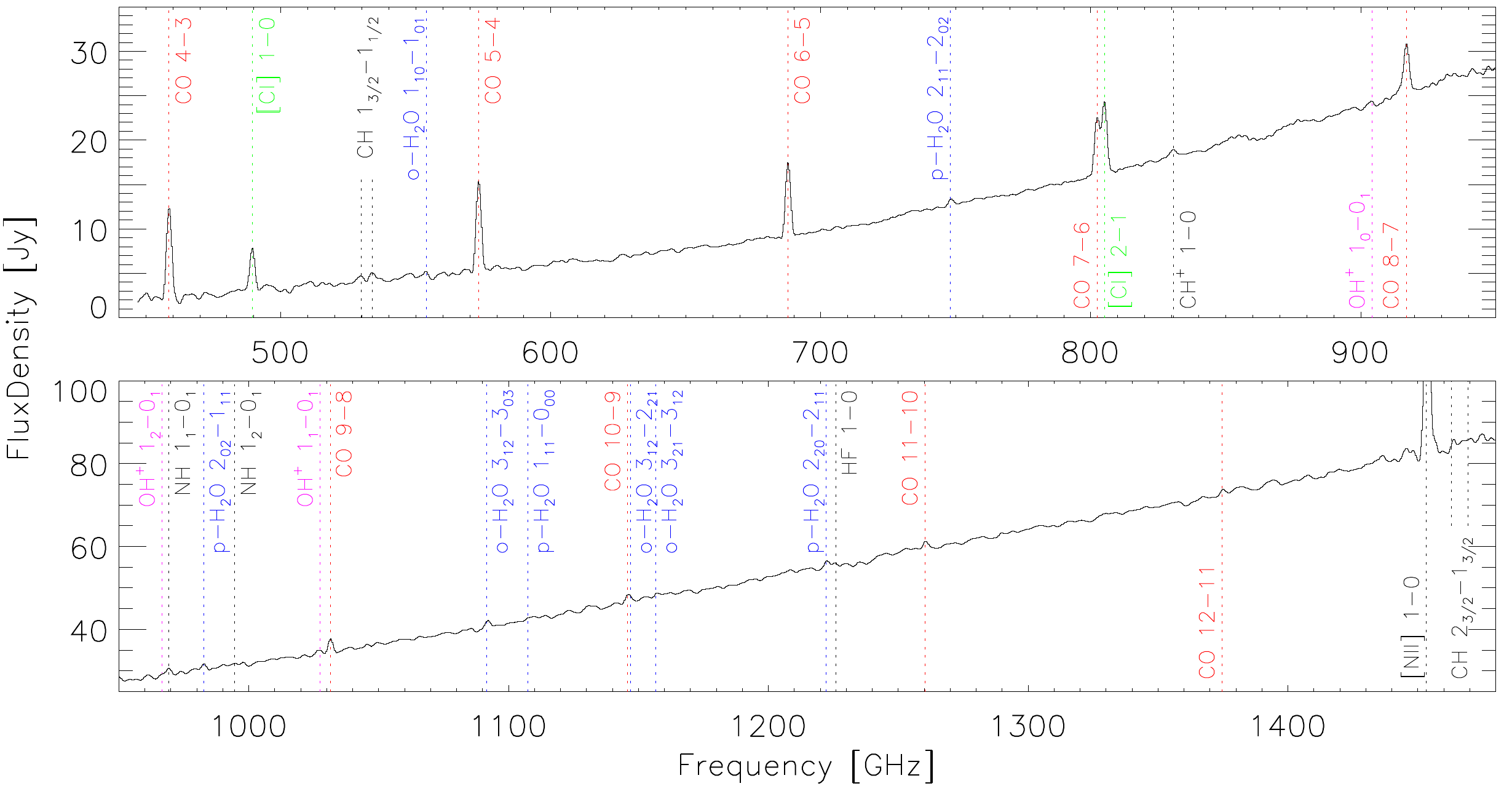}}}  
  \caption{{\it Herschel} SPIRE apodized spectrum toward NGC\,1365
    NE, corrected to a continuum source size of 14\as\ in a 40\as\
    antenna beam. The abcissa is the observed
      frequency. Semi-extended source calibration is applied according
      to the SPIRE
      Handbook (http://herschel.esac.esa.int/Docs/SPIRE/spire\_handbook.pdf).}
  \label{3}
\end{figure*}

\begin{table}[htb]
   \caption{Parameters of the detected SPIRE spectral lines (Fig. 3),
     corrected to a continuum source size of 14\as\ in 40\as\ antenna
     beam}
   \label{tab_spire_lines}
   \resizebox{\hsize}{!}{
   \begin{tabular}{l l c c c c}
   \hline\hline\noalign{\smallskip}
   Species  & Transition & Frequency & $E_{\rm u}$ & Flux & Flux Error \\
      &  & [GHz] & [K] &  \multicolumn{2}{c}{($10^{-17}$ W m$^{-2}$)} \\
   \hline\noalign{\smallskip}  
    $^{12}$CO & J = 4-3 & 461.041 & 55.3  &  19.11 & 0.38 \\
    $^{12}$CO & J = 5-4 & 576.268 & 83.0  &  19.27 & 0.38 \\
    $^{12}$CO & J = 6-5 & 691.473 & 116.2  &  15.87 & 0.38 \\
    $^{12}$CO & J = 7-6 & 806.652 & 154.9  &  11.95 & 0.38 \\
    $^{12}$CO & J = 8-7 & 921.800 & 199.1  &  10.16 & 0.38 \\
    $^{12}$CO & J = 9-8 & 1036.91 & 248.9  &  6.86 & 0.75 \\
    $^{12}$CO & J = 10-9 & 1151.99 & 304.2  &  3.57\tablefootmark{a} & 0.79 \\
    $^{12}$CO & J = 11-10 & 1267.01 & 365.0  &  3.50 & 0.75 \\
    $^{12}$CO & J = 12-11 & 1382.00 & 431.3  &  (1.85) & 0.75 \\
    $^{12}$CO & J = 13-12 & 1496.92 & 503.1 &   &  0.75 \\
    \noalign{\smallskip}
    o-H$_2$O & 1$_{10}$-1$_{01}$ & 556.936 & 60.9  &  1.10 & 0.38 \\
    o-H$_2$O & 3$_{12}$-3$_{03}$ & 1097.36 & 249.4  &  2.88 & 0.75 \\
    o-H$_2$O & 3$_{12}$-2$_{21}$ & 1153.13 & 249.4  &  (1.18)\tablefootmark{a} &
0.79 \\
    o-H$_2$O & 3$_{21}$-3$_{12}$ & 1162.91 & 305.2  &  2.21 & 0.75 \\
    o-H$_2$O & 5$_{23}$-5$_{14}$ & 1410.62 & 642.4  &   &  0.75\\
    p-H$_2$O & 2$_{11}$-2$_{02}$ & 752.033 & 136.9  &  1.49 & 0.38 \\
    p-H$_2$O & 4$_{22}$-3$_{31}$ & 916.172 & 454.3  &  (0.66) & 0.38 \\
    p-H$_2$O & 2$_{02}$-1$_{11}$ & 987.927 & 100.8  &  3.27 & 0.75 \\
    p-H$_2$O & 1$_{11}$-0$_{00}$ & 1113.34 & 53.4  &  (0.92) & 0.75 \\
    p-H$_2$O & 4$_{22}$-4$_{13}$ & 1207.64 & 454.3  &   &  0.75 \\
    p-H$_2$O & 2$_{20}$-2$_{11}$ & 1228.79 & 195.9  &  2.86 & 0.76 \\
    \noalign{\smallskip}
    CH & 1$_{3/2}$-1$_{1/2}$ & 532.7 & 25.7   &  1.19 & 0.38 \\
    CH & 1$_{3/2}$-1$_{1/2}$ & 536.8 & 25.8  &  1.81 & 0.38 \\
    CH & 2$_{3/2}$-1$_{3/2}$ & 1470.7 & 96.3  &  2.53 & 0.75 \\
    CH & 2$_{3/2}$-1$_{3/2}$ & 1477.3 & 96.7  &  2.23 & 0.75 \\
    \noalign{\smallskip}
    NH & 1$_2$-0$_1$ & 974.48 & 46.8  &  2.82 & 0.76 \\
    NH & 1$_1$-0$_1$ & 999.97 & 48.0  &  (1.06) & 0.76 \\
    \noalign{\smallskip}
    CH$^+$ & 1-0 & 835.137 & 40.1  &  1.47 & 0.38 \\
    \noalign{\smallskip}
    $[$CI$]$ & $^3$P$_1$-$^3$P$_0$ & 492.161 & 23.6  &  9.06 & 0.38 \\
    $[$CI$]$ & $^3$P$_2$-$^3$P$_1$ & 809.342 & 62.5  &  15.06 & 0.38 \\
    $[$NII$]$ & $^3$P$_1$-$^3$P$_0$ & 1461.128 & 70.1  &  87.60 & 0.75 \\
     \hline\noalign{\smallskip}
   \end{tabular}
    }
    \tablefoottext{a}{Blended CO and H$_2$O}
\end{table}

We have made map images of the combined NGC\,1365 torus NE and SW {\it
  Herschel} SPIRE observations using the 557 GHz o-\hto\ 
and 752 GHz p-\hto\ lines, as well as the $J=6-5$ and $9-8$ CO lines,
537 GHz CH and 835 GHz CH$^{+}$ lines. These images are shown in
Fig. 4 where they are 
superimposed upon contours of the PACS 70 $\mu$m observations (also
seen in Fig. 1), which show the distribution of warm dust. We note that
the NE torus region is much stronger in warm dust emission than
the SW torus peak. The ``hot spot'' \HII\ regions L1, L2, L3, L4, L11,
and L12 (Alloin et al. \cite{all81}) are also indicated in these
figures as is the position of the optical nucleus. Their equatorial
offsets from the optical nucleus are listed in Table 2, as are the
offsets of the compact radio sources detected by Sandqvist et
al. (\cite{san95}), and the co-located compact MidIR sources, detected
and studied in detail by Galliano et al. (\cite{gal05}, \cite{gal08}
and \cite{gal12}) using adaptive optics on the ESO 3.6 m telescope and VLT. The
co-location with CO($2-1$) intensity peaks in the 2\as\
resolution SMA aperture synthesis mapping by Sakamoto et
al. (\cite{sak07}) is also convincingly illustrated in their Figs. 9 and
10, and is confirmed in the ALMA CO($3-2$) higher resolution mapping
by Combes et al. (\cite{com19}).

The SEST maximum entropy method (MEM) map of the $J=3-2$
CO line with 5\as\ resolution (Sandqvist \cite{san99}) showed that a
central molecular torus exists around the nucleus with the NE and SW
peaks being of similar magnitudes. This is in clear contrast to the
appearance of the 557 GHz o-\hto\ and 752 GHz p-\hto\ lines where the
NE torus peak is considerably stronger than the SW peak, as can be
seen in Fig. 4. This is also illustrated in Fig. 6 
where the SPIRE profiles for these lines are presented. Since the
torus peak positions NE and SW are separated by only 15\as while the
beam size is 35\as, Fig. 6 in fact tells us that the \hto\ emissions are
very weak in the SW torus peak.

\begin{figure*}[ht]
\includegraphics[angle=0, width=.33\textwidth]{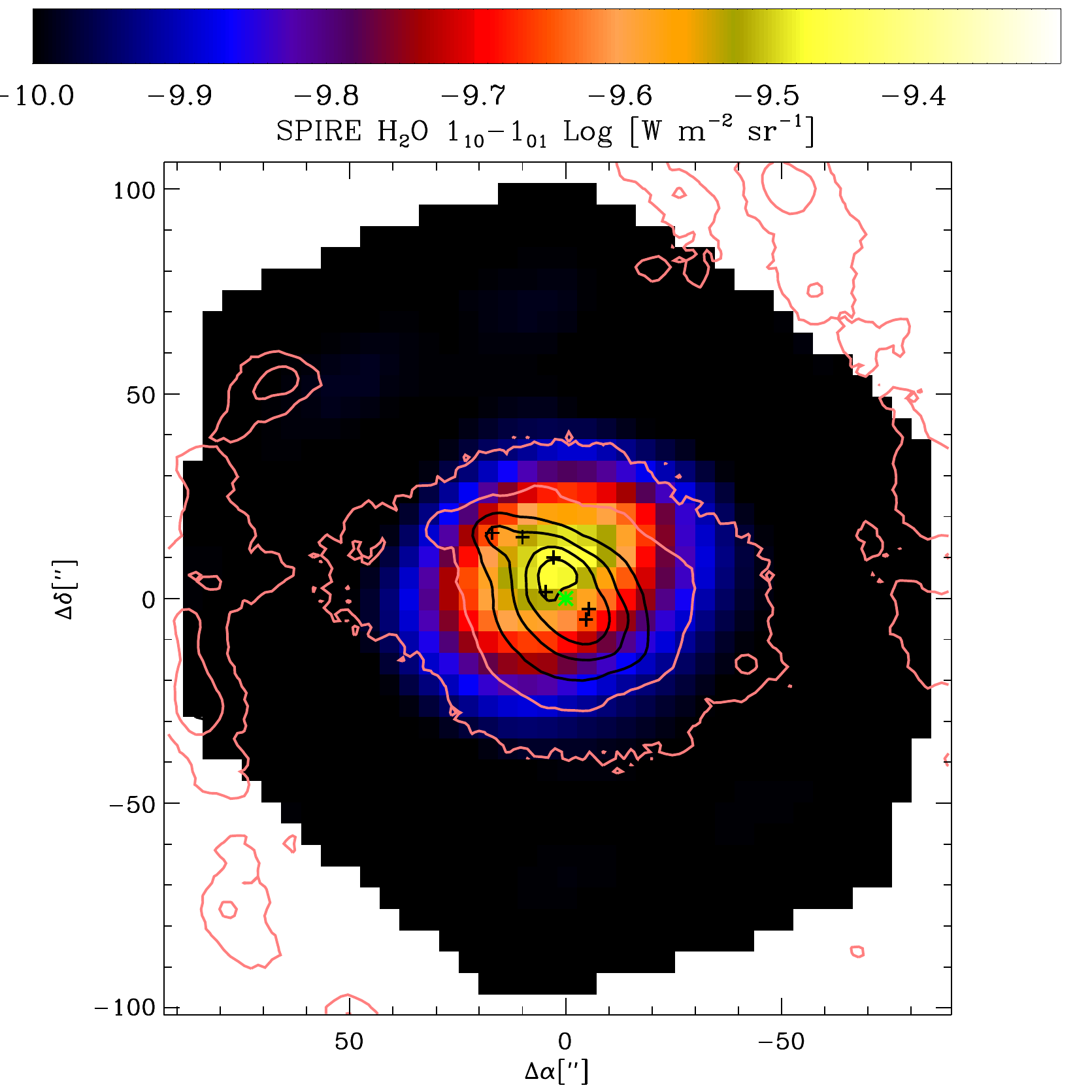}
\includegraphics[angle=0, width=.33\textwidth]{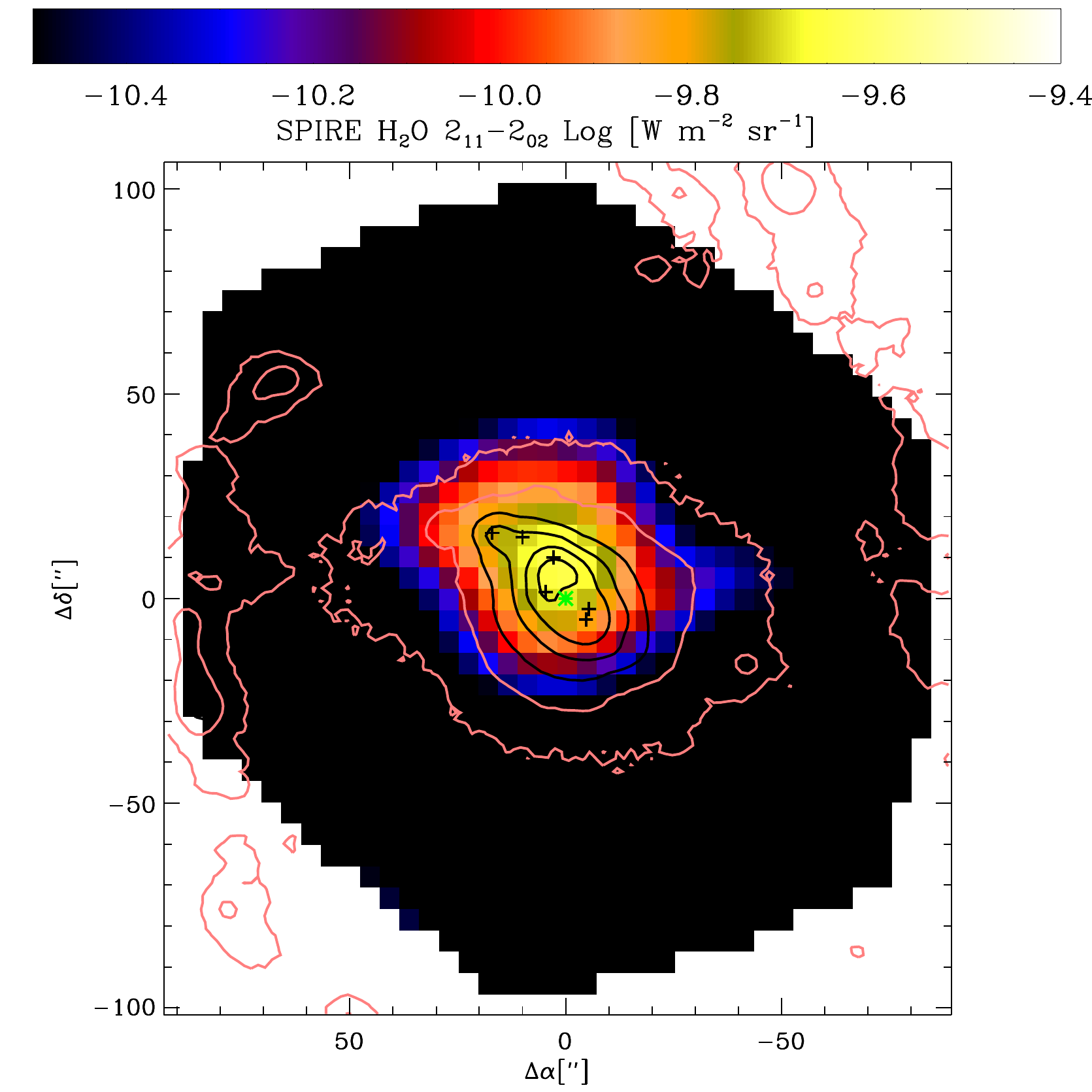} 
\includegraphics[angle=0, width=.33\textwidth]{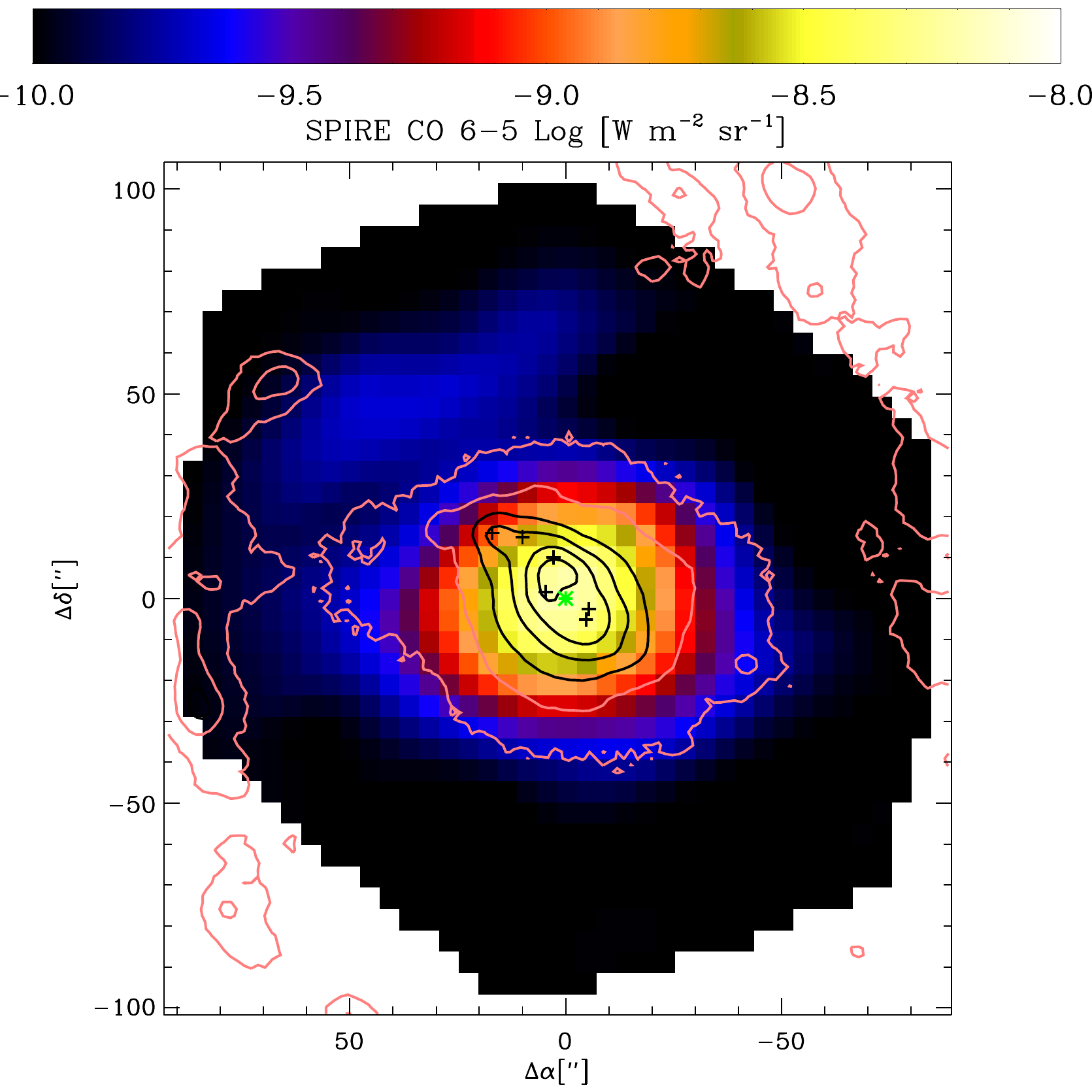}
\includegraphics[angle=0, width=.33\textwidth]{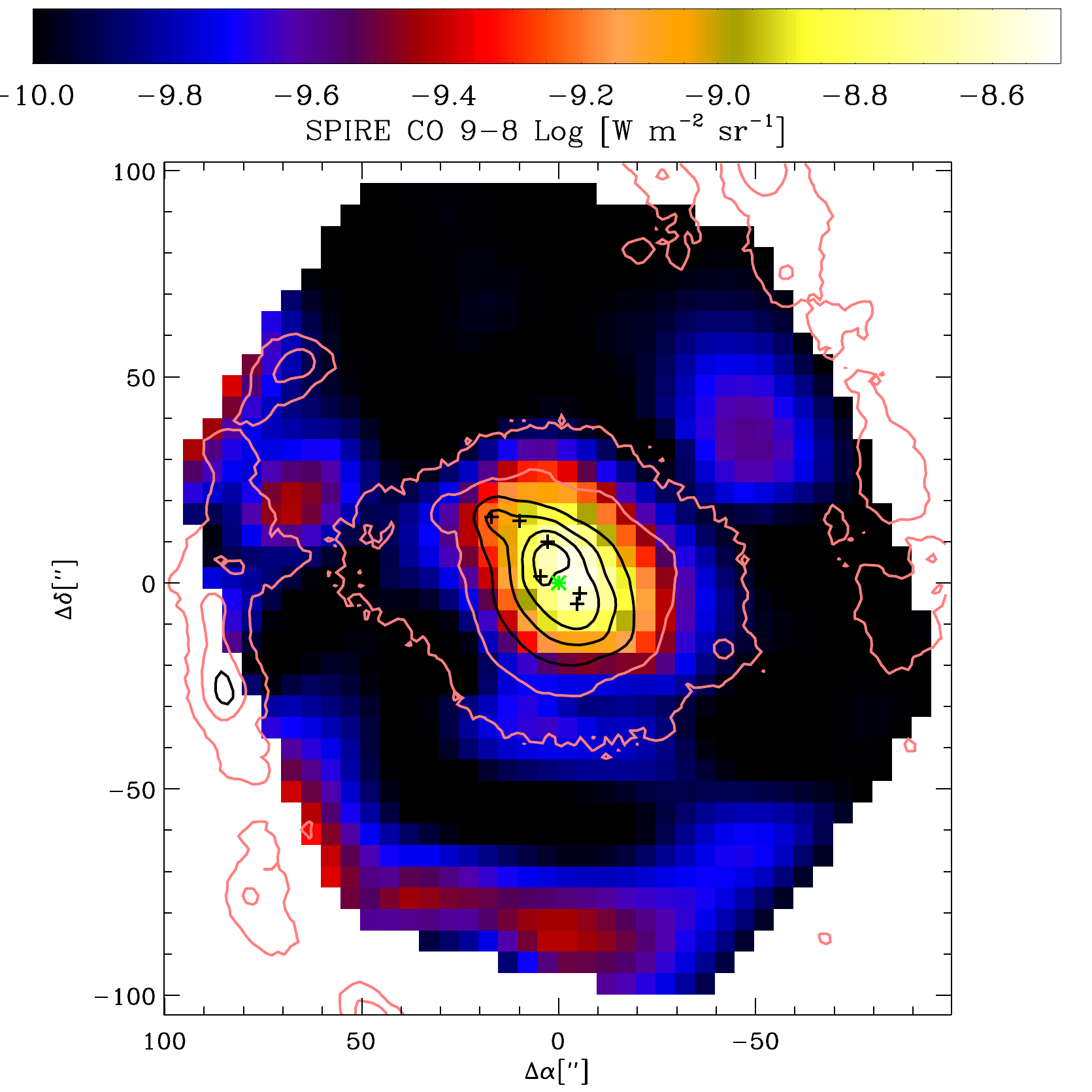} 
\includegraphics[angle=0, width=.33\textwidth]{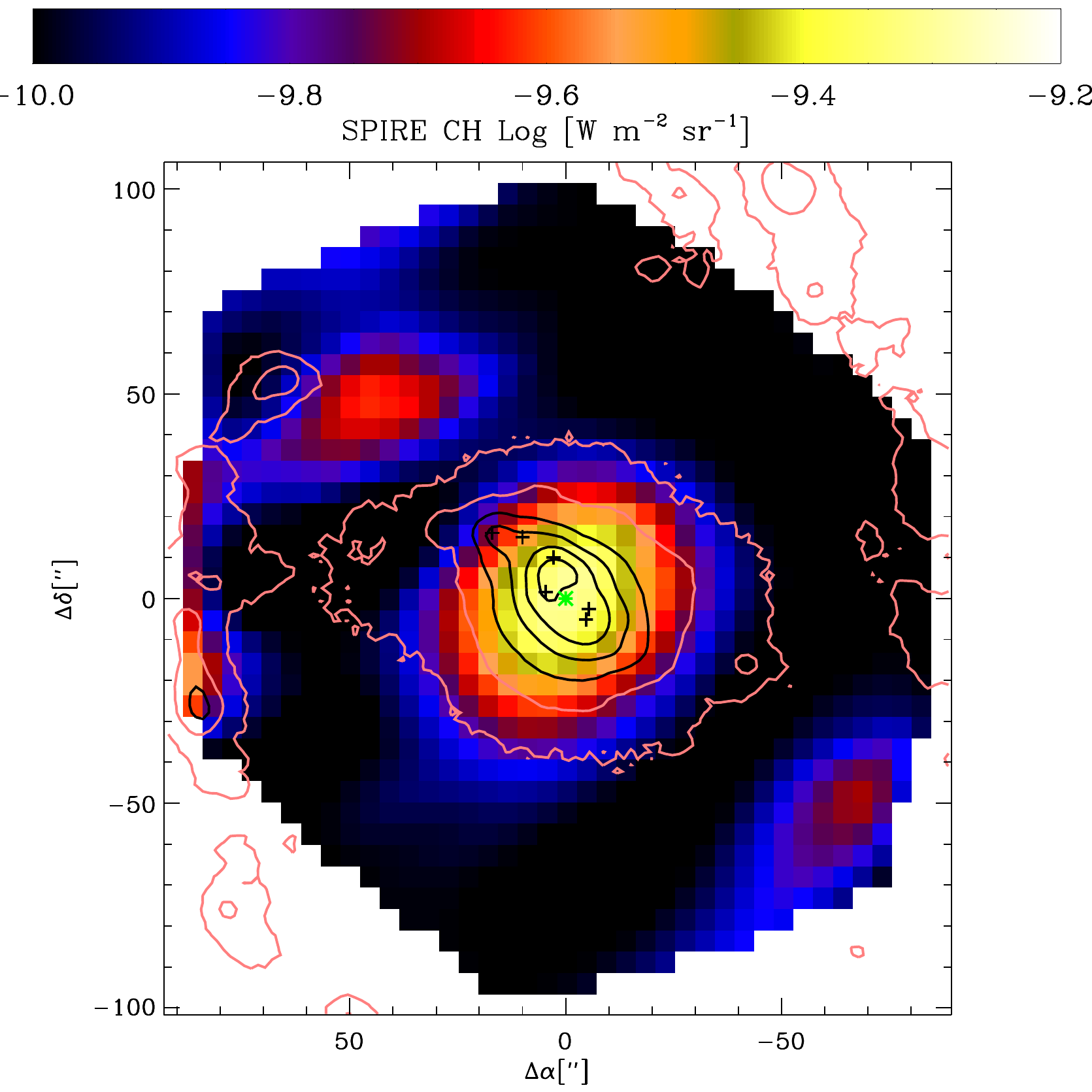}
\includegraphics[angle=0, width=.33\textwidth]{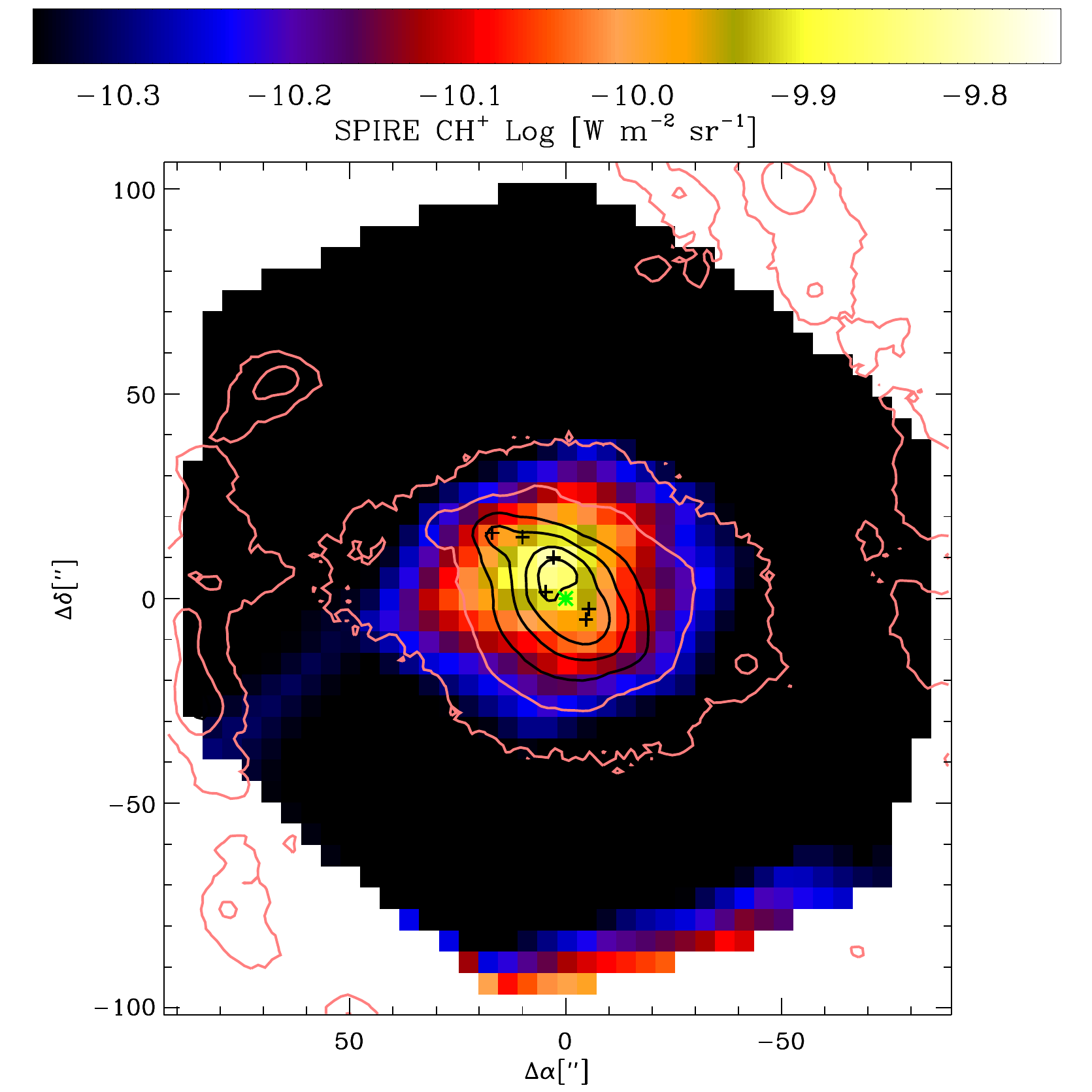} 
\caption{SPIRE maps of 557 GHz o-\hto\ ($\approx35\as$), 752 GHz
  p-\hto\ ($\approx35\as$), $J=6-5$ CO\ ($\approx30\as$), $J=9-8$ CO\
  ($\approx20\as$), 537 GHz CH\ ($\approx35\as$), and 835 GHz\ \CHp
  ($\approx35\as$), the beamwidths are in parentheses. The contours
  are from the PACS 70 $ \mu$m observations seen 
  in Fig. 1 and reflect the distibution of warm dust, the maximum
    being near the NE torus component. The crosses,
  from top to bottom (in decreasing declination), represent the ``hot
  spot'' \HII\ regions L4, L11, L12, L1, L3, and L2, the latter
    two being near the SW torus component. The equatorial
  offsets are with respect to the optical nucleus which is marked
  with a green asterisk.}      
  \label{4}
\end{figure*}

\subsection{{\it Odin} high spectral resolution \hto\ results}
The {\it Odin} observations of NGC\,1365 in the 557 GHz \hto\ line for the four
different periods were integrated yielding a resultant profile with a
total ON-source integration time of 81 hours. In Fig. 5 ``{\it Top}'', we
present the resultant profile with a channel resolution of 0.54
\kms\ and a linear baseline subtracted. In the ``{\it Upper Middle}'' part of
Fig. 5 we binned the {\it  Odin} profile  
to a velocity resolution of 5 \kms\ and compare it with several
CO($3-2$) profiles – from mapping observations obtained with SEST
(Sandqvist \cite{san99}). The blue profile is the total SEST
{\it Odin}-beam-convolved CO($3-2$) profile. Two 5\as -resolution SEST
statistical image 
deconvolution (SID) CO($3-2$) line profiles (see Sect. 3.4), observed
close to the NE torus peak, are also presented in this
panel. The green CO profile is in the direction of
$(\Delta\alpha,\Delta\delta)=(+6\as,+8\as)$; the red CO profile is in
the direction of 
$(\Delta\alpha,\Delta\delta)=(+7\as,+3\as)$. In the ``{\it Lower
    Middle}'' panel of Fig. 5, we binned the {\it Odin} \hto\
  profile to a velocity resolution of 30 \kms\ and indicate the 1$\sigma$
  rms noise level of 0.9 mK as a red line. Two emission features are
  above the 3$\sigma$ level, one near 1450 \kms, the other near 1570
  \kms. There may also be an absorption feature with a 2$\sigma$ intensity
  near 1500 \kms. It is interesting that near these three velocities
  there exist corresponding CO($3-2$) components in the green profile,
  presented in the {\it Upper Middle} panel of Fig. 5, a profile
  observed near the radio contiunuum source G (see Table 2). The
intensity of the SPIRE data from Fig. 3, convolved to an {\it
  Odin} beam, has an intensity similar to the {\it Odin} data convolved
to the 1.4 GHz SPIRE resolution. This is shown in Fig. 5 ``{\it
  Bottom}'' where these two profiles have been superimposed. From the
evidence presented in Fig. 5, we draw the conclusion that {\it Odin}
has indeed made a marginal detection of \hto\ in NGC\,1365 – two \hto\
emission features near 1450 and 1570 \kms\ and a possible absorption
feature near 1500 \kms.

\subsection{VLA \HI\ observations of the central regions of NGC\,1365}
We have used the VLA \HI\ data cube of J\"ors\"ater  \&\ van Moorsel
(\cite{jor95}) to study the central region of NGC\,1365 in
greater detail, using both the emission and absorption data. The total
integrated line intensity map is shown in Fig. 7. Emission is dominant at the
\HII\ region L4, absorption at the compact radio sources D, E
and G, but a marked absorption also is seen at F. Four velocity-integrated
maps, spaced at 100 \kms, are presented in Fig. 8. The dominant
\HI\ emission is in the ($-$228 to $-$104) \kms\ velocity interval and
originates in the L4 region at the upper left part of that figure near
offsets (+17\as,+16\as). The major
\HI\ absorption is seen in the velocity interval ($-$104  to 0) \kms\
and comes from the D, E, G region near offsets (+1\as,+7\as), as can
be seen in Fig. 8. And, finally, Fig. 8 shows that the major
absorption region in the velocity interval of (0 to +104) \kms\ comes
from the F region at offsets near (+3\as,$-$3\as),  and in the
velocity interval of (+104 to +228) \kms\ comes from the region
around sources A, L2 and L3, in the lower right of that \HI\ figure. One may
compare these features with the three velocity components seen in the
{\it Odin} \hto\ profile presented in Fig. 5 ``{\it Upper Middle}'', namely the
two emission peaks near 1460 
($-$150) and 1560 ($-$50) \kms\ and the absorption component at 1510
($-100$) \kms, where the velocities in brackets are
relative to the systemic velocity, all three originating northeast of
the nucleus. The \HI\ position-velocity (P-V) maps 
are presented in Appendix A. In Fig. A.1 (for $y=0$\as),
the map is along the major axis of NGC\,1365, through L4 (wide
emission), G (strong absorption), the nucleus (empty), and finally A
and L3 (absorption). The \HI\ map in Fig. A.1 (for $y=+2.5$\as) is
parallel to the major axis but $2\farcs5$ southeast of it, thus
passing through F – the absorption close to +50 \kms\ appears at the
position of F.

\subsection{Statistical image deconvolution analysis of our $J=3-2$ CO
  SEST observations} 
The SEST CO($3-2$) observations of NGC\,1365, published by Sandqvist
(\cite{san99}), were performed over an
$\approx 120\as \times 60\as$ region centered on the optical nucleus. A
grid spacing of 10\as\ was used for the outer parts of the region whereas
a grid spacing of 5\as\ was used for the inner $\approx 70\as \times 40\as$
region. We have applied a statistical image deconvolution (SID)
analysis (Rydbeck \cite{ryd08}) to the SEST $J=3-2$ CO observations to
obtain high resolution (5\as) maps at different velocities and P-V
maps. These maps will help us to identify the 
origins of the different {\it Odin} \hto\ components. The intensity
unit in the maps is antenna temperature. The main
beam efficiency of 0.26 has been applied in the P-V maps
to yield main beam brightness temperature. Examples of CO
integrated intensity maps for the total and selected
velocity intervals are shown in Figs. 7 and 8; examples of CO
P-V maps are shown in Fig. A.1.

A comparison of the integrated intensity \HI\ absorption and CO
emission maps, seen in 
Fig. 7, shows that the main features in both species are oriented in
a northeast-southwest direction along the circumnuclear torus. There
is an additional somewhat enticing feature seen in the \HI\ map, namely,
an approximately 20\as -long \HI\ absorption ridge emanating from the
nucleus in a south-southeast direction. This \HI\ ridge may have a CO
counterpart as shown by the low-level extension in the same direction
in the CO map. A careful study of the \HI\ and CO maps in Fig. 8
shows that this ridge has a very wide velocity dispersion. This ridge
will be discussed further in Sect. 4.3.1.

\begin{figure}
\resizebox{\hsize}{!}{\rotatebox{0}{\includegraphics{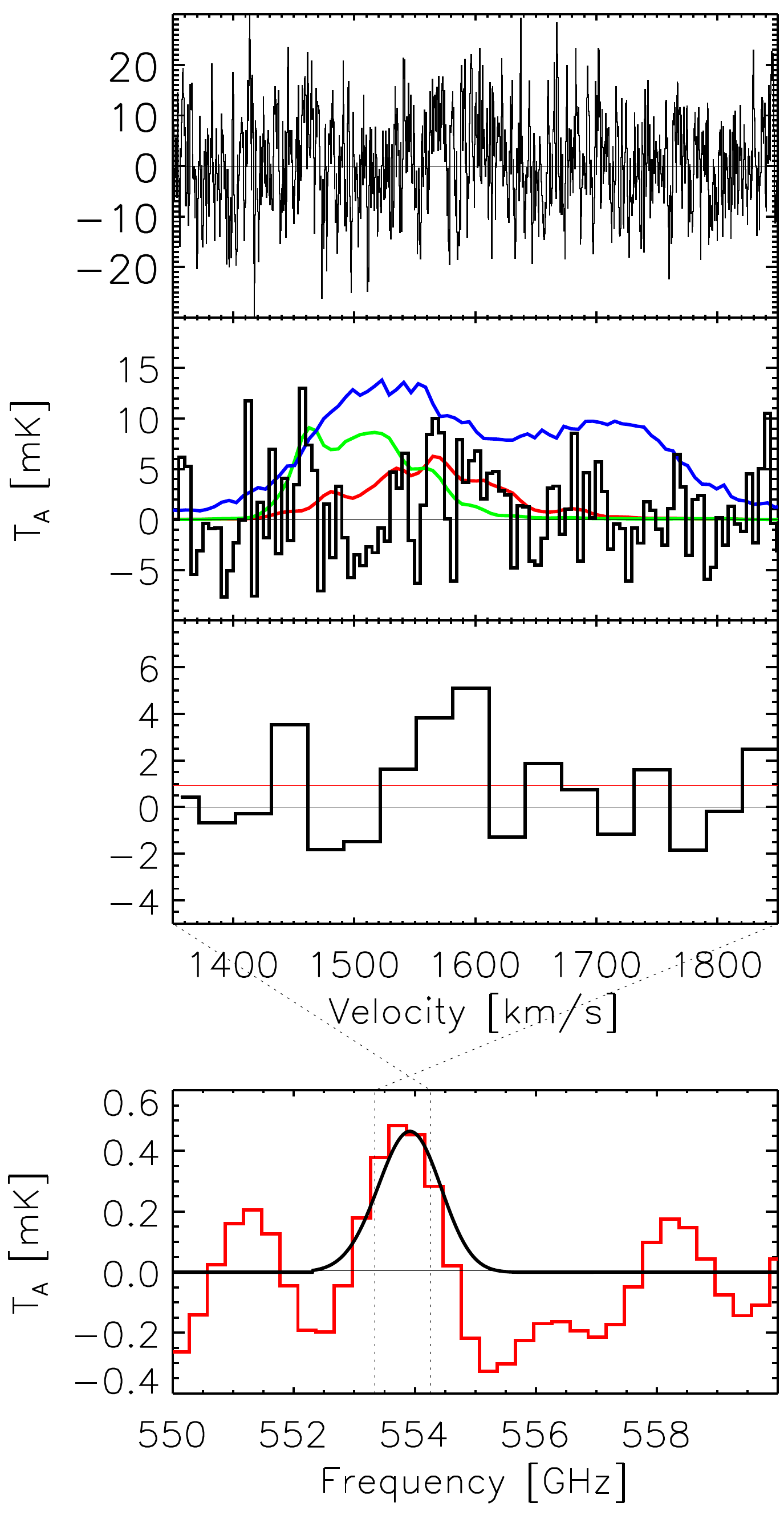}}}  
  \caption{{\it Odin} 557 GHz \hto\ profile observed
    toward NGC\,1365. {\it Top}: The unsmoothed profile with a
    resolution of 0.54 \kms. {\it Upper Middle}: The {\it Odin} \hto\ profile
    binned to a resolution of 5.0 \kms\ (black); the  SEST
    Odin-beam-convolved CO($3-2$) profile (blue); the green and red SID CO
    profiles are observed near the NE torus component (Sect. 3.2). The CO
    profiles should be multiplied by a factor of 100. {\it Lower
      Middle}: The {\it Odin} \hto\ profile binned to a resolution of
    30 \kms. The 1$\sigma$ rms noise level of 0.9 mK is indicated in
    red. {\it Bottom}: The {\it Odin} \hto\ profile convolved to a
  1.4 GHz (SPIRE) resolution (black); the SPIRE data convolved to an
  {\it Odin}-beam resolution (red).}    
  \label{5}
\end{figure}

\begin{figure}
  \resizebox{\hsize}{!}{\rotatebox{0}{\includegraphics{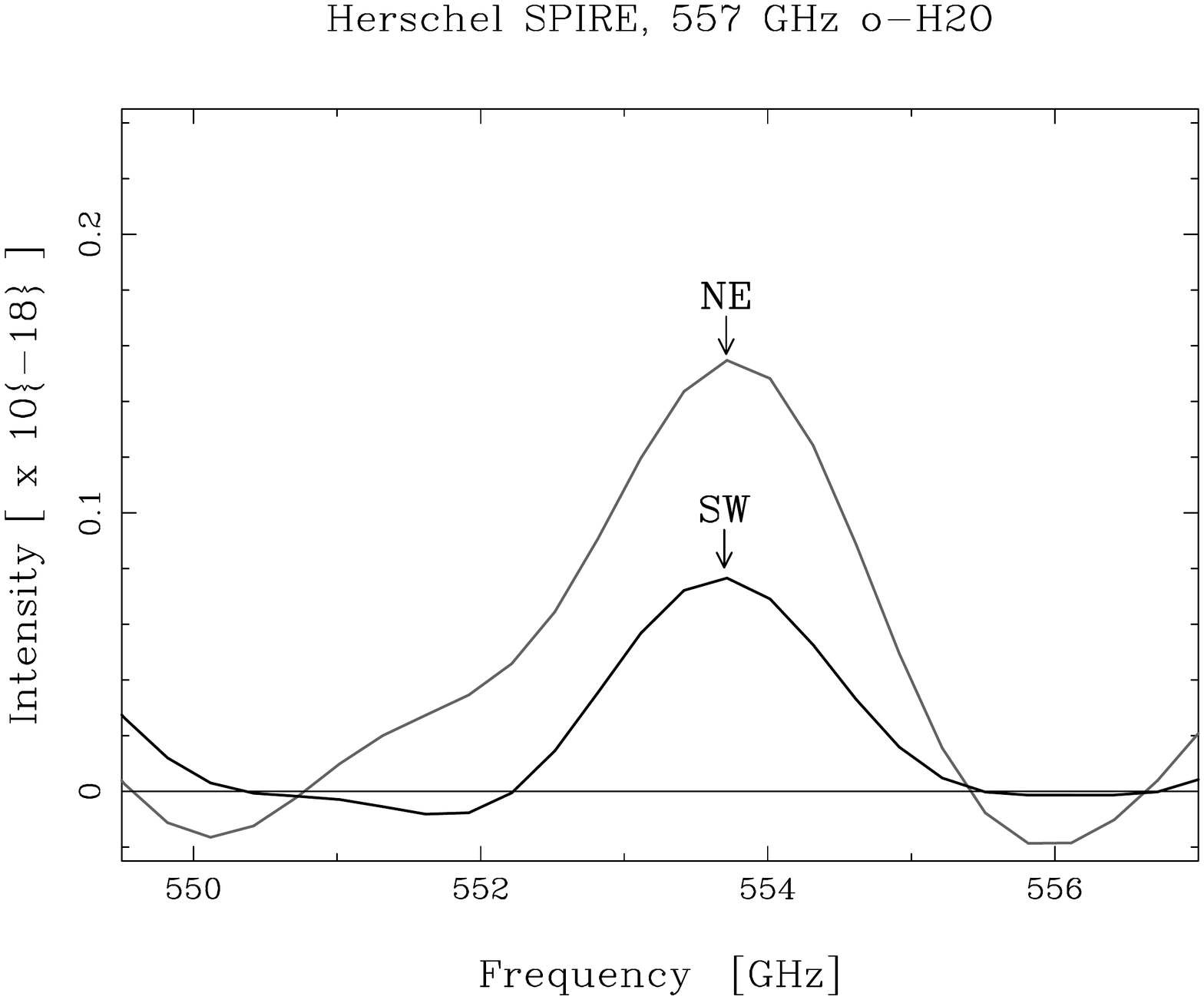}}
  \rotatebox{0}{\includegraphics{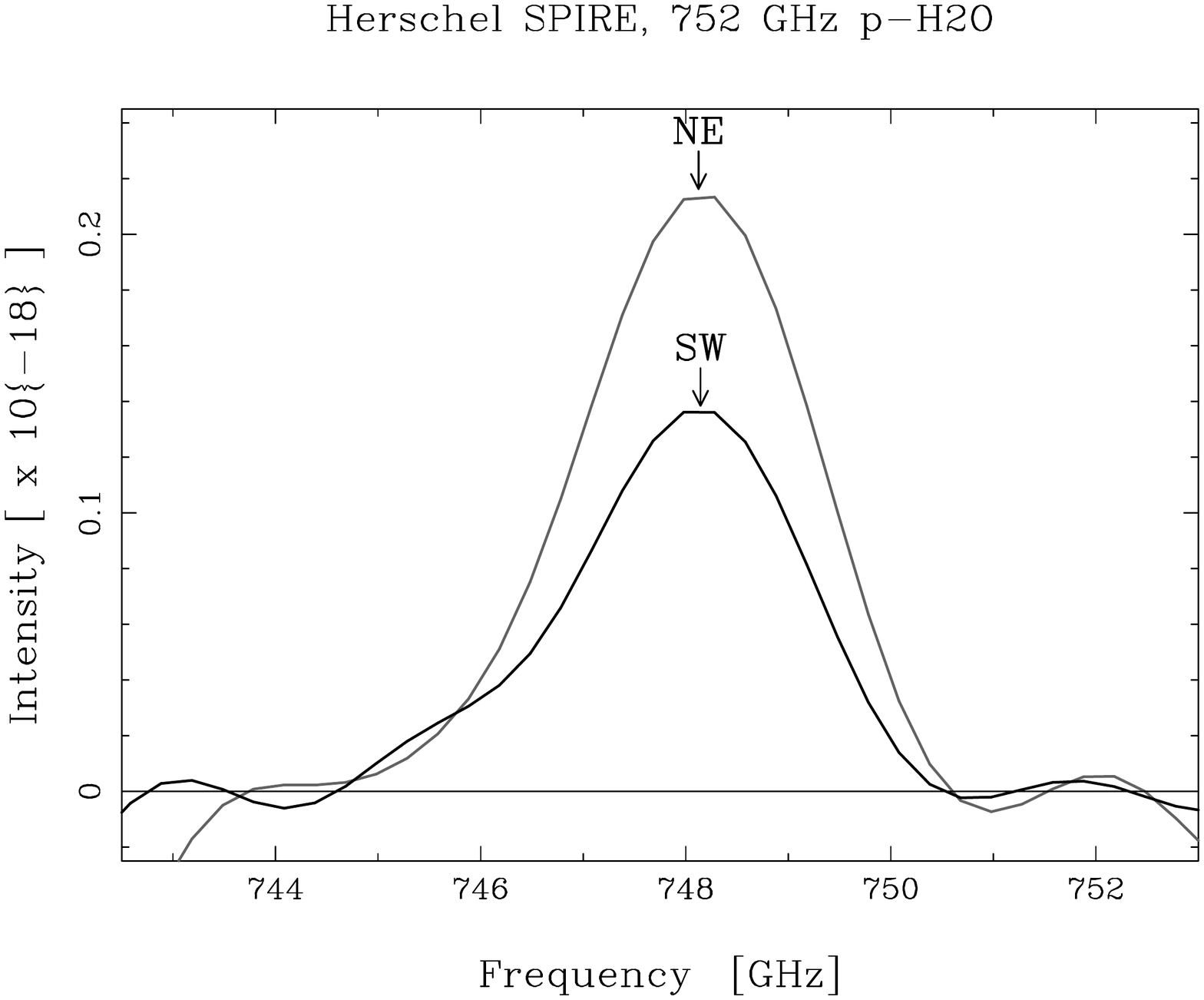}}
  }
  \caption{{\it Herschel} SPIRE 557-GHz o-\hto\ ({\it left}) and
    752-GHz p-\hto\ ({\it right})
    profiles observed toward the NGC\,1365 NE ({\it gray}) and SW
    ({\it black}) torus positions. Intensity units are  W m$^{-2}$ Hz$^{-1}$
    sr$^{-1}$. The HPBW beamwidth is 35\as\ and the separation between
    the NE and SW positions is 15\as.}       
  \label{6}
\end{figure}

\begin{table}
\caption{Equatorial offsets from the optical nucleus ($3^h33^m36\fs37,
-36\degr08\arcmin25\farcs4$ ) of hot-spot \HII\  regions (Alloin et
al. \cite{all81}) and compact radio sources (Sandqvist et al. \cite{san95}).}
\begin{flushleft}
\begin{tabular}{lllll}
  \hline\hline\noalign{\smallskip}

  Sources & $\Delta$RA  & $\Delta$Dec  &  Line  &  Velocity (Width)\tablefootmark{a}\\ 
          &  (\as)     &  (\as)   &       &     (\kms)  \\

\hline\noalign{\smallskip}

  L4  &  +17   &  +16  \\
  L11 &  +10   &  +15  \\
  L12 &   +2   &  +11   \\
  L1  &   +4   &  +2   \\
  L3  &   $-$6 &  $-$3 \\
  L2  &   $-$5 &  $-$5 \\ \\

  D (= M4)   &    0   &  +7   &  CO($2-1$)    &  +1560 (70)  \\
             &        &       &  Br${\gamma}$ &  +1555 (145) \\
             &        &       &  \htwo        &  +1560 (?)   \\
  E (= M5)   &   +3   &  +10  &  CO($2-1$)    &  +1540 (60)  \\
             &        &       &  Br${\gamma}$ &  +1520 (145) \\
             &        &       &  \htwo        &  +1530 (?)   \\
  G (= M6)   &   +5   &  +7   &  CO($2-1$)    &  +1480 (80)  \\
             &        &       &  Br${\gamma}$ &  +1475 (200) \\
             &        &       &  \htwo        &  +1470 (?)   \\
  H (= M8?)  &   +5   &  +3   &  CO($2-1$)    &  +1510 (70)  \\
  F\tablefootmark{b}          &   +3   &  $-$3 &  CO($2-1$) &
                                                              +1590
  (90)
  \\ 
  A          &   $-$4 &  $-$3  &  CO($2-1$) &  $\approx 1730-1750$ \\
                                                                  
  \noalign{\smallskip}\hline\end{tabular}
 \tablefoottext{a}{Velocities and line widths are from Sakamoto et
   al. (\cite{sak07}) and Galliano et al. (\cite{gal12}).} 
 \tablefoottext{b}{The steep spectrum of this compact radio source
   indicates emission from a synchrotron radiation outflow jet
   originating in the nuclear engine (Sandqvist et al. \cite{san95}; Lindblad
   \cite{lin99}).} 
 \end{flushleft}

\end{table}

\begin{figure*}[htbp]
 \centering
 \resizebox{17cm}{!}{
 \resizebox{!}{5cm}{{\includegraphics{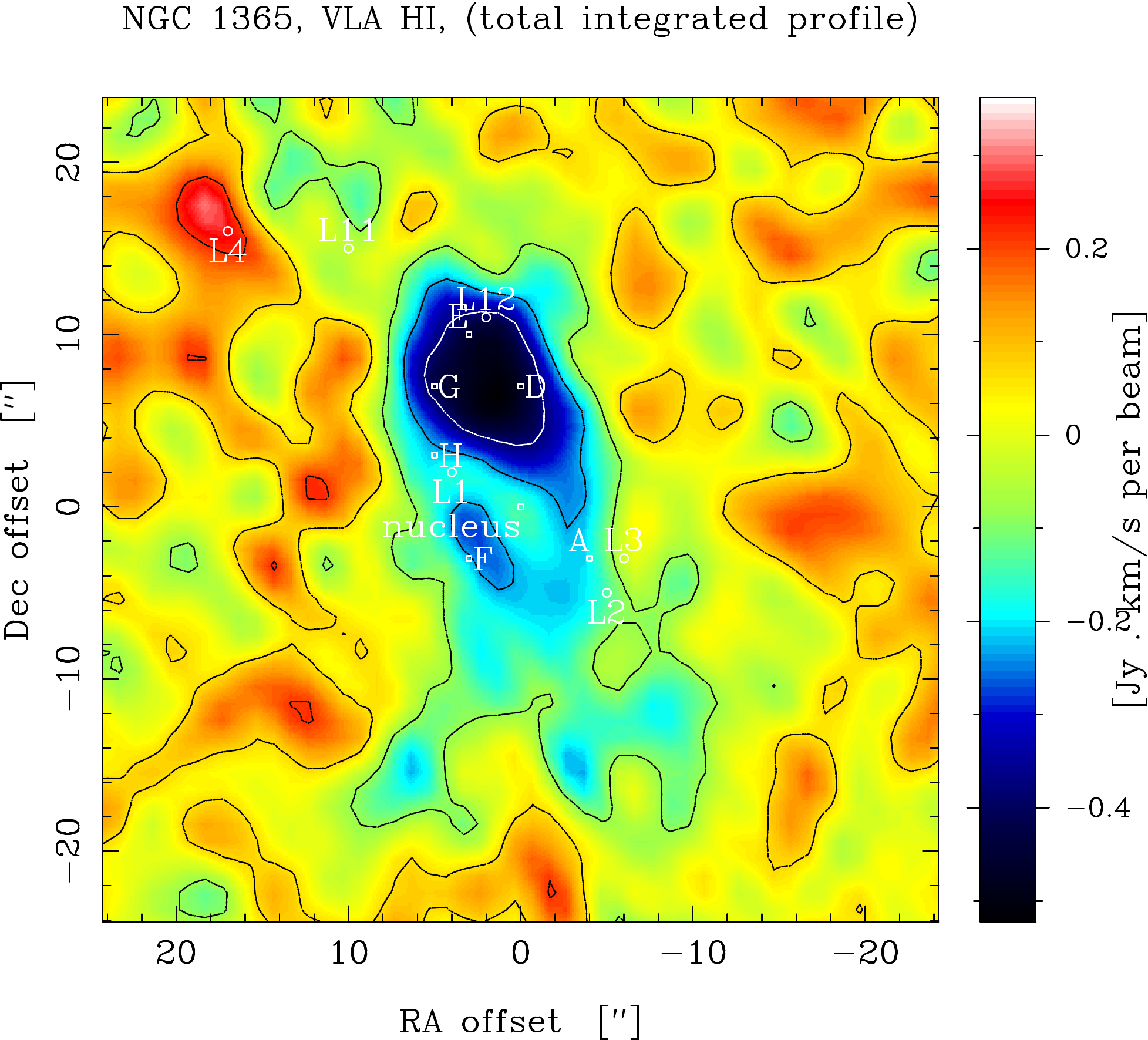}}}
 \resizebox{!}{5cm}{{\includegraphics{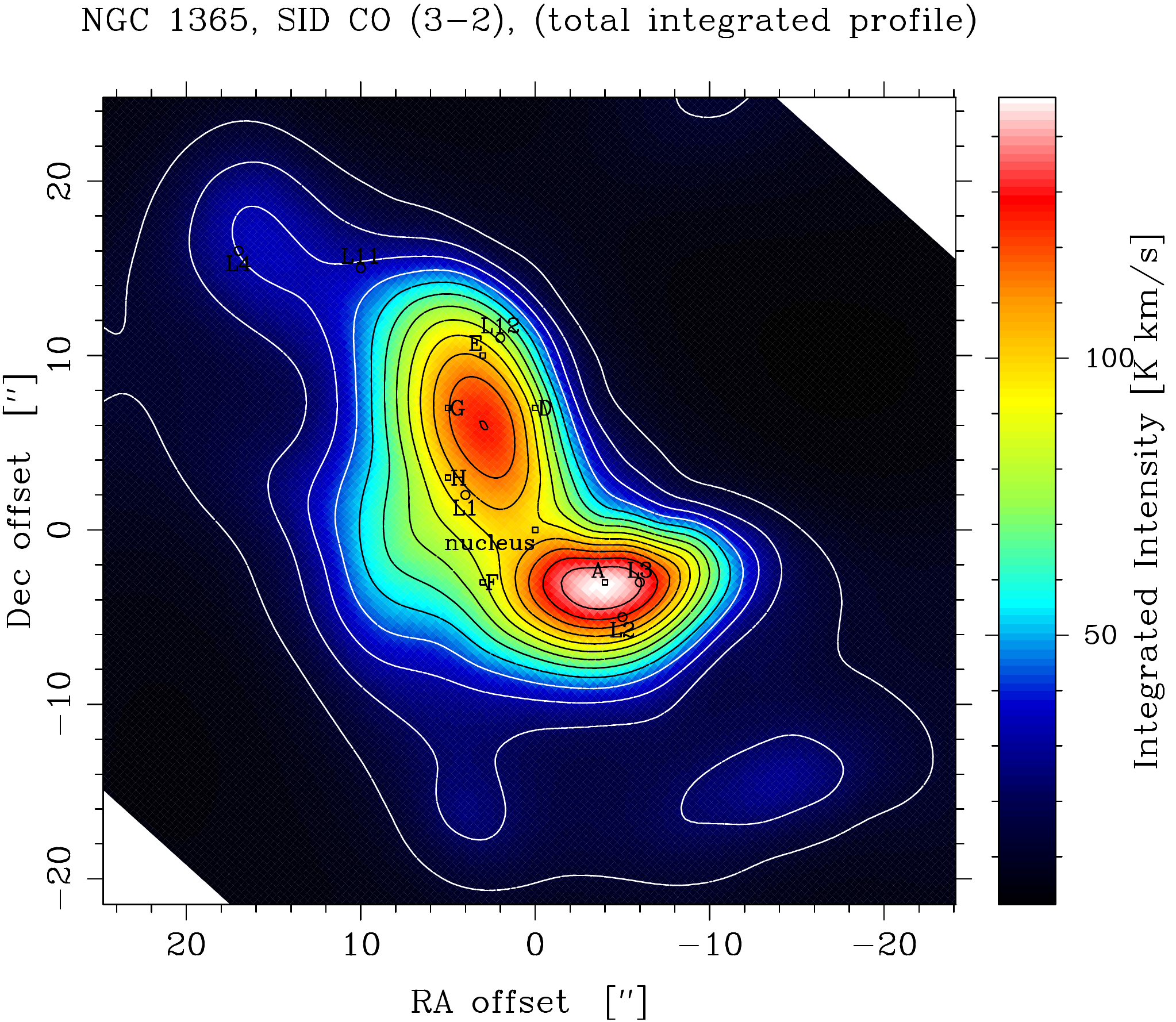}}}
 \resizebox{!}{5cm}{{\includegraphics{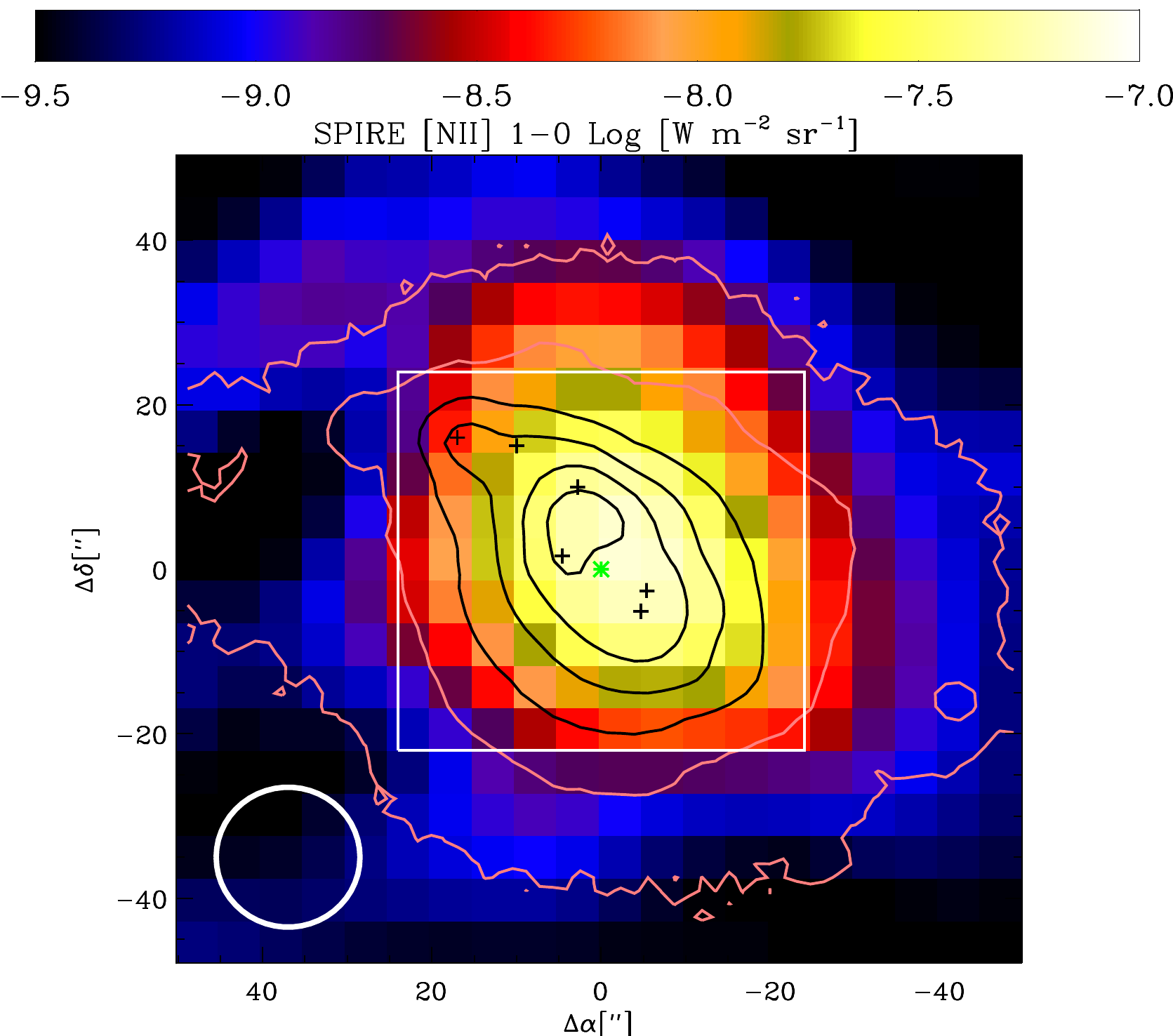}}}
 \resizebox{!}{5cm}{{\includegraphics{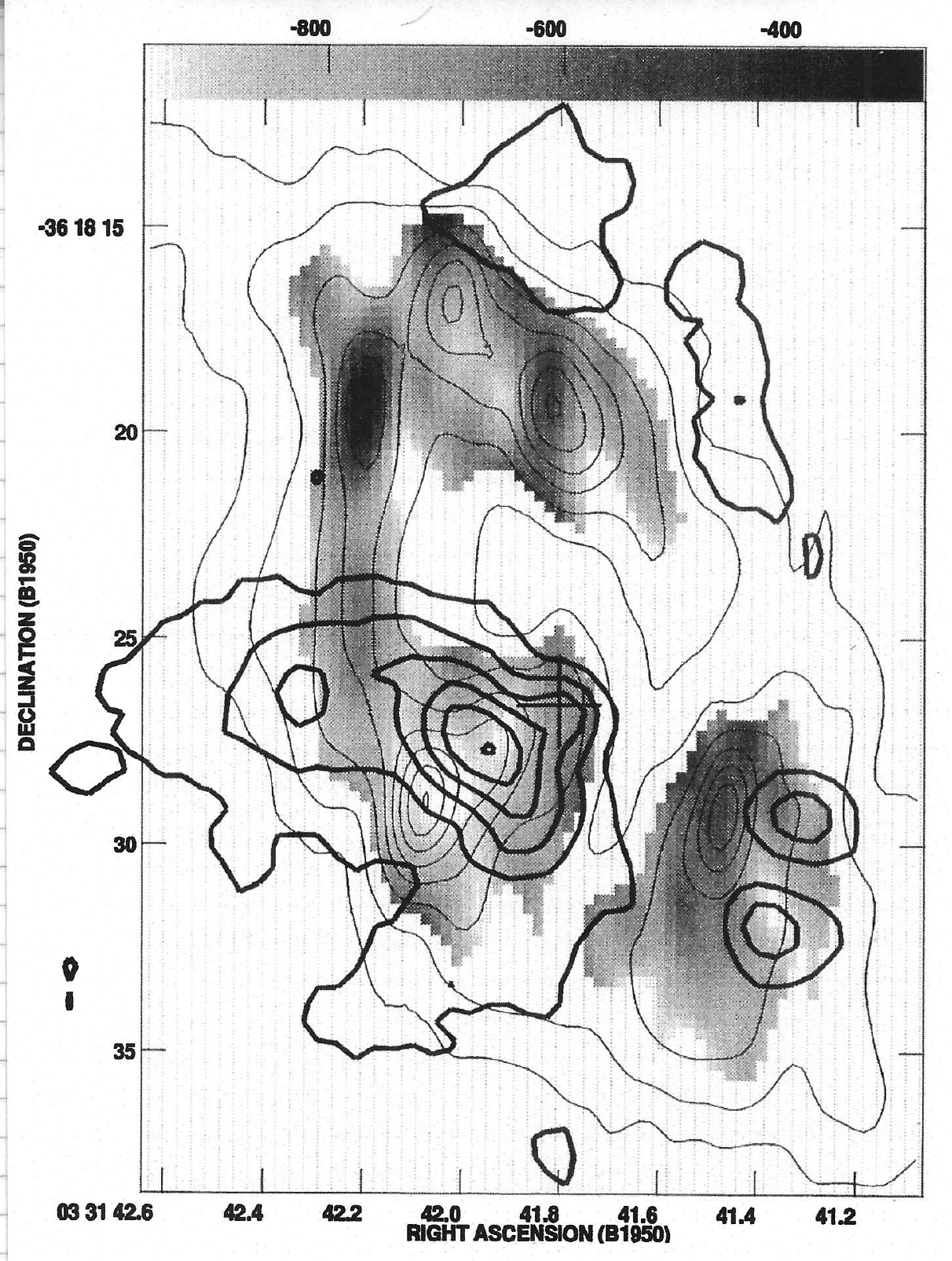}}}
}
  \caption{{\it Panel Number from the Left}: {\it (1)}: VLA
    \HI\ line total integrated intensity 
    distribution, emission and absorption, is shown in the central region of
    NGC\,1365. The circles are hot-spot \HII\ regions and the squares
    are compact radio sources as defined in Table 2. The offsets
    are with respect to the optical nucleus, which is marked by a
    square. {\it (2)}: The SEST SID CO($3-2$) total integrated line
    intensity (antenna temperature – multiply by 3.8 to obtain main
    beam brightness temperature). The equatorial offsets are with
    respect to the optical nucleus. {\it (3)}: The SPIRE map of 1461
    GHz \NII; the beamwidth is $\approx 17\as$. The white square
      marks the area covered in the \HI\ and CO maps. Remaining features
    are as defined in Fig. 4. {\it (4)}: The \OIII\ emission line 
    region (thick lines), and the VLA 20-cm radio continuum emission
    (thin lines) from the radio jet and the radio continuum
    circumnuclear ring. The gray scales, ranging from $-0.3$ (black)
    to $-1.0$ (white), indicate the 6/20 cm radio spectral index. The
    cross marks the position of the optical nucleus (from Sandqvist et
  al. \cite{san95}).}         
  \label{7}
\end{figure*}

\section{Analysis and discussion}

\subsection{Gas distribution, dynamics, and star formation in the
  central region of NGC\,1365} 

The distribution of molecular gas in the central region of NGC\,1365
can be understood in the framework of bar-driven gas dynamics, as
described in the Introduction. This implies that the rapid inward gas flows
along the bar dust lanes are transferred into an oval circumnuclear
torus of streaming gas clouds (see Sakamoto et al. \cite{sak07}, especially
their Fig.12).

\subsubsection{Cold molecular gas clouds causing intense star formation}

According to our {\it Herschel} SPIRE mapping (Fig. 4), the ground
state o-\hto\ (1$_{10}$ - 1$_{01}$) emission at 557 GHz, observed by
{\it Odin} in NGC\,1365 (in the velocity range $1400-1650$ \kms), is
mainly emanating from within a $10-15$\as\ size (i.e., $\approx 1$ kpc) region,
including the NE hot spot (of CO, etc.) of the circumnuclear torus and
some very nearby hot spots east thereof. However, we cannot
exclude that some of the water emission in the velocity range
$1400-1500$ \kms\  emanates from the L4 hot spot situated in the leading
dust lane of the eastern bar, about 25\as\ away from the center (see
e.g., Fig. 1), since this is also the velocity range observed at this
position in the CO($3-2$) line (see Fig. A.1). Our SPIRE map of the 752
GHz line in fact may support such a contribution, although not the 557
GHz line map (see Fig. 4). 

It may seem peculiar that we observe little or no ground state o-\hto\
emission from the SW torus region (i.e., in the velocity range
$1613-1800$ \kms), since the CO($2-1$) as well as the CO($3-2$) emissions
are observed to be similarly strong in the NE and SW torus regions
(Sakamoto et al. \cite{sak07}; Sandqvist \cite{san99}; Fig. A.1 of this
paper). However, in addition to the lack of sensitivity in our {\it
  Odin} observations, a likely explanation is offered by the recent
mapping of the dust emission at 870 $\mu$m in NGC\,1365 usíng LABOCA
(the Large APEX Bolometer Camera; Tabatabaei et
al. \cite{tab13}). These authors estimate that the eastern bar is more
massive than the western one by at least a factor of four. Hence the
bar-driven inflow of gas into the NE torus region may be expected to
be much more massive than the similar inflow into the SW torus
region. Tabatabaei et al. also find that the dust (and presumably the
gas) in the western bar is warmer (about 40 K) than that in the
eastern bar (about 20 K) by a factor of two. Even the cold dust
temperature map, presented in our Fig. 2, shows a tendency for the
western bar to be warmer. This temperature
difference may indeed explain why the (optically thick) CO($3-2$)
emissions remain similarly strong in the eastern and western torus and
bar regions (Fig. 8 [$-102$ to 0 \kms] and [$+102$ to $+228$ \kms])
in spite of the apparently very different gas masses.

Using the careful SMA aperture synthesis CO (2-1) mapping at
2\as\ resolution by Sakamoto et al.(\cite{sak07}), Elmegreen et
al. (\cite{elm09}) estimate that the gas inflow into the NE
torus region has a velocity of $\approx 80$ \kms\ and an accretion
rate of $\approx 22$ \msol /yr. They also estimate a star formation
rate (SFR) of $\approx 10$ \msol /yr in the central, circumnuclear
torus region (using the observed FIR luminosity). Tabatabaei et
al. (\cite{tab13}) find an SFR of $\approx 15$ \msol /yr in the
central 80\as\ region in a similar way. Based upon their CO mapping,
Sakamoto et al. (\cite{sak07})  
and Sandqvist (\cite{san99}) estimated the observed central molecular
cloud mass to be $M_{\rm mol} (R \lid 1 {\rm kpc}) \approx 1 \times
10^9$ \msol\ and $M_{\rm mol} (R \lid 2 {\rm kpc}) \approx
5.4 \times 10^9$ \msol, respectively, however relying on very different
$X_{\rm CO}$-factors, $0.5 \times 10^{20}$ vs. $2.3 \times 10^{20}$ (K
\kms)$^{-1}$. If we instead use a common conversion
value of $1.2 \times 10^{20}$  cm$^{-2}$ (K \kms)$^{-1}$ in both
cases, as suggested 
by Tabatabaei et al., we arrive at $M_{\rm mol} (R \lid 1 {\rm kpc})
\approx 2.4 \times 10^9$ \msol\ and $M_{\rm mol} (R \lid 2 {\rm kpc}) \approx
2.8 \times 10^9$ \msol\ – mass estimates likely to be accurate to within
a factor of two. 

Our CO($3-2$) P-V diagram (Fig. A.1 [$y=-2.5$\as])
nicely demonstrates an overall solid body rotation 
(i.e., a rotational velocity increasing linearly with the distance
from the center) of the cicumnuclear molecular cloud  torus region
inside $R \lid 1$ kpc, but with considerable velocity dispersion
caused by individual cloud motion, cloud-cloud collisions, shocks and
outflows (also causing increased turbulence in the clouds). For $R$ >
1 kpc we observe the expected differential rotation. In fact, we may
also (in Fig. A.1 [$y= -2.5$\as] see a
hint of the NE bar-driven inflow in terms of two velocity components
(at $-150$ and $-110$ \kms, at a major axis offset of +10\as, or 0.9
kpc). Correcting for a galaxy inclination of 40\degr, this velocity
difference becomes close to the aforementioned inflow velocity of 80
\kms. The P-V diagram also allows us to calculate the dynamical mass
$M_{\rm dyn}$  (of gas, stars plus central AGN) inside a specified radius
(using the balance between the centrifugal force caused by rotation at
a velocity $V_{\rm rot}(R_o)$ and the gravitational attraction of the mass
inside $R_o$ – valid in the low-order approximation of circular
motion), namely, \\  

     $M_{\rm dyn}$ ($R \lid R_o$) = $R_o \times V_{\rm rot}^2(R_o)$/G,
     \ \ \ \ \ \    (1) \\

where G is the gravitation constant, which in more useful parameters
can be rewritten \\

$M_{\rm dyn} (R \lid R_o) \approx 1.2 \times  10^{10}$ \msol $ \times
R_o{\rm [kpc]} \times (V_{\rm rot} / 230$)$^2$.    \ \ \ \ \        (2) \\

From our P-V-diagrams (Fig. A.1), we may now estimate  $V_{\rm rot}
\approx 150$ \kms /sin 40\degr = 233 \kms\  and  $V_{\rm rot} \approx
$ 200 \kms /sin 40\degr = 311 \kms\ at
$R_o = 1$ and 2 kpc, and hence calculate  dynamical masses of $1.2
\times 10^{10}$ \msol\ and $4.4 \times 10^{10}$ \msol, respectively. Our
previously observationally determined gas masses  amount to about 20\%
of the dynamical masses in the center versus an average of 6\% in the 
more extended region (including the center). Hence we may see a trend
of increasing gas mass fraction toward the center, but not
unexpectedly the stellar mass dominates. 

\subsubsection{Warm ionized medium, created by the intense star
  formation}

We now move from the discussion of the bar-driven, outer
circumnuclear molecular gas torus, causing the formation of
supermassive stellar clusters, to the creation of the warm ionized
medium (WIM), the ISM component surrounding the molecular cloud
ensemble and the stellar superclusters, which is caused by the intense
UV radiation of the multitude of newly formed stars. 

Nitrogen has a higher ionization
potential (14.5 eV) than does Hydrogen (13.6 eV) and will stay atomic
N where hydrogen is atomic H or molecular \htwo, and 
becomes ionized N$^+$ where hydrogen is almost completely ionized
H$^+$. The 205 $\mu$m \NII\ line is easily collisionally excited in warmer gas
since its upper state energy is 70 K and its critical density is only
44 cm$^{-3}$. Hence it is an excellent probe of the WIM, consisting of
moderately dense \HII\ regions and ionized 
boundary layers of clouds, that is to say, a useful measure of the star
formation rate (SFR) (Langer et al. \cite{lan16}; Zhao et
al. \cite{zha16}). Our SPIRE map of the 205 $\mu$m \NII\  line
(Fig. 7), observed with a 
17\as\ beam, shows that this emission is extended beyond the SW torus
peak and here is as intense as in the NE peak. The high SFR observed
also at and beyond the SW torus peak provides a nice explanation of
the higher dust, gas temperature (40 vs 20 K) and lower dust, gas mass
(by a factor four) in the western bar region versus the eastern bar, as
estimated by Tabatabaei et al. (\cite{tab13}). This observed mass and
temperature asymmetry of the bar may be understood if the western bar
region resides in a, somewhat later, high SFR stage, where a larger
amount of the star forming molecular clouds has been consumed and
where the intense UV radiation from the newly formed stars has heated
the dust.  

\begin{figure*}[ht]
\includegraphics[angle=0, width=.24\textwidth]{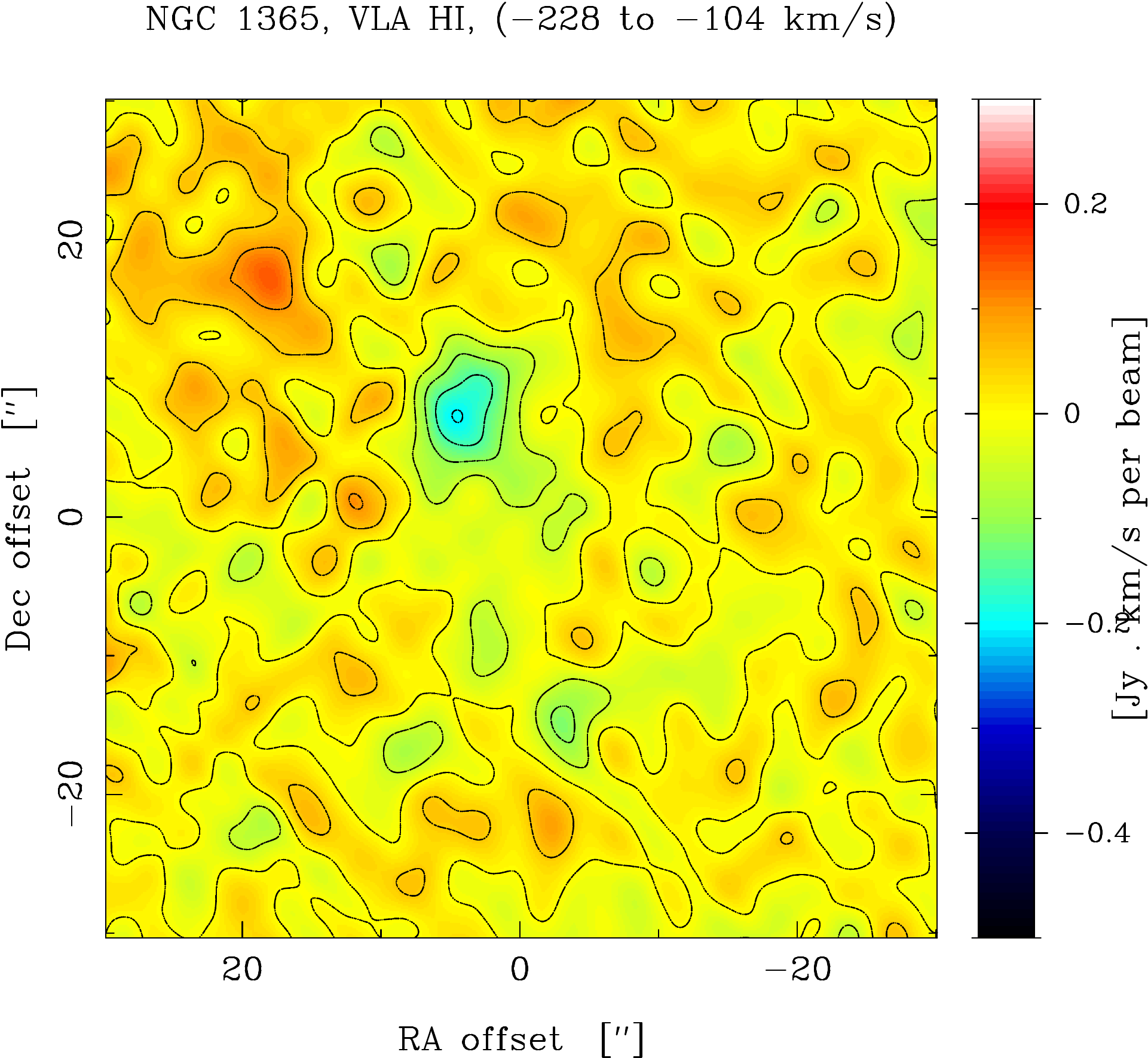}
\includegraphics[angle=0, width=.24\textwidth]{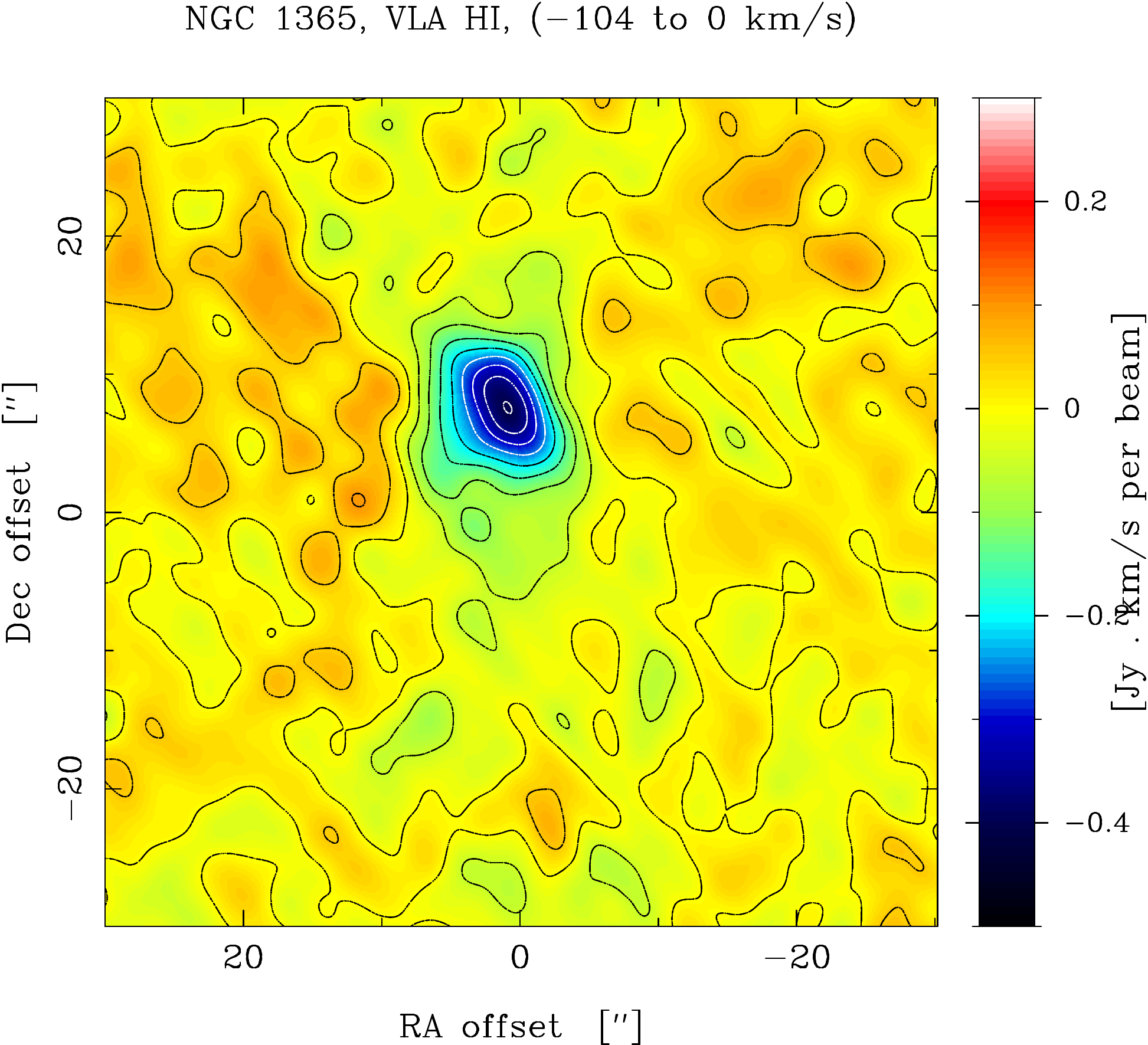} 
\includegraphics[angle=0, width=.24\textwidth]{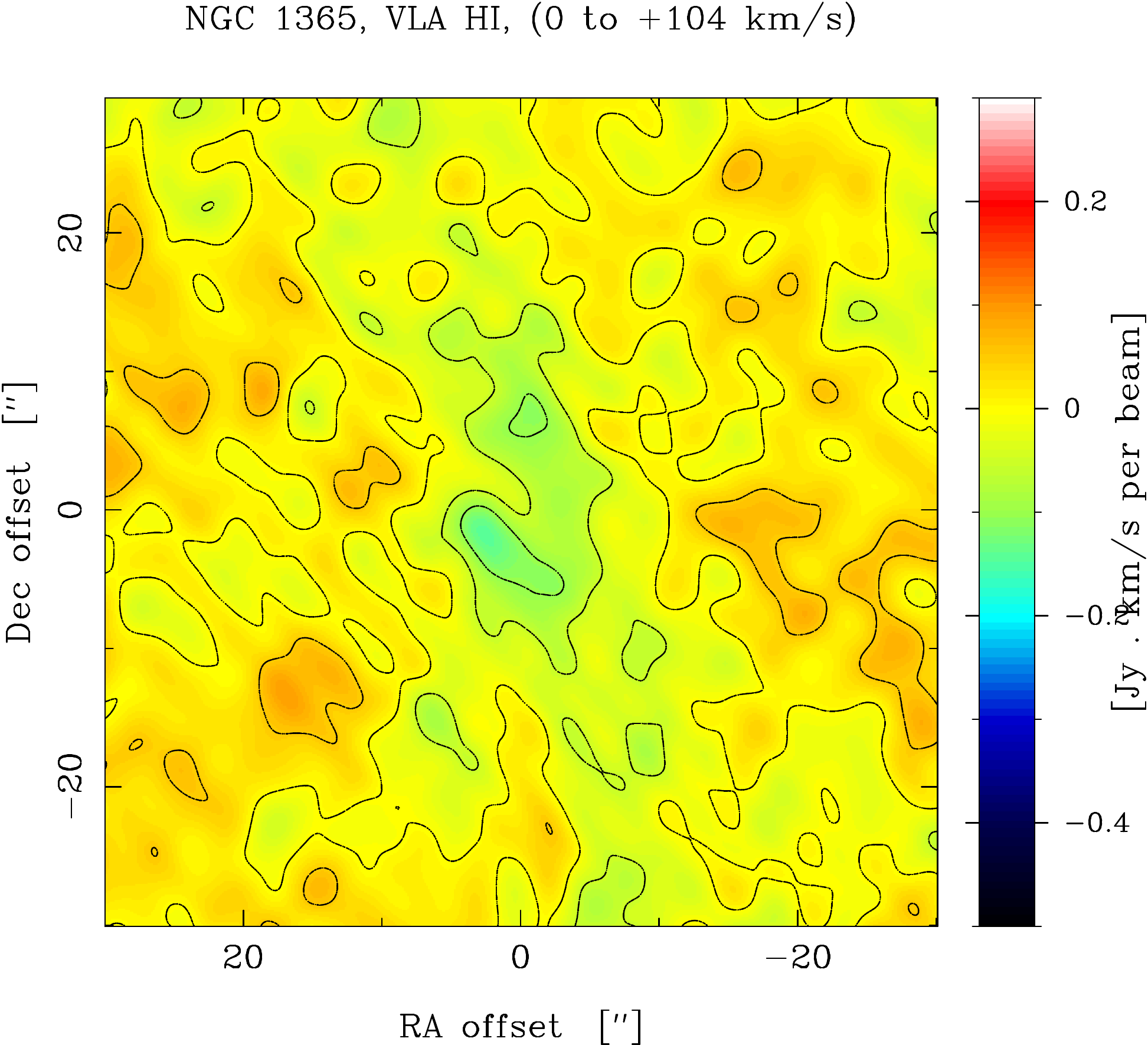}
\includegraphics[angle=0, width=.24\textwidth]{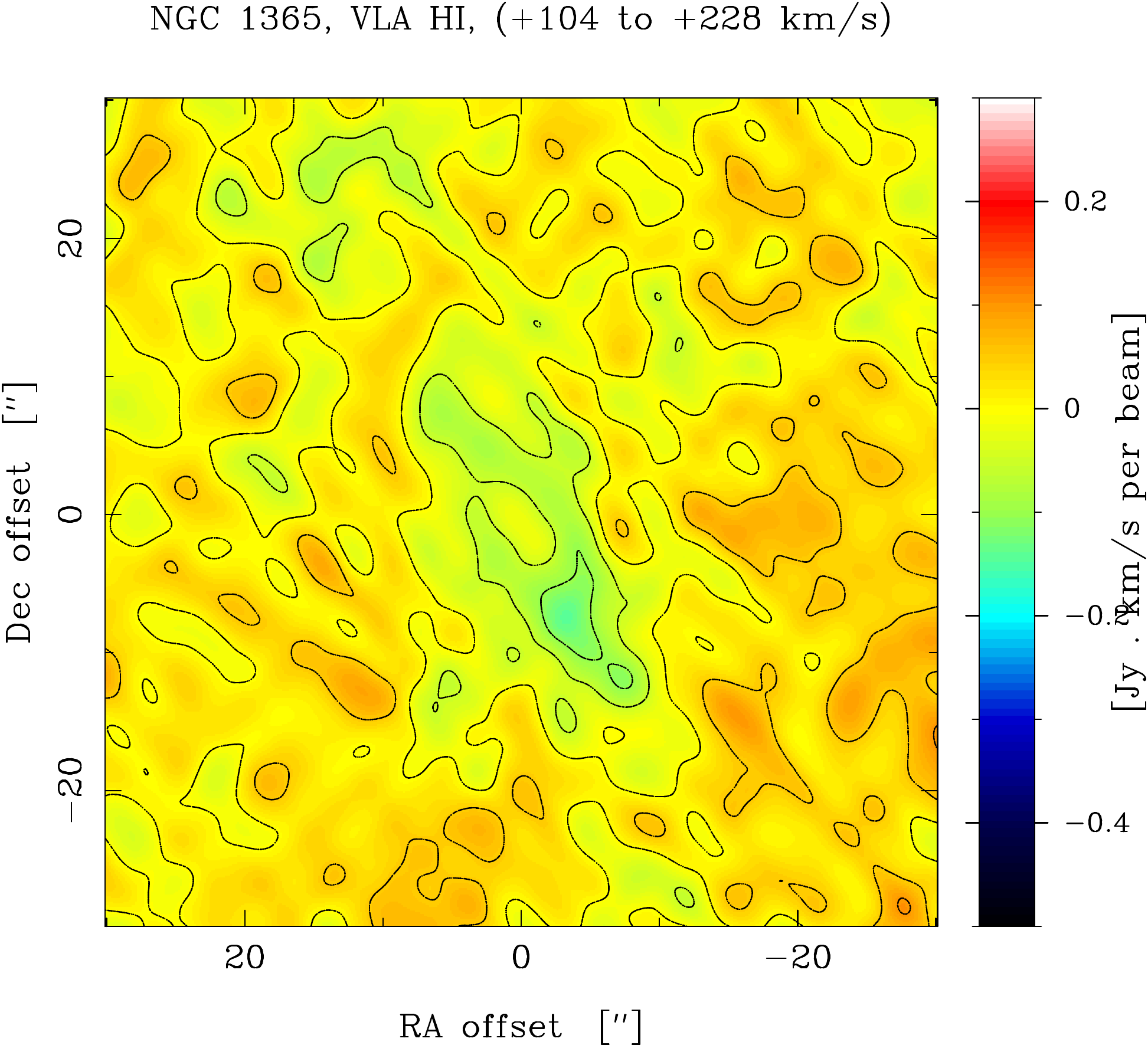}
\includegraphics[angle=270, width=.24\textwidth]{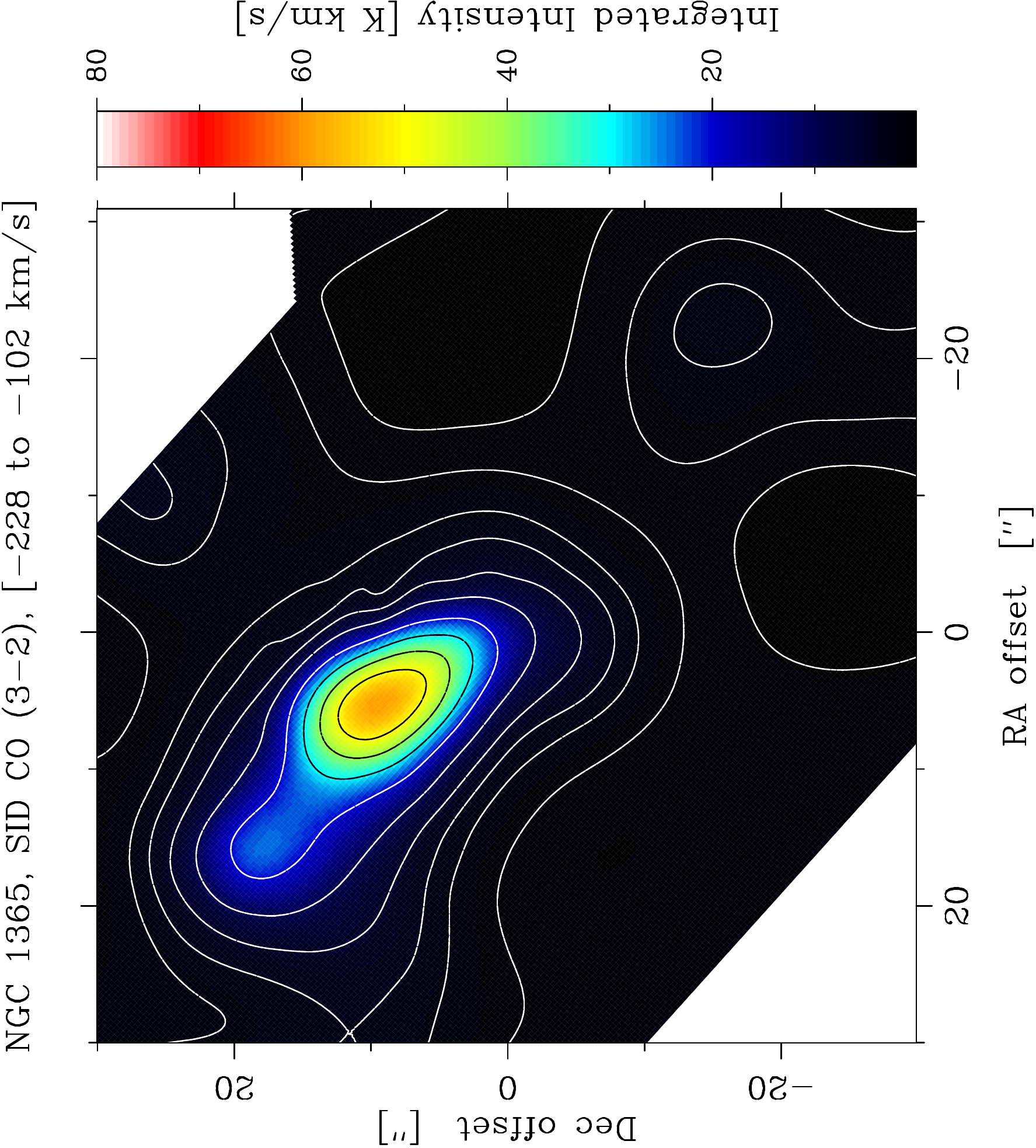}
\includegraphics[angle=270, width=.24\textwidth]{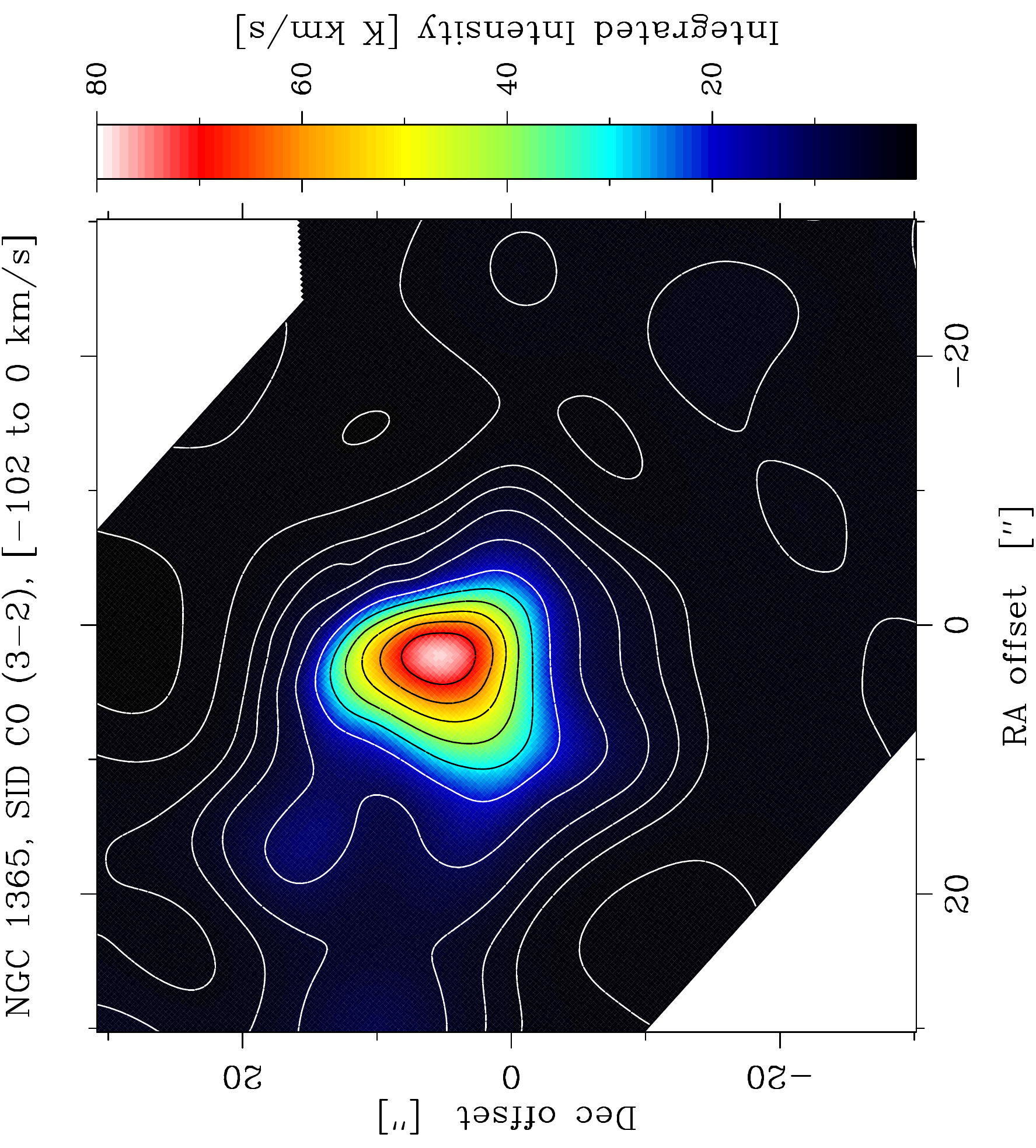}
\includegraphics[angle=270, width=.24\textwidth]{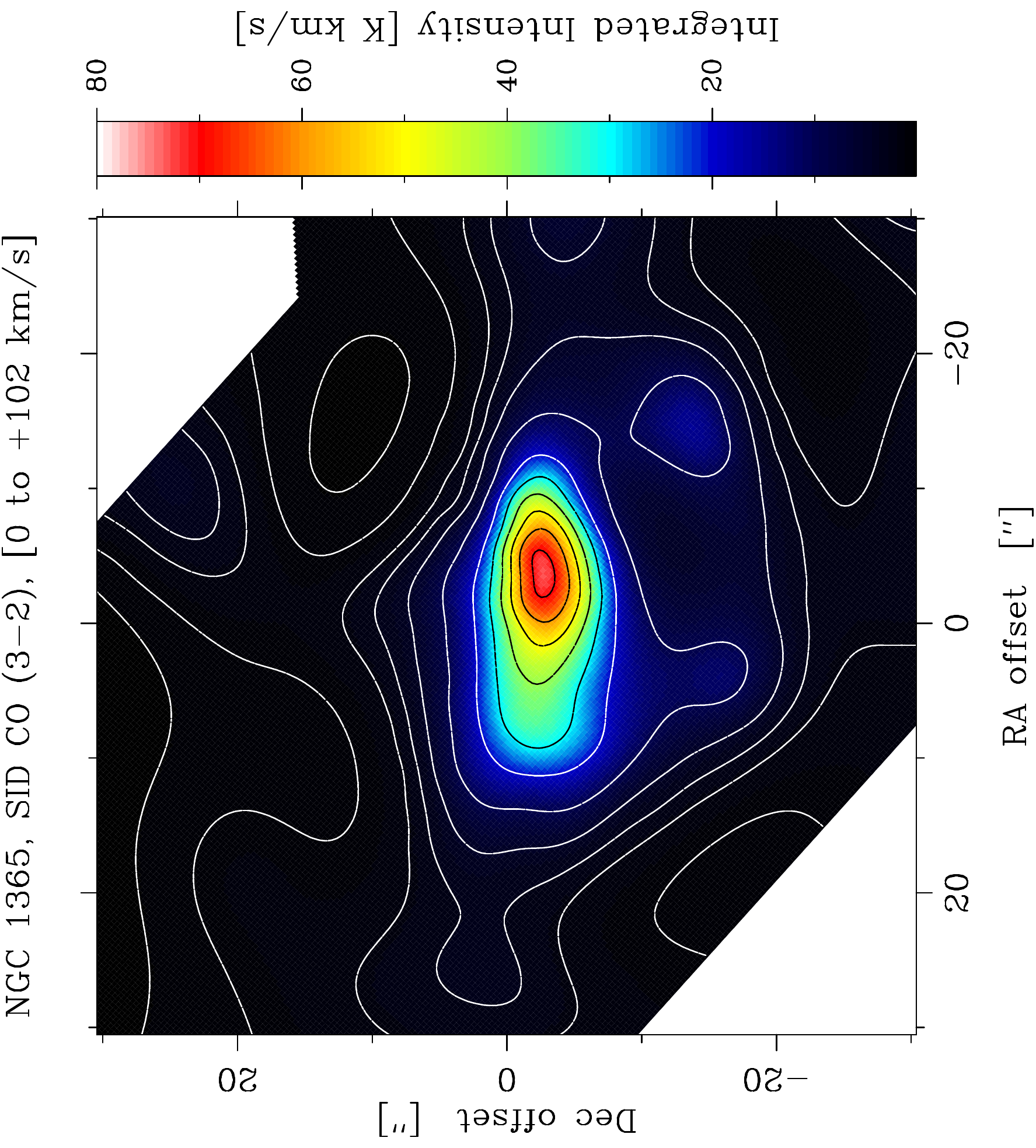}
\includegraphics[angle=270, width=.24\textwidth]{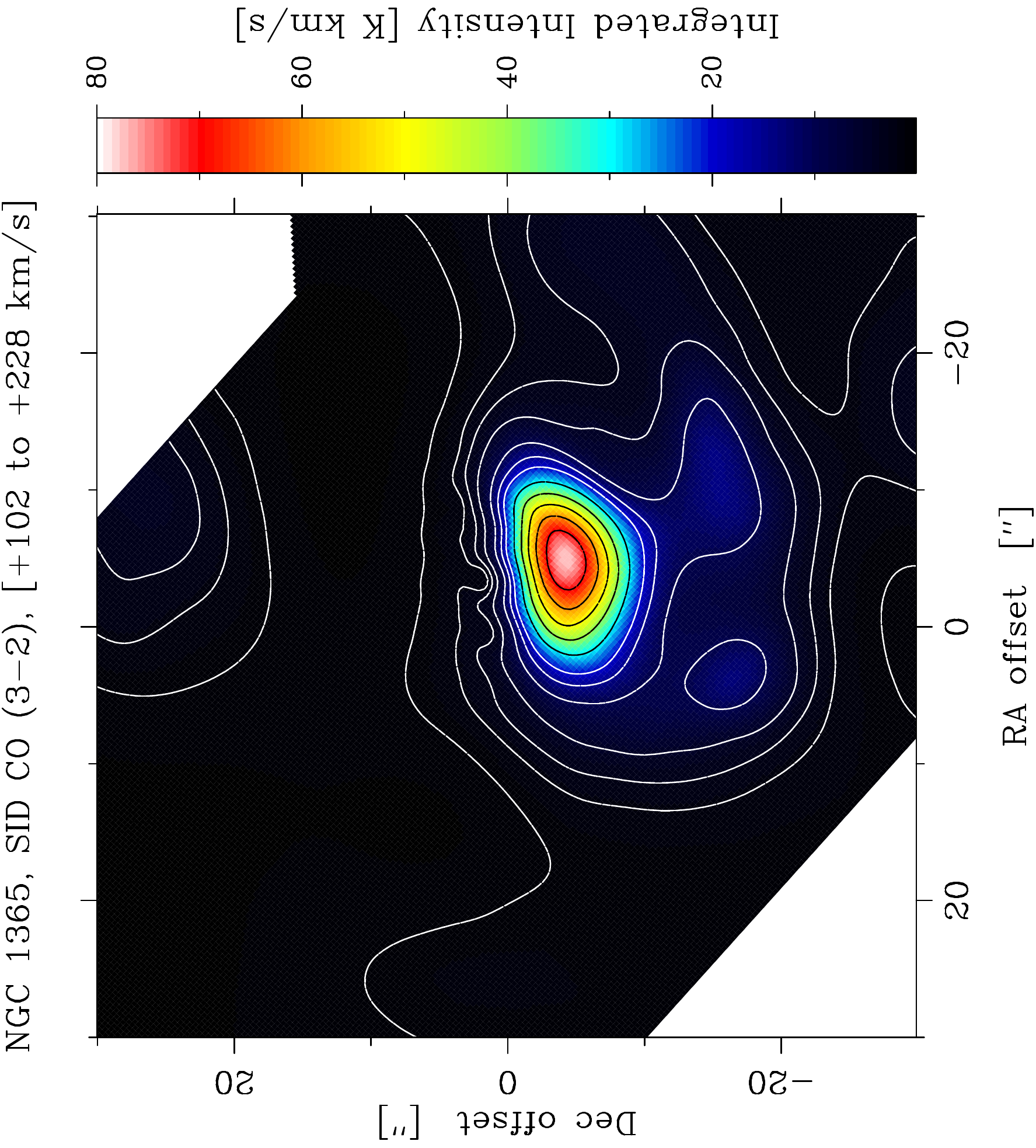}
\caption{VLA \HI\ [{\it top}] and SEST SID CO($3-2$) [{\it
    bottom}] line intensity maps integrated over the velocity
    range indicated in the header [($-228$ to $-102$), ($-102$ to 0), (0
    to +102), (+102 to +228)] \kms. The CO integrated intensity is
    antenna temperature – multiply by 3.8 to obtain main beam
    brightness temperature. The equatorial offsets are with
    respect to the optical nucleus. The velocity offsets are with
    respect to the systemic LSR velocity of 1613 \kms.}         
  \label{8}
\end{figure*}

\subsection{Analysis of the {\it Odin} and {\it Herschel} SPIRE
  observations}

\subsubsection{{\it Odin} results}

A Gaussian fit with two emission and one absorption components has
been applied to the {\it Odin} \hto\ profile shown in Fig. 5. The
values for $T_A$, $V$, and $\Delta V$ are 7.1 mK, 1456 \kms, 42 \kms\
and 5.1 mK, 1582 \kms, 81 \kms\ for the emission components and $-1.5$ 
mK, 1511 \kms, 40 \kms\ for the absorption component. The two emission
features appear at velocity offsets of $-157$ and $-31$ \kms\ with
respect to the systemic LSR velocity of 1613 \kms. The absorption
feature has a velocity offset of $-102$  \kms. Although the Gaussian fit
may be somewhat uncertain – due to the very low signal-to-noise ratio,
and since the lower excitation foreground
\hto\ regions may absorb not only the thermal dust continuum but also
some of the velocity ranges of the background \hto\ emission from
higher excitation regions – it is worthwhile to use the velocity
information in an attempt to determine the origin of the these three 557
GHz \hto\ regions. In the following, we use our own SEST SID CO($3-2$) and
VLA \HI\ results as well as the CO($2-1$) SMA
observations by Sakamoto et al. (\cite{sak07}) and new ALMA
observations of the CO($3-2$) line by Combes et al. (\cite{com19}).

While the large {\it Odin} beam is covering most of
the bar, the \hto\ region has already been limited in extent to
the central region by our SPIRE 557 GHz \hto\ map presented in
Fig. 4. The \hto\  emission is dominant near the NE torus component where the
aforementioned hot spot L-regions and compact radio sources
gather.

Our CO($3-2$) P-V map in Fig. A.1 ($y=-2\farcs5$) shows a distinct feature
at a velocity offset of $-150$ \kms, which is close to one of the
\hto\ emission features velocity of $-157$ \kms. That occurs at a
major axis offset $x=10$\as. This position is very close to the
compact radio source G (see Table 2) and is also the region in the ALMA
velocity map where the velocity offset is most negative, namely
$\approx -150$ \kms. Near the position of E, the ALMA velocity offset
is around $-90$ \kms, which is comparable to the {\it Odin} \hto\
absorption feature value of $-102$ \kms. Near D the ALMA velocity
offset is closer to $-50$ \kms, which can be compared to the {\it
  Odin} emission feature value of $-31$ \kms. The VLA \HI\ maps in
Fig. 8 show major absorptions toward G and D in the velocity offset
ranges of ($-228$ to $-104$) \kms\ and ($-104$ to 0) \kms,
respectively, agreeing reasonably well with the above-mentioned
comparisons. Our conclusion is thus that the three {\it Odin} \hto\
features originate in the NE shock region, giving rise to the violent
star formation activity caused by the interactions of the $x_1$- and
$x_2$-like streamlines in the inner bar and center, as proposed by
Sakomoto et al. (\cite{sak07}).

An estimate of
the expected background continuum antenna temperature, convolved to the
{\it Odin} effective beam of 140\as\ by 160\as, can be obtained from
the {\it Herschel} SPIRE spectrum seen in Fig. 3. The flux density
continuum level at 555 GHz is 4.5 Jy which converts to 1.5 mK antenna
temperature, convolved to the {\it Odin} beam.

We now make an effort to estimate \hto\ abundances based upon 
our three-component Gaussian fit to the 557 GHz ground state o-\hto\ spectrum
observed by Odin. The absorption feature around 1511 \kms\ has a width
$\Delta V = 40$ \kms\ and appears to be saturated since the absorption
is similar to, or larger than, the continuum level determined from our
{\it Herschel} SPIRE spectrum (Fig. 3). A column density estimate for
this lower density, lower excitation foreground gas, or rather a lower
limit thereof, then may be obtained from the relation (derived by
Karlsson et al. \cite{kar13}), \\

$N$(\hto)$ = 4/3 \times 4.89 \times 10^{12} \times \tau$(\hto) $
\times \Delta V$,   \ \ \ \         (3) \\ 

where we assume an optical depth $\tau \approx 2 - 3$ and an
ortho-to-para ratio OPR $= 3$. The resulting total water column
density becomes $N$(\hto) $ \gid (5-8) \times 10^{14}$ cm$^{-2}$. The
corresponding \htwo\ column density, in front of the source(s) of the
thermal dust continuum emission (and \hto\ emission) being absorbed,
may be estimated from the visual extinctions of the mid/near IR (and
radio) sources M4, M5 and M6 (see Table 2), from their observed
Br${\gamma}$/Br${\alpha}$ ratios, estimateded to be $A_{\rm v} = 13.5$, 3.2
and 8.5 mag, respectively (Elmegreen et al. \cite{elm09}). If we use
an “average” $A_{\rm v} \approx 10$ mag in the scaling relation found from
observations by the Copernicus satellite (Bohlin et al. \cite{boh78}),
that is,\\

$N$(\htwo)/$ A_{\rm v} = 0.94 \times 10^{21}$ cm$^{-2}$,   \ \ \ \ \ \
\ \ \ (4) \\

our estimate would be $N$(\htwo) $ \approx 1 \times 10^{22}$
cm$^{-2}$, with a considerable uncertainty range (for a discussion and
references see e.g., Hjalmarson and Friberg 1988). However, this value 
is in fact about half of the average total column density found from
our SID CO($3-2$) analysis (soon to be discussed), 
which is a useful criterion of correctness. Finally, the water abundance can be
estimated to be $X$(\hto) = $N$(\hto)$/N$(\htwo) $ \gid 5 \times
10^{-8}$ – an abundance level which is well accommodated by PDR
chemistry in lower density, translucent and diffuse regions
(e.g., Hollenbach et al. 2009, 2012; see also our Tables 5 and B.1).

Our previous absorption line analysis must be considered comparatively
accurate – since the method itself relies on the realistic assumption that the
absorbing molecular population is mainly residing in the lowest possible
energy state – although the optical depth of the 557 GHz o-\hto\
absorption tentatively observed by {\it Odin} is uncertain. In
case of the two weak emission features apparent in the same {\it Odin}
spectrum, the analysis uncertainty stems to a large 
extent from the relatively poor knowledge of sizes, densities, and
temperatures of the emission regions, which by nature also are bound to be
inhomogeneous. Their approximate sizes of $\approx 10$\as\ are
estimated from our SID CO($3-2$) maps at 5\as\ resolution (for the
relevant velocity ranges shown in Fig. 8) and the SMA CO($2-1$) maps
at 2\as\ resolution by Sakamoto et al. (\cite{sak07}). Guidance on
temperatures and densities is drawn from the analysis of our
multi-transition CO observations by SEST and {\it Herschel} SPIRE,
presented in Sect. 4.2.2. Two-component fits to the CO
rotation diagram, as well as the CO SLED, reveal a hot ($T_{\rm kin}
\approx 350$ K) 
component, most likely caused by shock excitation, together with a lukewarm
($T_{\rm kin} \approx 40$ K) component having an average \htwo\ density of
10$^4$ cm$^{-3}$. The average \htwo\ column density of the latter
component is estimated to be $6.3 \times 10^{21}$ cm$^{-2}$ across a
10\as\ $\times$ 30\as\ region. According to Eq. (4) this corresponds to
a visual extinction $A_{\rm v} \approx 7$ mag. However, the surface
filling factor of CO emission in this region may be less than 100\%. The
characteristics of this lukewarm region are very similar to what is
expected from PDR-models for high FUV illumination (such as that in
the Orion molecular cloud; cf. Fig. 7 of Hollenbach 
et al. \cite{hol09}). This interpretation is also supported by the
best two-component fit of the SED of the thermal dust emission from
the central regions of NGC\,1365 – one component at $25 \pm 2$ K and a
necessary second one at $55 \pm 5$ K, emitted from $\approx 1$\% of the
total dust mass (Alonso-Herrero et al. \cite{alo12}; Tabatabaei et
al. \cite{tab13}). 

With this information at hand, we assume a cloud temperature $T_{\rm
  kin} = 55$ K and density $n$(\htwo) $= 10^4 - 10^5$ cm$^{-3}$ in our
RADEX (van der Tak et al. \cite{tak07}) molecular excitation and radiative
transfer model fitting of the 557 GHz o-\hto\ emission features, each
having an estimated size of only $\approx 10$\as\ in the
very large {\it Odin} antenna beam ($140\as \times 160\as$, resulting
in a beam filling factor of 1/225). Including the beam 
filling correction, we get as input parameters $T_{\rm b} \approx 1.6$
K and $\Delta V = 42$ \kms\ 
for the emission around $-157$ \kms\ and $T_{\rm b} \approx 1.1$ K and
$\Delta V = 81$ \kms\ for the emission around $-31$ \kms. If we
assume an OPR = 3, the corresponding total ortho- and para-\hto\ column
densities are RADEX modeled to be $N$(\hto) $ \approx 15 \times
10^{15}$ cm$^{-2}$ and $19 \times 10^{15}$ cm$^{-2}$, respectively, for
$n$(\htwo) $= 10^5$ cm$^{-3}$, and increasing by an order of magnitude
for the lower cloud density of $n$(\htwo) $= 10^4$ cm$^{-3}$. The
comparison \htwo\ column densities may be roughly estimated from our
SID CO($3-2$) maps for the velocity ranges $-200$ to
$-100$ \kms\ and $-100$ to 0 \kms\ (Fig. 8) using the scaling
relation\\

$N$(\htwo) $= 1.2 \times 10^{20} \times I$(CO) cm$^{-2}$, \ \ \ \ \
(5) \\

where $I$(CO) is the integrated intensity [in K \kms] (Tabatabaei et
al. \cite{tab13}; for a 
discussion of the scaling relation method, see e.g., Hjalmarson
\&\ Friberg 
\cite{hja88}), resulting in $N$(\htwo) $ \approx 2.3 \times 10^{22}$
cm$^{-2}$ and $3.0 \times 10^{22}$ cm$^{-2}$, respectively, for the
two \hto\ emission regions. For an average cloud density of 10$^5$
cm$^{-3}$, the resulting \hto\ abundance versus \htwo\ can be calculated as
$X$(\hto) $=  N$(\hto)$/N$(\htwo) $ \approx 6 \times 10^{-7}$ for both
regions – an abundance increasing by an order of magnitude in case the
average density is 10$^4$ cm$^{-3}$. The lower abundance value appears
to be consistent with PDR modeling results (Hollenbach et
al. \cite{hol09}, \cite{hol12}), especially since additional
ionization support may exist in terms of cosmic ray focusing by the
observed magnetic field alignment (Beck et al. \cite{bec05}). However,
substantial contributions to the ground state \hto\ emission, caused by
shock excitation and shock chemistry, are also expected from our modeling
of the {\it Herschel} SPIRE data presented in Sect. 4.2.2 (see Fig. 9).

\subsubsection{{\it Herschel} SPIRE results}

Before entering a detailed presentation of the multi-transition
excitation and radiative transfer analysis of our SPIRE CO and \hto\
observations (Fig. 3 and Table 1), it may be helpful to examine the
spatial distributions of the 557 GHz ground state o-\hto\ line ($E_u =
27$ K in the ortho ladder; $E_u = 61$ K vs the para ground state) and the
752 GHz excited state p-\hto\ line $(E_u = 137$ K vs the para ground
state), illustrated in Figs. 4 and 6. The two maps show that both
lines are strongest in the NE torus peak area. However, it is apparent
from Fig. 6 that the 557 GHz line has little or no emission at
the SW torus peak, while the 752 GHz excited state transition
definitely exhibits emission here (taking into account the beam size
of 35\as\ and the 15\as\ separation of the observed NE and SW
positions). Our subsequent analysis of the CO, as well as the
\hto\ SLEDs, will lead us to the 
conclusion that shock excitation and shock chemistry must be main
players in case of the higher energy lines. Therefore, it is useful to know
beforehand that such conditions have indeed been observed in all of
the “hot spot” \HII, CO and stellar supercluster regions in the
circumnuclear torus, and also in the presumptive synchrotron radiation
jet source “F” (see our Table 2; VLT SINFONI high resolution near IR
observations by Galliano et al. \cite{gal12} and Fazeli et
al. \cite{faz19}, to be discussed in some detail in Sect. 4.4). 

We now present an analysis of the {\it Herschel} SPIRE apodized
spectrum obtained toward the NE component of the central torus, which
was presented in Fig. 3. As has already been pointed out, the NE
component is the dominant one and so of major interest. It is also the
component having a velocity field closer to that implied by the {\it Odin}
\hto\ profile. We shall first treat the CO lines and then the \hto\
lines.

The rotation diagram for the CO lines is presented in Fig. 9({\it
  Upper left}). The three lowest-level CO lines (blue dots) are
obtained from SEST observations 
presented by Sandqvist et al. (\cite{san95}), Sandqvist (\cite{san99}), and
Curran et al. (\cite{cur01}), while the remaining higher-level CO lines
(red dots) are from this paper. The results of RADEX analysis applied
to these CO lines (also presented in Fig. 9({\it Upper left})) show
that the high $J$
CO lines can be best fit by a single kinetic temperature of 400 K and an
\htwo\ density of $n$(\htwo) $=3 \times 10^3$ cm$^{-3}$, yielding a CO column
density of $N({\rm CO}) = 2.5 \times 10^{15}$ cm$^{-2}$.

A two-temperature RADEX fit to the CO SLED is
presented in Fig. 9({\it Upper right}). The best fit is obtained for
an extended cold 
component with a temperature of $T_{\rm Ext} = 26$ K, density
$n$(\htwo) $=7 \times 10^3$ cm$^{-3}$, yielding a CO column
density of $N({\rm CO}) = 1 \times 10^{17}$ cm$^{-2}$, and a hot
component from a 
more compact region having a temperature of $T_{\rm Hot} = 350$ K, density
$n$(\htwo) $=10^4$ cm$^{-3}$, yielding a CO column
density of $N({\rm CO}) = 4 \times 10^{15}$ cm$^{-2}$.

Perhaps it would be more realistic to study the NE region as one
involving shocks (see Fig. 9({\it Lower left})). In this case we have
used the C-shock model of 
Flower \&\ Pineau des For\^ets (\cite{flo10}) where we have scaled
their values to an equivalent (accumulated) source size of
10\as - in order to obtain a best fit. For the low
$J$ CO lines, we obtain here a slightly warmer extended region with a
temperature of 
$T_{\rm Ext} = 40$ K and RADEX values of $n$(\htwo) $=10^4$ cm$^{-3}$ and
$N({\rm CO}) = 7 \times 10^{16}$ cm$^{-2}$ (represented by the blue
curve) and a slow velocity C-type shock with a shock speed of 10 \kms\
and a pre-shock density of $n$(H+2\htwo) $= 2 \times 10^4$ cm$^{-3}$
(represented by the red curve). This model yields a CO abundance of
$X$(CO) $\approx 10^{-4}$.

Our RADEX modeling fit to the intensities of the lower energy lines
yields $N$(CO) 
$ = 1 \times 10^{17}$ cm$^{-2}$ for a homogeneous cloud at a
temperature $T_{\rm kin} \approx 40$ K and 
density $n$(\htwo) $ = 10^4$ cm$^{-3}$. From the mapping of the CO($2-1$) 
emission by Sakamoto et al. \cite{sak07}) and our SID CO($3-2$)
deconvolution, it appears that the half power size of the emission
region is only about 30\as\ $ \times$ 10\as, which
means a beam filling of only 16\% in a 40\as\ beam-size. The resulting CO
column density hence becomes $6.3 \times 10^{17}$ cm$^{-2}$, which for
a CO/\htwo\ abundance ratio of $\approx 1 \times 10^{-4}$ means an \htwo\ column
density $ \approx 6.3 \times 10^{21}$ cm$^{-2}$. According to the
scaling relation shown in Eq. (4), this corresponds to a 
visual extinction of $ \approx 7$ magnitudes. The estimated temperature and
density, combined with this visual extinction, most likely tells us
(cf. Hollenbach et al. \cite{hol09}, \cite{hol12}) that this CO
emission must originate in the extended PDR surfaces of the molecular
complexes, created by the intense UV illumination from the observed,
co-located, newly formed, massive stellar superclusters (Elmegreen et
al. \cite{elm09}; Galliano et al. \cite{gal12}; Fazeli et al. \cite{faz19}). 

We now turn our attention to the ortho- and para-\hto\ lines in the SPIRE
spectrum of Fig. 3.  RADEX models were performed for two different temperatures,
namely 400 and 1000 K, and model values for density $n$(H$_2$)
cm$^{-3}$ – and column density $N$(H$_2$O) cm$^{-2}$, respectively,
were $10^3 - 3 \times 10^{14}, 10^4 -  3 \times 10^{13}, 10^5
- 3 \times 10^{12}$ for o-\hto; $10^3 - 10^{14}, 10^4 -
10^{13}, 10^5 - 10^{12}$ for p-\hto. None of these models were found
to be applicable to the SPIRE \hto\ data. 

We then attempted water line modeling, using the same slow-velocity
C-type shock as above in the CO line analysis, and these results are
shown in Fig. 9({\it Lower right}). In this figure the blue and red
points and curves 
represent o-\hto\ and p-\hto, respectively. The  ortho-to-para ratio
is assumed to be 3. A shock speed of 10 \kms\ and a pre-shock density
$n$(\htwo) $= 10^4$ cm$^{-3}$, nicely matching the SPIRE CO data,
do not at all produce a fit to the observed \hto\ lines. An improvement of
the fit is obtained by increasing the pre-shock density by an order of
magnitude to $n$(H+2\htwo) $= 2 \times 10^5$ cm$^{-3}$ and retaining
the same  shock speed of 10 \kms. But the best fit is obtained by
using the original pre-shock density of $n$(H+2\htwo) $= 2 \times 10^4$
cm$^{-3}$ and increasing the shock speed to 40 \kms. It is, however,
still somewhat difficult to fit all the para-\hto\ lines. An
approximate \hto\ abundance for this model is similar to the CO
abundance, that is to say, $X$(\hto) = $X$(CO) $\approx 10^{-4}$. In studies of
the SPIRE and PACS results for the  ULIRGs  Mrk\,231, NGC\,4418, and
Arp\,220, it has been found that FIR pumping  provides an excellent fit
to the observed \hto\ line intensities (Gonzalez-Alfonso et al. \cite{gon10},
\cite{gon12}, \cite{gon14}; cf. Yang et al. \cite{yan13}). The estimated
  high \hto\ abundances of $10^{-6} - 10^{-5}$ here were suggested to
  result from shock chemistry, (multiple) hot cores, cosmic ray
  enhancement, XDRs, and, or ``undepleted chemistry'' where the icy grain
  mantles were evaporated. In our case of NGC\,1365 FIR pumping,
  paired with an enhanced \hto\ abundance caused by slow velocity shock
  sputtering of icy grain mantles (supplemented by any of the above
  processes), – all expected from observational data – this provides a
  viable alternative to ``our poor man'' fast shock
  modeling. Although no PACS observations of FIR pumping \hto\
  absorption lines are available in case of NGC\,1365, we note that the
  observed intensity of the o-\hto\ ($3_{21} - 3_{12}$) line, having
  an upper state energy of 305 K, is much higher than can be reached
  by the shock modeling  (see Fig.9({\it Lower right})). This strongly suggests that
  we see the effect of FIR pumping  by o-\hto\ ($2_{12} - 3_{21}$)
  absorption at 75 $\mu$m (Gonzalez-Alfonso et al. \cite{gon14}).

\begin{figure*}[ht]
\includegraphics[angle=90, width=.49\textwidth]{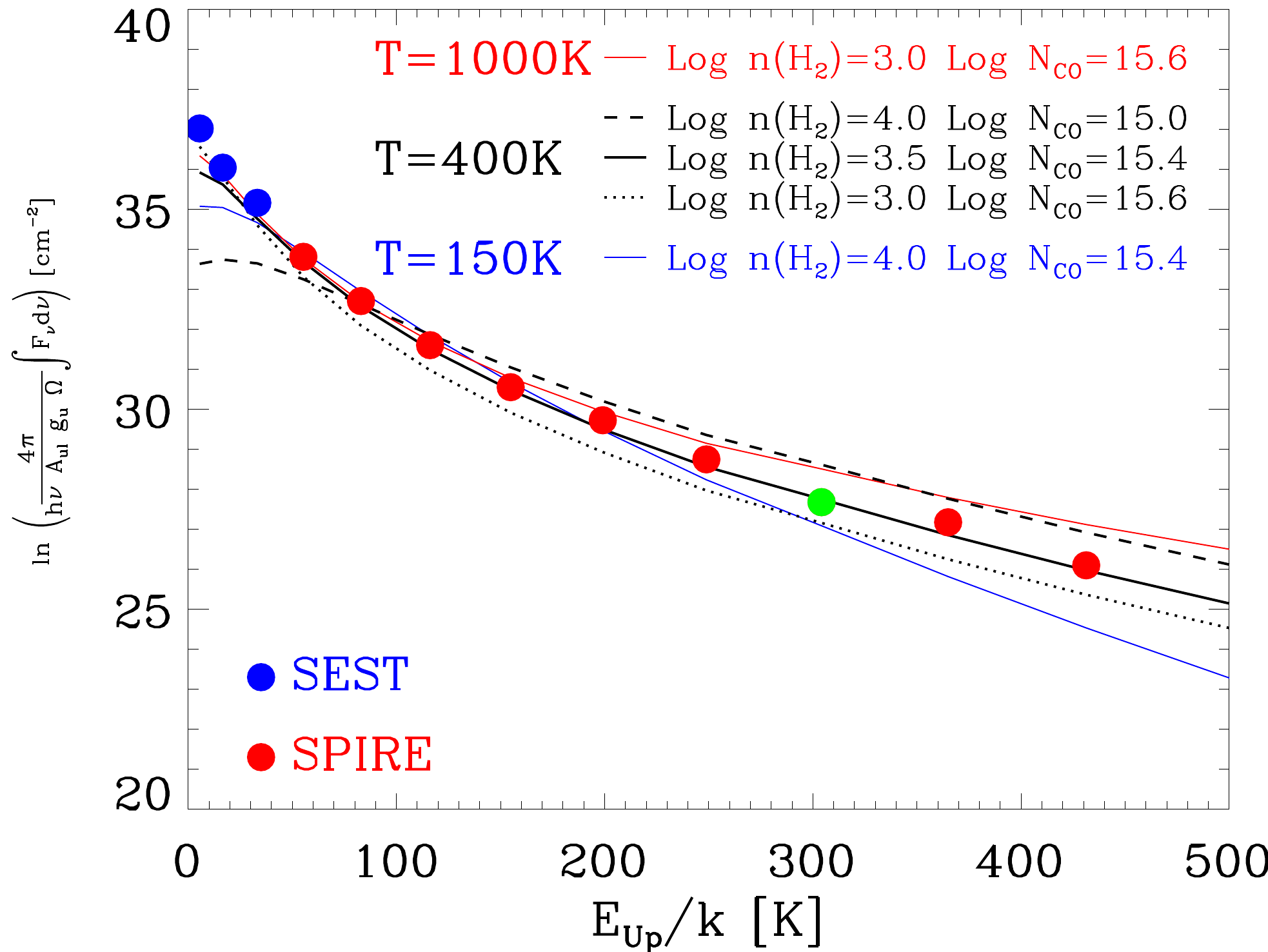}
\includegraphics[angle=90, width=.49\textwidth]{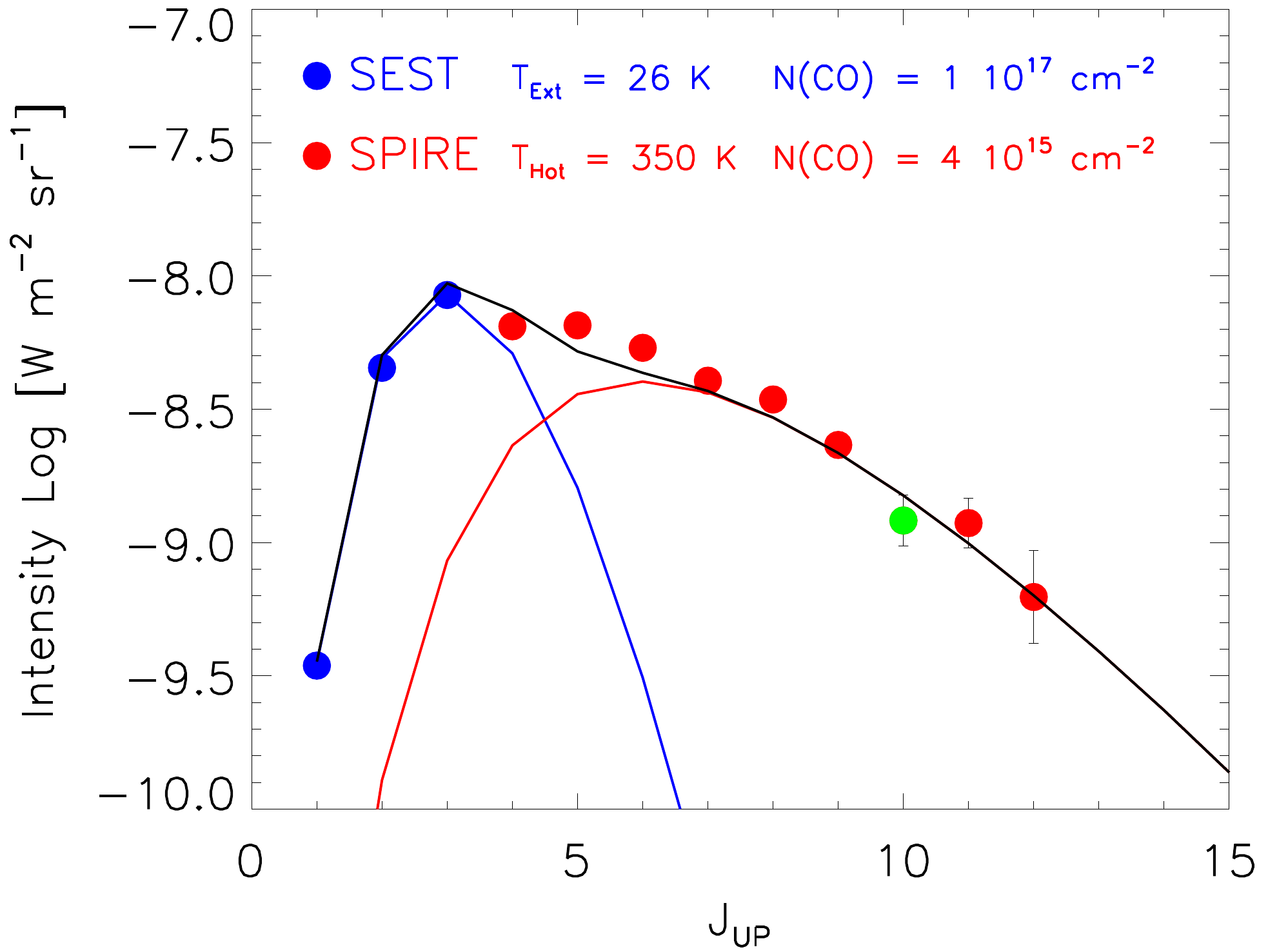}
\includegraphics[angle=90, width=.49\textwidth]{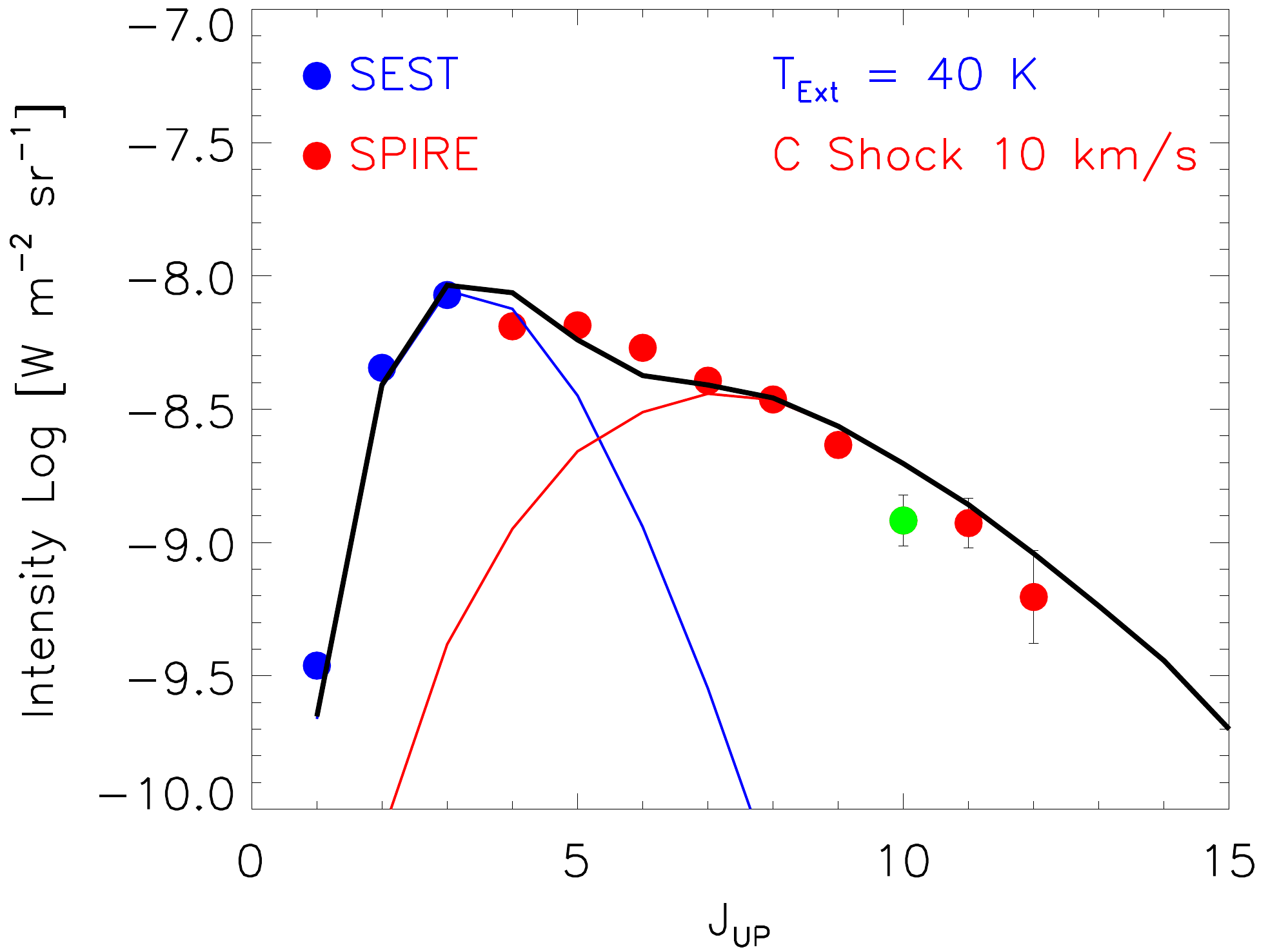}
\includegraphics[angle=90, width=.49\textwidth]{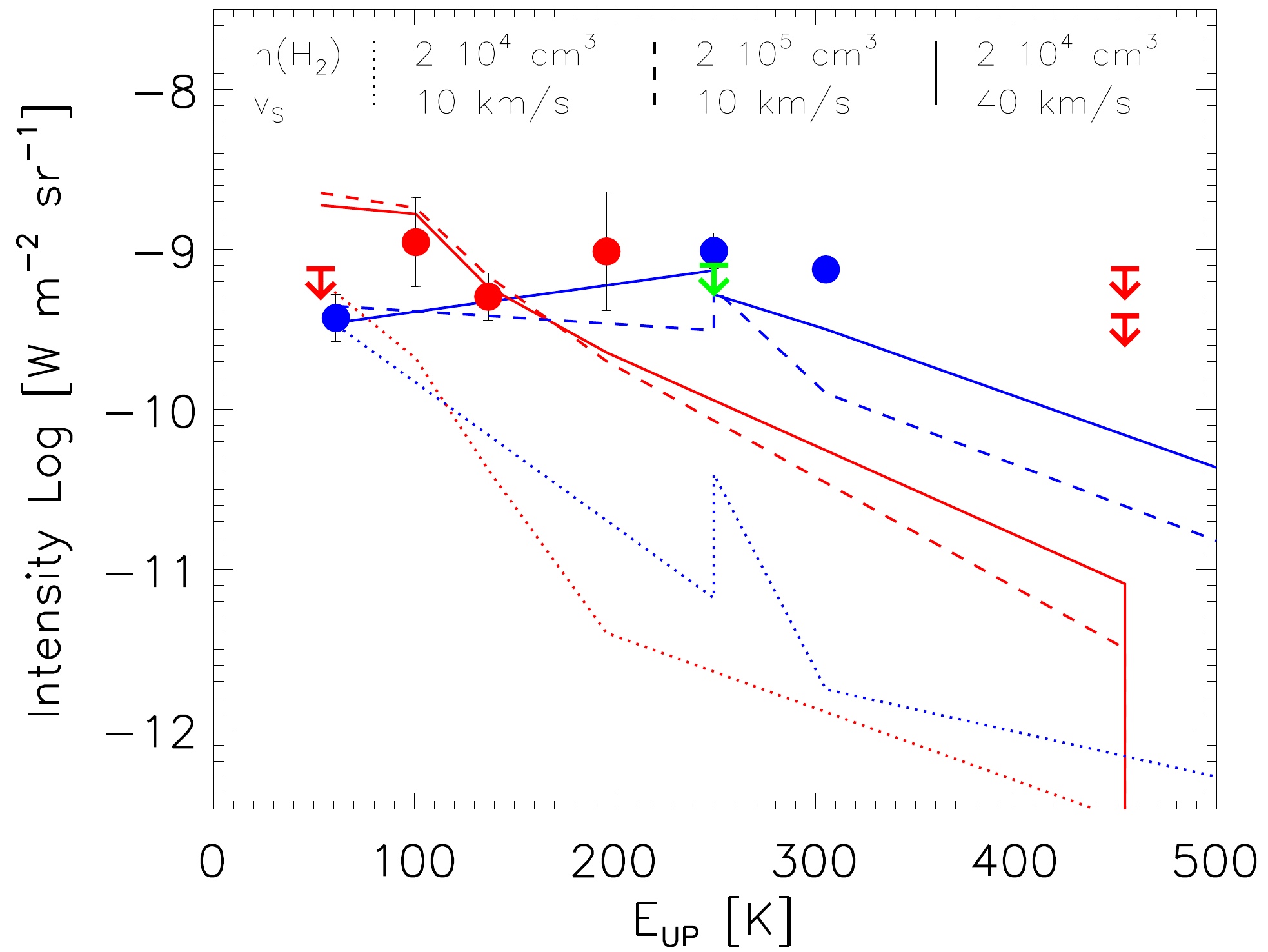}
\caption{{\it Upper left}: Rotation diagram for the CO lines. Blue dots: SEST 
  observations from the literature, Red dots: SPIRE data, Green dot:
  SPIRE CO line blended with a water line. The full, dashed, and dotted
  lines show different single temperature RADEX models. {\it Upper right}: Two
  temperature (RADEX) fits to the CO lines. One extended cold
  component (blue) with $n$(H$_2$) = $7 \times 10^3$ cm$^{-3}$ and
  $N$(CO) = $1 \times 10^{17}$ cm$^{-2}$, and one hot component (red)
  from more compact regions $n$(H$_2$) = $10^4$ cm$^{-3}$ and $N$(CO)
  = $4 \times 10^{15}$ cm$^{-2}$. {\it Lower left}: A model
  consisting of a slightly warmer extended component 
    with RADEX values of $n$(H$_2$) = $7 \times 10^3$ cm$^{-3}$ and
    $N$(CO) = $7 \times 10^{16}$ cm$^{-2}$ (blue), and a slow
 velocity C-type shock with a shock speed of 10 km s$^{-1}$ and a
 pre-shock density of $n$(H +2H$_2$) = $2 \times 10^{4}$
 cm$^{-3}$ (red). The C-shock model comes from Flower \&\ Pineau des
 For\^ets (\cite{flo10}) where we have scaled their values to a
 10\as-source in order to obtain a best fit. {\it Lower right}: Water lines
 modeling. Water emission lines observed by 
    SPIRE – Blue dots: o-\hto, Green arrow: Blended
    ortho-line, Red dots: p-\hto. The same slow-velocity C-type shock
    as for CO but the model with a shock velocity of 10 km
    s$^{-1}$ and a pre-shock density of $n$(H +2H$_2$) = $ 2 \times
    10^{4}$ cm$^{-3}$ underestimates the line intensities. Increasing
    the speed or the pre-shock density will give a better fit but it's
    still hard to fit all para lines.} 
\label{9}
\end{figure*}

\subsection{Nuclear gas accretion and outflows} 

The gas inflow processes from the outer torus region onto the inner
torus, nuclear accretion disk, of NGC\,1365 were discussed already in
the Introduction of this paper. We have in Table 3 collected a
number of relevant parameters for the galaxy spiral arm disk, the bar,
the bar-driven outer torus region, the inner cold gas torus, the
central rapidly spinning SMBH, and the various observed nuclear outflows
We have also included in Table 3
some useful comparison parameters for the lens-shaped spiral galaxy
NGC 1377, seen edge on, where rather convincing signs of a precessing
nuclear bi-polar molecular outflow were observed at ALMA (Aalto
et al. \cite{aal16}).  

\subsubsection{Atomic, molecular and ionic outflows, and a synchrotron
  radiation jet from the nuclear engine} 

It was mentioned already in the Introduction, that among
the radio hot spots of the circumnuclear ring of gas in NGC\,1365,
ionized by recent star formation, the outer torus of radius 800 pc,
a region “F” with a very steep radio spectrum was “hiding”, see
  Fig. 7(4) (Sandqvist et al. \cite{san95}). This was further
illustrated in our Table 2. The envisioned synchrotron radiation
relativistic outflow jet from, and presumably powered by, the nuclear
engine, the black hole and its 
accretion disk, therefore is expected to be be oriented along the
rotation axis of the inner circumnuclear molecular gas torus. However,
the projected jet orientation instead appears to be along the
similar rotation axes of the galaxy, its prominent bar, and the outer
molecular gas torus, and along the symmetry axis of the wide-angle ionized gas
outflow observed in optical \OIII\ lines, as illustrated in
  Fig. 7(4) (Sandqvist et
al. \cite{san95}; Hjelm \&\ Lindblad \cite{hje96}; Sandqvist
\cite{san99}; Lindblad \cite{lin99}; Venturi et al. \cite{ven18}; see
our Table 3). This observational fact is very important in our
forthcoming scenario interpretation (Sect. 4.3.2). The observation of a
synchrotron radiation jet, having a de-projected size of some 400 pc,
depending upon its somewhat uncertain inclination, see 
Table 3, in fact means that we are probing the radiative
acceleration losses in a plasma beam where the electrons are spiraling
at relativistic velocities along a strong co-aligned magnetic
field. Here it is interesting to point out that the observed magnetic
field, which according to the VLA mapping by Beck et al. (\cite{bec05}) is
closely aligned with the bar just in the central regions, is turning
its direction toward the nuclear engine.

\begin{sidewaystable*}
\caption{Physical parameters of NGC\,1365 compared with those of NGC\,1377.}
\begin{flushleft}
\begin{tabular}{lllllllll}
  \hline\hline\noalign{\smallskip}

  Object  & Rotation & {\it or} &  Radius & Rotation {\it or} &  Mass &  &
                                                                 Period
                                                                 &
                                                                   Remarks
  \\
     &  Outflow  & Axis  & {\it or} Size & Outflow Vel. &
                                                   Dyn.\tablefootmark{a} &
                                                            Gas\tablefootmark{b}
                                                            &  {\it or} Age &  \\
      & P.A.[\degr]  &  Incl.[\degr] &   [{\it kpc}/pc] & [\kms] &
                                                          [$\times10^9$
                                                          \msol] &   &
                                                                       [Myr]
                                                                 &  \\

  \hline\noalign{\smallskip}

  {\bf NGC\,1365} \\
  
  \hline\noalign{\smallskip}

 Disk\tablefootmark{c} & 130 & 50 & {\it 22}  &  $\approx$270 & 360 & 13 &
                                                               $\approx$500
                                                               & \HI \ 
                                                                 mass
  \\
  Bar\tablefootmark{c,d} &   & 50 & {\it 10} & 180 & 13\tablefootmark{d} & & 
                                                                     340
                                                               &
                                                                 Stellar
                                                                 mass
  \\
  Disk & & 50 & {\it 10} & 310 & 220 & 9 & 200 & \htwo\ mass \\
  
  Disk & & & {\it 2} & 310 & 44 & 2.8 & 40 &  \\
  \hline\noalign{\smallskip}
  Disk/torus\tablefootmark{b} & & & 1 & 230 & 12 & 2.4 & 27 & ``Star
                                                              formation
                                                              gas ring'',\\
  Outer torus & $\approx$130 & $\approx$50 & 800 & 185 & 6  &  & 27 &
                                                                      with
                                                                      solid
  body rotation\\
  Outer torus &  & & 500 & 115 & 1.5 &  & 27 & as visualized in
                                               Fig. A.1 \\
 \hline\noalign{\smallskip} 
{\it Nuclear engine:} & & & & & & & \\
  Inner torus\tablefootmark{d} & $\approx$160 & $\approx$63 & 26 &
                                                                  $\approx$100
                                                    & 0.06 & 0.007 &
                                                                     1.6
                                                                 &
                                                                   Note
                                                                   torus
                                                                   axis
                                                                   tilt\tablefootmark{d},
                                                                   caused
                                                                   by
                                                                   $\approx$
                                                                   30\degr\
                                                                   precession angle 
  \\ 
  {\bf Black Hole}\tablefootmark{d} & & & & & 0.004 & &  & Rapidly spinning
                                                        SMBH \\
  \hline\noalign{\smallskip}
  {\it Nuclear outflows}: \\
  Radio Jet\tablefootmark{c,e} & $\approx$135 & $\approx$20 & $\approx$400
                                                    & 0.7c? & & & &
                                                                   Table
                                                                    2;
                                                                    Inclination
                                                                    caused
                                                                    by
                                                                    nuclear
                                                                    engine
                                                                    tilt

  \\
  CO outflow\tablefootmark{f} & $\approx$ 135 & $\approx$20 & $\approx$400 &
                                                                     $\approx$70
                                                    & & & 5.9 &
                                                                Sect. 4.3.1
                                                                \&\
                                                                   ref
                                                                   d);
                                                                   Inclination
                                                                   by
                                                                   nuclear
                                                                   engine tilt
  \\
    &   &  ($\approx$80) & ($\approx$2200) & ($\approx$23)&  &   &
                                                                   (90)
                                                                 &
                                                                   Unlikely
                                                                   inclination
                                                                   alternative;
                                                                   See
                                                                   Sect. 4.3.2
  \\ 
  \OIII\ bi-cone\tablefootmark{g} & 130 & 50 & $\pm 2500$ & $\pm 200$ &
                                                                       $-$
                                                            & 0.001 &
                                                                     $\approx$12
                                                                 &
                                                                  Hollow,
                                                                   wide
                                                                   angle
                                                                   ($\pm
                                                                   50\degr$)
                                                                   ionized
                                                                   gas
                                                                   ``fountain''
  \\
  \HI\ outflow & $\approx$160 &  & 2000 & $\pm200$ & & & $\approx$10 &
                                                                      Fig. 7;
                                                                         Desorption/dissociation 
                                                                         from
                                                                         icy
                                                                         grains?
  \\
  
  \hline\noalign{\smallskip}
 {\bf NGC1377} \\
  \hline\noalign{\smallskip}
  Nuclear disk\tablefootmark{h} & 15 & $\approx$0 & 29 & 75 & 0.04 &
                                                                     0.017
                                                               & & \\
  CO outflow\tablefootmark{h} & 15 & $\approx$0 & 200 & 140 & $-$ &
                                                                    $0.01-0.05$
                                                               & 1.4 &
                                                                       Opening
                                                                       angle
                                                                       $\approx
                                                                       60
                                                                       -
                                                                       70\degr$
  \\
  Map area\tablefootmark{j} & 15 & $\approx$0 & $-$ & & $-$ & 0.16 & &
                                                                       Size:
                                                                       $4\as
                                                                       \times
                                                                       4\as$
  \\
Accretion disk\tablefootmark{j} & 15 & $\approx$0 & $\approx$10 & 110 & 0.03 &
                                                                      0.018
                                                                      &
                                                                      $\approx$1
                                                                      &
                                                                        Rotation
                                                                        period
 \\
Bipolar CO Jet\tablefootmark{j} & 15 & $\pm (10-25$)\tablefootmark{k} &
                                                                       $\pm
                                                                        150$
                                     & $240-850$\tablefootmark{m} & $-$ & $0.002 -
                                                         0.02$\tablefootmark{m} &
                                                                 $0.2-0.7$
                                                                 &
                                                                   Precession Period:
                                                                   $0.3-1.1$
  \\
\hline\noalign{\smallskip}
  
\end{tabular}

    \tablefoottext{a}{This paper, Eq.(2).}
    \tablefoottext{b}{This paper, Sect. 4.1.}
    \tablefoottext{c}{Lindblad (\cite{lin99}).}
    \tablefoottext{d}{Combes et al. (\cite{com19}), who (in their
      Table 4) list an inner torus tilt, with an uncertainty of $\pm 10\degr$.}
    \tablefoottext{e}{Sandqvist et al. (\cite{san95}).}
    \tablefoottext{f}{This paper, Table 2 \&\ Sect. 4.3.2.}
    \tablefoottext{g}{Hjelm \&\ Lindblad (\cite{hje96}); Lindblad
      (\cite{lin99}); Venturi et al. (\cite{ven18}).}
    \tablefoottext{h}{Aalto et al. (\cite{aal12}).}
  \tablefoottext{j}{Aalto et al. (\cite{aal16}).}
  \tablefoottext{k}{in Ref. (j) estimated precession angle of CO jet
    and nuclear engine.}
  \tablefoottext{m}{interval depending upon precession angle.}
\end{flushleft}
\end{sidewaystable*}

According to the CO($2-1$)
SMA mapping by Sakamoto et al. (\cite{sak07}), there is in the outer torus also
a “molecular cloud hot spot” co-located with the synchrotron jet
source “F”, which has an intensity peak at a velocity that is
inconsistent with the expected streaming motion in the 
torus. If the CO($2-1$) hot spot at F were a  “mirror image” of
   the one at D – moving along the circumnuclear torus – its line of
   sight (LOS) velocity, according to Table 2, is expected to be
   $\approx 1613  + (1613 - 1560) = 
   1666$ \kms. However, its observed peak emission LOS velocity is 1590
   \kms,  that is, $\approx$ 76 \kms\ “too low” and “blue shifted” versus
   the systemic velocity by $1613-1590 = 23$ \kms. This points at the
   existence of a molecular gas outflow from the nuclear engine,
   oriented along the direction of the nuclear synchrotron radiation
   jet. The peak emission is LOS blue-shifted by about 25 \kms\ 
with respect to the systemic velocity. This suggests the presence
of a rather collimated molecular outflow, having a de-projected
velocity of about 70 \kms\ in the same direction as the synchrotron
jet, and which perhaps is driven by the same launching process (see Table
3). In fact, also Combes et al. (\cite{com19}) propose the possibility
of a nuclear outflow, observed in the ALMA CO($3-2$) velocity field,
similar to that found for NGC\,613 (Audibert et
  al. \cite{aud19}). We here compare the CO($2-1$) profile observed at
2\as\ resolution in radio source F (Fig. 11 of Sakamoto et
al. \cite{sak07}), displaying a pronounced rather narrow peak at 1590
\kms\ on top of a broader range (1500 – 1750 \kms) pedestal, with our
5\as\ resolution CO($3-2$) spectrum in the same direction. This is
shown in Fig. 10, where the arrow indicates a narrow feature. The
weakness of the presumptive molecular jet emission at $\approx -25$
\kms, observed at 5\as\ resolution may suggest that the extent
of this emission is $\lid  2$\as, equivalent circular size in
the plane of the sky, that is, indicating a collimated, molecular gas
outflow of size 4\as\ $\times$ < 1\as.

\begin{figure}
  \resizebox{\hsize}{!}{\rotatebox{0}{\includegraphics{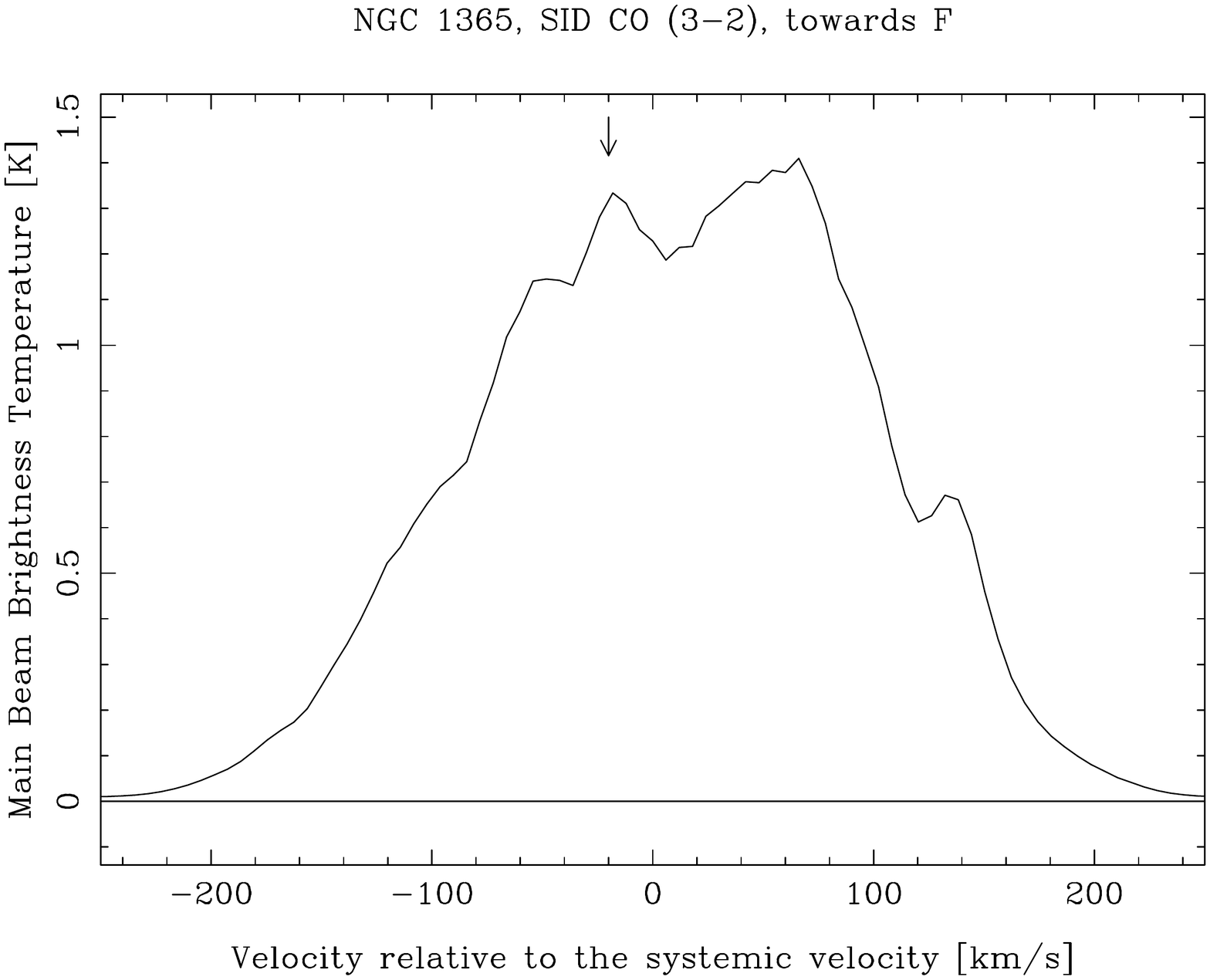}}}
  \caption{SID CO($3-2$) profile toward the continnum source
    ``F''. The arrow indicates a weak and narrow feature,
      possibly indicative of the molecular outflow, discussed
      in Sect. 4.3.1}.       
  \label{10}
\end{figure}

The envisioned $\approx$400 pc size relativistic synchrotron jet and collimated
molecular outflow are enshrouded by a kpc size fan- or, rather,
cone-shaped ionized gas outflow, revealed by its visual \OIII\ emission
(Hjelm \&\ Lindblad \cite{hje96}, discussed in considerable detail in
the review paper by Lindblad \cite{lin99}). More recently, this
hollow wide-angle ($\approx$100\degr) outflow of ionized gases has
been studied in considerable detail by Lena et al. (\cite{len16}) and Venturi et
al. (\cite{ven18}). From their 
optical high spectral resolution mapping of the central 5 kpc (60\as
$\times$ 60\as) region of NGC\,1365, at a spatial resolution of 70 pc
(0.76\as), by means of VLT MUSE (multi-unit spectroscopic explorer), paired with
Chandra X-ray data and elaborate analyses, Venturi et
al. (\cite{ven18}) revealed the very inhomogeneous and
anisotropic structure 
of the mass outflow rate and the velocity field of this bi-conical,
wide angle, $\pm 2.5$ kpc size, nuclear outflow of ionized gases, having
velocities up to 200 \kms. Their Fig. 7, displaying the estimated
ionized gas mass outflow rate at 5\as\ (450 pc) resolution, demonstrates
very pronounced radial as well as angular variations in the range
($10^{-5} - 10^{-2}$) \msol/yr.
From a consideration of the kinetic energy rate and the momentum rate
of the extended outflow of ionized gas and the mass outflow rate
available from the AGN, obtained from their Chandra X-ray 
data, Venturi et al. argue that the extended ionized gas outflow must be
momentum powered, that is, driven by the X-ray pressure (i.e., by X-ray
photon collisions with the dust component of the gas). The authors
also point out that the expected multiple supernova explosions (SNe)
in the “star formation ring”, the outer torus in our nomenclature, may
also contribute to the powering and temporal variation of
the very extended ($\pm 2.5$ kpc), wide angle ($\pm 50$\degr), bi-conical
nuclear outflow. This is especially so since the observed
outflow extent is as large as the diameter ($\lid$2 kpc) of the outer
torus, while the diameter of the AGN accretion disk (the inner torus)
is only $\approx$50 pc (see our Table 3). A main reason for the radial
variations of the ionized gas may be found in the observed variable
nuclear X-ray emission causing (temporal) variations in the outflow
launching radiation pressure. We return to the likely cause of
the angular variations in Sect. 4.3.2. While SNe-induced multiple “chimney
processes” presumably are in action in the outer torus (as modeled by
e.g., Norman \&\ Ikeuchi \cite{nor89}; Melioli et al. \cite{mel09}),
the larger scale “nuclear fountain” scenario may be the main driver
(as modeled by e.g., Wada \cite{wad12}, \cite{wad15}; Wada et
al. \cite{wad16}).  

In Fig. 7(1) it may be seen that there is an $\approx 20\as$ \HI\
absorption ridge emanating from the nucleus, in a south-southeast
direction. This \HI\ ridge can also be followed in the four individual
maps shown in Fig. 8, implying that the ridge has a very
wide velocity dispersion ($\approx 400$ \kms). This \HI\ ridge
coincides in position with the southern edge of the \OIII\ cone and
may have arisen by the action of collisions of the outflow-driving
X-ray photons with the dust grains. This interaction most likely would
cause the water ice covering the cold dust grain surfaces to be
desorbed and subsequently cause the water vapor to be dissociated
into 2H + O. It is here interesting to note that, according to Fig. 7
of Venturi et al. (\cite{ven18}), the mass outflow rates of ionized
gas are observed to be especially high along P.A. $\approx$ 135\degr\
and $\approx$ 160\degr, that is, in the direction of the synchrotron
radiation and molecular gas jet as well as along the elusive \HI\
absorption ridge just discussed. Here it may be interesting to note
that the \HI\ ridge may well be accompanied by a low level wide velocity
CO($3-2$) emission ridge at P.A. $\approx 160$\degr, as hinted in
Fig. 7(2).  

The very wide velocity dispersion of the \HI\ absorption ridge,
mentioned above, with both negative and positive velocity components
is perplexing, but not completely surprising. Hjelm \&\ Lindblad (\cite{hje96})
found that the southeastern edge of their \OIII\ cone had also shown a
”split” structure with both negative and positive velocity components
in a velocity range similar to the \HI\ ridge (see their observations
with slits 6, 7, 10, and 11 in their Fig. 6). This was interpreted as
contributions from both the outflow cone and the  disk. A first
approximation to the \HI\ column density, $N$(\HI), in the \HI\ ridge can be
obtained from $N$(\HI)/$T_{\rm s} = 1.823 \times 10^{18} \times \tau
\times \Delta V$ (cm$^{-2}$K$^{-1}$), where 
$T_{\rm s}$ is the spin temperature, and the optical depth of the \HI\
absorption is $\tau = -$ln$[1+(I_{\rm \HI})/(I_{\rm c})]$,  $I_{\rm
  \HI}$ (a negative number) and 
$I_{\rm c}$ are the intensities of the \HI\ absorption line and continuum,
respectively, in mJy/beam (Sandqvist \cite{san74}). The beam for the
  \HI\ observations is 11\farcs 6 $\times$ 
6\farcs 3, and the continuum has been estimated from Fig. 6 of Sandqvist et
al. (\cite{san95}) and corrected to the \HI\ resolution. A second approximation
has been made using the continuum observations in Fig. 7 of Beck et
al. (\cite{bec02}), also adjusted to the \HI\ resolution. The results
for two positions in the \HI\ ridge are presented in Table 4.

\begin{table*}
\caption{\HI\ column densities in the \HI\ absorption ridge.}
\begin{flushleft}
\begin{tabular}{llllllllll}
  \hline\hline\noalign{\smallskip}

Source & Offsets & $I_{\rm \HI}$ & $\Delta V$ & $I_1$ & $I_2$ &
                                                                $\tau
                                                                _1$ &
                                                                      $\tau_2$
  & $N_1/T_{\rm s}$ &  $N_2/T_{\rm s}$ \\
     &  ($\Delta \alpha$,$\Delta \delta$) & [mJy/beam] & [\kms] &
                                                                 [mJy/beam]
                                                      & [mJy/beam] & &
  & [cm$^{-2}$K$^{-1}$] & [cm$^{-2}$K$^{-1}$] \\

  \hline\noalign{\smallskip}

  \HI\ ridge top & ($+2\as,-9\as$) & $-0.45$ & 380 & 5.2 & 14.5 &
                                                                  0.091
                                                                    &
                                                                      0.032
  & $6.3 \times 10^{19}$ & $2.2 \times 10^{19}$ \\
  
 \HI\ ridge bottom & ($+7\as,-15\as$) & -0.65 & 340 & 5.2 & 14.5 &
                                                                   0.13
                                                                    &
                                                                      0.045  & $8.3 \times 10^{19}$ & $2.8 \times 10^{19}$ \\

\noalign{\smallskip}\hline\end{tabular}
\end{flushleft}

\end{table*}

Using the canonical \HI\ spin temperature of 100 K, we may now estimate
an \HI\ column density for the \HI\ ridge of a few $\times 10^{21}$
cm$^{-2}$. For the maximum \HI\ integrated emission region in Fig. 7,
located near the position of the \HII\ hot spot L4, we determine an
\HI\ column density of $6 \times 10^{22}$ cm$^{-2}$, using the
conversion factor given by J\"ors\"ater \&\ van Moorsel (\cite{jor95})
– a column density value similar to that found for the +50 \kms\ cloud
in the Sgr\,A complex in the center of our own galaxy (Sandqvist \cite{san74}). 

In support of our own forthcoming scenario proposal in Sect. 4.3.2, we
now will turn to the very interesting SMA and ALMA results presented by
Aalto et al. (\cite{aal12}, \cite{aal16}). Their CO($2-1$) SMA
observations at a resolution of $\approx$ 
0.6\as\ of the lenticular spiral galaxy NGC\,1377, at a
distance of 21 Mpc where 1\as\ $\approx$102 pc, – very similar to the parameters
of NGC\,1365 – revealed an extended molecular outflow and an
orthogonal nuclear disk – outflow extent $\approx$200 pc, opening angle
of $60\degr - 70\degr$, outflow velocity $\approx$140 \kms\ and mass
$\approx (1-5) \times 10^7$ \msol; dynamical versus molecular nuclear disk
masses $\approx 4 \times 10^7$ \msol\ versus $1.7 \times 10^7$
\msol). 

From their
analysis of the ALMA CO($3-2$) observations at 0.2\as\ resolution, and
maximum recoverable structure size of 5\as, Aalto et
al. (\cite{aal16}) discovered a very collimated molecular
jet, counter-jet system of 
size $\pm 150$ pc, displaying velocity reversals along the jets with a
swing of $\pm 150$ \kms. This was interpreted and satisfactorily
modeled as a precessing bi-polar jet system, having a precession
angle $\approx 10 - 25$\degr\ around a direction along the plane of
the sky, precession period $\approx$ 0.3 – 1.1 Myr, outflow velocity
$\approx$ 240 – 850 \kms, and outflow mass $\approx$ (0.2 – 2)$ \times 10^7$
\msol. The outflow rate was estimated to be $\approx$10 \msol /yr – a value
they considered to be very uncertain. The molecular mass in the entire
$4\as \times 4\as$ map was estimated to be $16 \times 10^7$ \msol. The
rotational velocity of an accretion disk of size $\approx$10 pc was
estimated to be $\approx$110 km/s, which implies a rotation period of
$\approx$1 Myr. This in fact provides a lower limit to the jet
precession period, which must be larger than the rotation period of
its launching accretion disk. The more relevant physical parameters
have been collected in our Table 3, for comparison with our scenario in
NGC\,1365. 

Aalto et al. (\cite{aal16}) propose that their
elusive, gaseous precessing  jet, counter-jet system is likely to be
powered by faint bipolar “radio jets”, that is, by relativistic plasma
outflows suffering from energy losses by synchrotron radiation, as was
described earlier, or alternatively by an accretion disk-wind similar
to those found toward proto-stars. The latter case also includes
discussions of analogous MHD models of precessing bi-polar outflows,
powered by the accretion onto a proto-star – a jet precession which
has been observed, for example, in L1551 IRS5 and IRAS 16293-2422 (Fridlund
\&\ Liseau \cite{fri94}; Kristensen et al. \cite{kri13}). Aalto et
al. (\cite{aal16}) also carefully discuss the various 
scenarios which may cause the observed molecular jet precession, with
at-the-point references. Their favored alternatives are: i)
misalignment between the spin orientation of the SMBH and the rotation
axis of the surrounding accretion disk, or ii) an accretion inflow of
gas having misaligned angular momentum. 

\subsubsection{Tilted massive inner torus of cold gas – a precessing
  nuclear engine?} 

Scrutinizing the various observed and calculated physical
parameters collected in Table 3, we note two problematic issues 
where nature, however, has provided elegant and efficient solutions. The first
one is the “angular momentum problem” at the formation of a compact
object – in our case a central black hole, while a very similar problem
appears in proto-stellar formation – as illustrated by the rotation
time column. While the rotation time of the galaxy disk falls in the
range $500-40$ Myr, the bar-induced outer torus, having a radius of $400
-1000$ pc and a cold gas mass of $\approx 2 \times 10^9$ \msol, is
rotating almost as a solid body with a revolution time of 27
Myr. This outer torus is encircling a much smaller inner
torus, accretion disk of radius 26 pc and estimated cold gas mass of
$\approx 7 \times 10^6$ \msol\ and a rotation time of $\approx 1.6$
Myr – in its turn surrounding a rapidly spinning SMBH having a mass
of $\approx 4 \times 10^6$ \msol\ (Combes et al. \cite{com19}; Risaliti
et al. \cite{ris13}). To overcome the “angular momentum problem”
nature has found a solution here to the energy balance problem in terms of
MHD launching of a relativistic plasma outflow, which is observable because of
its synchrotron radiation loss, and a co-existing collimated molecular gas
(CO) outflow. It has also launched the directly observable X-ray emission from
the nuclear engine, which in addition is causing the observed very
extended wide angle ionized gas outflow by the action of its radiation pressure.

With these facts in
mind, we arrive at the second problem in case of NGC\,1365. While the
symmetry axis of the hollow wide angle ionized gas outflow
appears to coincide with the rotation axis of the galaxy disk and the
outer torus (P.A. = 130\degr; Incl. = 50\degr), the rotation axis of
the inner torus, black hole nuclear engine (P.A. =
160\degr$\pm$10\degr; Incl. = 63\degr$\pm$10\degr) according to the
model fit of CO($3-2$) ALMA data by Combes et al. (\cite{com19}),
deviates by 
$\approx$ 30\degr $\pm$10\degr\ from the common direction. The direction of the
synchrotron radiation jet, collimated molecular gas outflow is
observationally known only in terms of its P.A. $\approx$ 135\degr.
However, to allow the previously outlined powering scenario of the
synchrotron radiation jet, collimated CO outflow, the rotation axis of
the nuclear engine must have agreed with the outflow direction at an
earlier time. If we consider these facts, the only realistic 
alternative seems to be that the inner torus is precessing, at an
assumed precession angle of $\approx$ 30\degr, around the common
rotation axis of the galaxy disk and the outer torus. The precession
period must be larger than the inner torus rotation period of 1.6 Myr and
presumably be lower than the rotation period of the outer torus, 27 Myr.
The inclination of the synchrotron radiation, CO jet in this scenario must be
$\approx$ 50\degr$\pm$30\degr, that is, Incl. = 20\degr, or 80\degr,
where the former case results in an outflow size of 400 pc, a
reasonable outflow velocity of 70 \kms\ and an outflow age of
$\approx$ 6 Myr. The other alternative must be considered unlikely
because of the resulting very large size and age (see Table 3). Since
the rotation axis of the accretion disk must have precessed (in the
same direction as the rotation of the outer and inner tori) by about
75\% of its period from its launching of the synchrotron jet, molecular
outflow to its current position, we may now estimate the precession
period of the nuclear engine to be about 8 Myr, which is well below
the rotation period of the outer torus, 27 Myr. 

This scenario solves not only the
launching of the synchrotron, CO jet but may also help to explain the
angular mass loss rate inhomogeneity of the wide angle ionized gas
outflow observed by Venturi et al. (\cite{ven18}). In addition the proposed
inner torus precession scenario provides a natural solution of the
problem that the bi-conical outflow is covering a very wide sector
(P.A. $\approx$ 80\degr\ $-$ 180\degr) even at a radius of 2.5 kpc,
which corresponds to a launch time of $25 - 50$ Myr at an estimated
outflow velocity of $100 - 50$ \kms\ and is hollow, that is, with little or
no outflow near the symmetry axis (cf. Hjelm and Lindblad
\cite{hje96}). This is not well accommodated by state of the art
theoretical modeling of nuclear outflow fountains (e.g., Wada
\cite{wad15}). A precession angle of $\approx$ 25 $-$ 30\degr\ and a bi-conical
  outflow sector of $\pm$25\degr\ should here be able to do the
  required job of supporting the observed ionized outflow, created
  during several precession periods. 

At first sight a precession angle as large as 20\degr\ $-$ 40\degr\ may seem
unrealistic. However, the ALMA observations of NGC\,1377 by Aalto et 
al. (\cite{aal16}), paired with their careful analysis, lead to a required
accretion disk precession angle of 10\degr\ $-$ 25\degr. The physical reason for
the suggested inner torus precession in NGC\,1365 could be “an
accretion inflow of gas having misaligned angular momentum”, which is the
previously mentioned second alternative for NGC\,1377 of Aalto et
al. (\cite{aal16}), since the outer torus of NGC\,1365 is asymmetric
and very inhomogeneous in its mass distribution. As shown in Table 3, the
accretion tori of NGC\,1365 and NGC\,1377 have similar sizes, rotation
times and cold gas masses.

\subsection{Discussion of the \hto\ emissions in NGC\,1365 versus
  those in other galaxies}  

The outer circumnuclear torus region and the nuclar region contain all
the ingredients necessary to locally raise the gas-phase water
abundance to levels which are several orders of magnitude higher than
the quiescent, cold cloud value of $8 \times 10^{-10}$ versus \htwo, assuming an ortho-to-para ratio of 3. Here most \hto\ is likely to
reside as water ice on cold grain surfaces (e.g., van der Tak et
al. \cite{tak10}).  

We have in Table 5 collected the results from our spatially unresolved
but spectrally resolved {\it Odin} observations of the 557 GHz ground state
o-\hto\ line. There are also the results from spectrally unresolved, but
spatially partly resolved, {\it Herschel} SPIRE
multi-transition \hto\ and CO mapping observations, for comparison with:
i) {\it Odin} ground state o-\hto\ and o-\htio\ results for our Galactic center
region, Sgr\,A (Karlsson et al. \cite{kar13}),  ii) {\it Herschel}
HIFI, SPIRE, and PACS results for the merger NGC\,6240 (with
galaxy-wide shocks; Meijerink et al. \cite{mei13}, Liu et
al. \cite{liu17}), and iii) the results from a detailed radiative
transfer and excitation analysis of a sample of other galaxies
extensively observed by {\it Herschel} HIFI, SPIRE, and PACS (Liu et
al. \cite{liu17}; Gonzalez-Alfonso et al. \cite{gon10}, \cite{gon12}).   

In addition to the comparisons between galaxies provided in Table 5,
we have in an Appendix B included a Table B.1 comparing our estimated
\hto\ abundances in NGC\,1365 with more detailed Sgr\,A results, as
well as with observed \hto\ abundances in a number of well-known types
of Galactic molecular cloud regions, where also detailed chemical models are
established. For the interested reader a short summary of current
chemical models of water formation and destruction is also included in
this Appendix.

\begin{table*}
\caption{Comparison of NGC\,1365 and other galaxies.}
\begin{flushleft}
\begin{tabular}{llllll}
\hline\hline\noalign{\smallskip}

Object & Region	&  Density & Gas / Dust	&  \hto\ abundance &  Comments
  \\
    &  &  (cm$^{-3}$) & Temperature (K)	&     vs \htwo\ ($\times
                                          10^{-8}$) & \\
\hline\noalign{\smallskip}
  
NGC\,1365 &	absorption\tablefootmark{a} &	  $10^2-10^4$ &
                                                                $\approx
                                                                20$\tablefootmark{c}
                                        & $5-8$   &  A \\
 (Barred & emission\tablefootmark{a,b} &	  $10^4-10^6$ &
                                                                $40-60
                                                                /
                                                                55$\tablefootmark{c}
                                        &  $6-600$\tablefootmark{j,k}	&              A	\\
    spiral) &	shocks\tablefootmark{b}	& $10^4-10^5$\tablefootmark{d}
                                          & $350 /
                                            22-32$\tablefootmark{c} &
                                                                      $200\tablefootmark{n}
                                                                      -10,000$\tablefootmark{l} 
                                                                      &
                                                                        B
  \\ \\
Sgr\,A\tablefootmark{e} &	absorption &
                                             $10^2-10^3$\tablefootmark{i}
                           &
                             $100-30$\tablefootmark{i}/15\tablefootmark{m}
                                        & 3 &	     C  
  \\
  &  emission &	 $10^4-10^5$\tablefootmark{f} &
                                                80\tablefootmark{f}/$20-30$\tablefootmark{m} 
                                        & 
                                                $2 -  7$    &
                                                               D \\
  &	line wings &	$3 \times (10^4-10^5)$	&
                                                 $80-160$/$20-30$\tablefootmark{m}
                                        & 
                                                              $80 - 570$
                                                           &
                                                                     E
  \\ \\
NGC\,6240 & warm emission &	     $10^6$	&  $60-70/60-70$  &
                                                                  10 &
                                                                       F
  \\
  (Merger\tablefootmark{g,h}) &	shocks &	$5 \times
                                         (10^4-10^6)$\tablefootmark{d}
                           &  $120-400/20-30$   &       10 &	G \\
  \\
  Galaxy &	absorption &	 $10^3 -10^5$ &	 $20-200/20-30 $ &
                                                                   $0.1-10$
                                                         &          H,
  J \\
sample\tablefootmark{h} &	cold  emission &	 $10^3 -10^5$	&
                                                                  $20-30/20-30$       &    $0.1 -1$	  &             H \\
  
 &	warm emission &	 $10^5-10^6$	&  $40-70/40-70$       &
                                                                 $1-
                                                                 10$ &
                                                                       H
  \\
  &   hot emision &	 $10^5-10^6$ &	$100-200/100-200$ &  $100 -1000 $   &
                                                                      K
  \\
\noalign{\smallskip}\hline\end{tabular}
\end{flushleft}
$^{(a)}$ This paper – {\it Odin} observations of ground state o-\hto\ (see text).
$^{(b)}$ This paper – {\it Herschel} SPIRE mapping observations plus
modeling (see text).
$^{(c)}$ Dust temperatures from Alonso-Herrero et al. (\cite{alo12})
and Tabatabei et al. (\cite{tab13}).
$^{(d)}$ Pre-shock density; The post-shock density becomes an order of
magnitude higher.
$^{(e)}$ from {\it Odin} observations of ground state o-\hto\ and
o-\htio,  assuming an ortho-to-para ratio of 3 (Karlsson et al. \cite{kar13}).
$^{(f)}$ from Walmsley et al. (\cite{wal86}).
$^{(g)}$ from {\it Herschel} SPIRE oservations (Meijerink et
al. \cite{mei13}).
$^{(h)}$ from {\it Herschel} HIFI, SPIRE and PACS observations (Liu et
al. \cite{liu17}).
$^{(i)}$ from the Sandqvist et al. (\cite{san15}) {\it Herschel} HIFI
487 GHz search for \ot.
$^{(j)}$ the abundance scales $\approx$ inversely with cloud density
$^{(k)}$ lower abundances may be accomodated by PDR models, while the
high ones indicate shocks (see Comments).
$^{(l)}$ fast shock model abundance (Flower \&\ Pineau de For\^ets
\cite{flo10}) in an effective accumulated 10\as\ size area (see
Comments).
$^{(m)}$ from Sandqvist et al. (\cite{san08}).
$^{(n)}$ from slow shock chemistry, in case FIR pumping dominates
(Gonzalez-Alfonso et al. (\cite{gon10}, \cite{gon12}, \cite{gon14}). 

\end{table*}

\subsubsection{Comments to Table 5}

{\it NGC\,1365}:

{\bf A)} {\it Odin} observes very weak ground state o-\hto\ emission from dense
molecular cloud complexes located in the circumnuclear torus region in
the velocity range $1400-1650$ \kms\ – an emission which in a velocity
range around 1500 \kms\ suffers from absorption caused by the low
excitation, foreground cloud envelopes of the cloud core emission
plus the thermal dust continuum background. Taking into account the
source size, approximately 15\as, and location of maximum emission apparent from
the {\it Herschel} SPIRE mapping, we find that the integrated \hto\
intensities observed by {\it Odin} and {\it Herschel} agree very well, and that
the velocity range of the {\it Odin} emission is as expected for emission
from the NE torus region.  

The existence of water absorption by lower excitation cloud envelopes
most likely can be ascribed to the PDR chemistry, including grain
surface reactions, in action in such regions. The water
emission from the massive complexes of cloud cores, co-located with
super starburst clusters and \HII\ regions residing in the NE torus
region of NGC\,1365 (cf. Elmegreen et al. \cite{elm09}), is bound to originate
from a combination of PDR chemistry, including grain surface
reactions, quiescent warm molecular cloud ion-molecule chemistry, and
also hot cores. There is possibly influence by an increased ionization level
caused by cosmic ray focusing by the aligned magnetic field, as
mapped by Beck et al. (\cite{bec05}), and locally also by X-ray emission from
the star forming superclusters. Outflow and shock chemistry may also
contribute to the ground state o-\hto\ emission (see next Comment). 
  
{\bf B)} The CO and \hto\ emissions from the higher energy states appear to be
best explained by shock excitation. While the CO-SLED can be nicely
matched by a two-component model 
using $40-60$ K for the lower energy lines and requiring a low velocity
($\approx 10$ \kms) shock at a pre-shock density of $10^4$ cm$^{-3}$
to reach the observed intensities of the higher energy lines. The \hto
-SLED seems to require a high velocity ($\approx 40$ \kms) shock at a
pre-shock density of about $10^4$ cm$^{-3}$, using the shock model by
Flower \&\ Pineau de Fôrets (\cite{flo10})( cf. Comment G for the
merger NGC\,6240). There is probably a requirement also for some
radiative Far IR (FIR)
excitation, which was used to model the {\it Herschel} SPIRE observations
of \hto\ in the ULIRGs Mrk\,231, NGC\,4418 and Arp\,220 by
González-Alfonso et al. \cite{gon10}, \cite{gon12}, as discussed
  in Sect. 4.2.2.  

If FIR excitation were dominating, the required \hto\ abundance
would be a factor of $10 - 100$ times lower than the fast shock (40
\kms) model abundance of $10^{-4}$ versus \htwo\ (Gonzalez-Alfonso et
al. \cite{gon10}, \cite{gon12}, \cite{gon14}; Flower \&\ Pineau de
For\^ets \cite{flo10}). This is an abundance level achievable by low velocity
shock chemistry including shock release of icy grain mantles. However,
no PACS observations of the FIR pumping water vapor absorption lines
are available in case of NGC\,1365.  
\\    

{\it Sgr\,A}:

{\bf C)} This is absorption in the low excitation $-30$ \kms\ foreground spiral
arm. The PDR chemistry, including grain surface reactions, is the likely
explanation for the observed water abundance. However, the possible
{\it Herschel }HIFI detection of the 487 GHz \ot\ line in this spiral arm
tells us that a ($\approx 20$ \kms) Galactic spiral arm density wave shock also
may be in action (cf. Sandqvist et al. 2015). 

{\bf D)} The emissions from the warm and dense +20 \kms, +50 \kms, and CND
(circumnuclear disk, torus) cloud cores are resulting from PDR
chemistry, including grain surface reactions, quiescent cloud ion-molecule
chemistry, and possibly hot cores. 

{\bf E)} These \hto\ line wings, not suffering from
self-absorption, associated with the +20 \kms, +50 \kms, and CND
clouds must be caused by shock chemistry as well as shock
desorption, sputtering from icy grain mantles. 
\\

{\it NGC\,6240}:

{\bf F)} This is emission from warm cloud complexes in the merger
interface region. 

{\bf G)} This is most likely emission from “galaxy-wide shocks”. Meijerink et
al. (\cite{mei13}) found that the CO-SLED, observed by {\it Herschel}
SPIRE, is best modeled by a ($\approx 10$ \kms) shock excitation and 
a pre-shock density of $5 \times 10^4$ cm$^{-3}$, leading to a
post-shock density of $4 \times 10^5$ cm$^{-3}$. They also point out
that the 2 $\mu$m \htwo\ emission, observed at the merger central region,
requires a shock in the velocity range $47 - 16$ \kms\ for a pre-shock
density of $5 \times 10^5 - 10^7$ cm$^{-3}$. See also the \hto\ SLED
analysis by Liu et al. (\cite{liu17}) and the discussion of NGC\,6240
in their Appendix.  
\\

{\it Other galaxies observed by HIFI, SPIRE and PACS aboard Herschel Space
  Observatory}:

{\bf H)} These results were based upon multi-transition CO and \hto\
SLEDs as well as dust SED observations by {\it Herschel }HIFI, SPIRE,
and PACS, paired with extensive radiative transfer and excitation
modeling, including collisional as well as radiative
excitation, see Liu et al. (\cite{liu17}).  

{\bf J)} Similarly to what is visualized by the {\it Odin} 557 GHz ground state
o-\hto\ spectrum of NGC\,1365, emission plus absorption is often seen in
the observed {\it Herschel }HIFI ground state o-\hto\ spectra, and are even
more frequent in the {\it Herschel }HIFI 1113 GHz ground state p-\hto\ spectra
(van der Tak et al. \cite{tak16}; Liu et al. \cite{liu17}). It may be
interesting to note that the 1113 GHz line is not clearly visible and
not claimed to be detected in our NGC\,1365 SPIRE data.   

{\bf K)} Only for the most infrared luminous galaxies Mrk\,231,
NGC\,4418, and Arp\,220, these compact hot gas, dust components were
required to fit the observational data. Here the high temperature,
possibly caused by shocks, leads to efficient release of \hto\ from the 
icy grain surfaces into the gas phase, hence explaining the high gas
phase abundance of water. In addition, X-ray emission from the AGN, or
from a nuclear star burst, may provide a contribution by XDR
chemistry. FIR excitation appears to be very important in these
galaxies (Liu et al. \cite{liu17}; Gonzalez-Alfonso et
al. \cite{gon10}, \cite{gon12}). 

\subsubsection{Some conclusions concerning the \hto\ emission and
  absorption regions in galaxies}

The ortho and para ground state rotational transitions of \hto\ at 557
and 1113 GHz often exhibit sharp absorption features caused by lower
excitation foreground gas – where the water vapor content of  $\approx
10^{-9} - 10^{-8}$ versus \htwo\ is resulting from PDR chemistry –
intersecting broad velocity emission from the denser molecular cloud
ensemble – caused by a mixture of PDR and “hot core” chemistries and
possibly also is influenced by shock sputtering and shock chemistry. These
processes all rely upon the release of icy grain mantles and
result in average \hto\ abundances of $\approx 10^{-8}-10^{-7}$.

The rotationally excited state \hto\ emissions, observed by {\it Herschel}
HIFI and SPIRE, appear to be best explained by a combination of FIR and
shock excitation, where FIR excitation is likely to be dominating in
the ULIRGs. Shock chemistry and shock release of icy grain mantles
are the likely processes causing the estimated very high \hto\
abundances of $\approx 10^{-6}-10^{-4}$, with possible influences from “hot core
chemistry” in regions of very extensive star formation.

The FIR \hto\ lines at shorter wavelengths, observed by {\it Herschel} PACS,
all have very high critical densities for collisional excitation and
therefore are bound to be “FIR pumped”. The  very high \hto\ abundances
of $10^{-6}-10^{-5}$, estimated from these observations, are likely to result
from efficient release of icy grain mantles, caused by varying
combinations of “hot cores” near massive young stars, shock
sputtering,  and XDR chemistry caused by intense X-ray emission, as
well as shock chemistry – all depending upon the environmental
conditions in the galaxy in question.

\subsection{Methylidyne radical, CH, and its ion \CHp\ – probable
  signatures of shocks and intense UV-illumination in the NGC\,1365
  circumnuclear torus region}

We here enter a discussion of the presence of CH as well as
\CHp\ in the circumnuclear torus region, as is revealed by the SPIRE
spectrum (Fig. 3 and Table 1) and by our CH and \CHp\ maps, shown in
Fig. 4. While an emission line at 835.1 GHz nicely matches
the $J = 1 - 0$ rotational transition of \CHp\ in its
$^1\Sigma$ electronic ground state, CH in its $^2\Pi$
electronic ground state appears in terms of two emission doublets, at
532.7/536.8 GHz and 1470.7/1477.4 GHz, classified as the ($1^+_{3/2}$ –>
$1^-_{1/2}/1^-_{3/2}$ –> $1^+_{1/2}$) and
($2^-_{3/2}$ –> $1^+_{3/2}/2^+_{3/2}$ –> $1^-_{3/2}$)
transitions, respectively (with quantum states denoted $N_J^{parity}$.
More details are given in Appendix C, and an energy level diagram can
be found in Rangwala et al. \cite{ran14}). These CH transitions are caused by a
step-wise internal energy decay of the CH molecular population from
the {$\Lambda$}-doubled second lowest ($N = 2, J = 3/2$)
rotationally excited state (at $E_u \approx 96.5$ K) to the
$\Lambda$-doubled lowest ($N = 1, J = 3/2$) rotationally excited state
(at $E_u \approx 25.7$ K), and subsequently to the $\Lambda$-doubled $^2\Pi$
($N = 1, J = 1/2$) CH ground state. Here the rather obvious, and
simplest, contributing excitation mechanism would be “direct radiative
pumping” causing population transport from the $N = 1, J = 1/2$ ground
state $\Lambda$-doublet state to the $N = 2, J = 3/2$
rotationally excited $\Lambda$-doublet state by means of CH
absorptions of the thermal dust emission at 2006.8/2010.8 GHz ($\approx 149$
$\mu$m), that is, close to the thermal emission maximum of dust at a
temperature of $\approx 20$ K, as is observed in NGC\,1365 (Figs. 1
and 2). However, no proof in terms of such observed 149 $\mu$m CH absorptions is
available in NGC\,1365, but appears to have been observed in the
($L_{\rm FIR} \approx 10^{12}$\lsol)
ULIRG, Arp\,220, by {\it Herschel} PACS, as reported in the SPIRE and HIFI
based CH analysis by Rangwala et al. (\cite{ran14}).

Arp\,220 is a
late-stage merger with two closely located counter-rotating disks and
also was included in the multi-line \hto\ analysis by Liu et al. (\cite{liu17}),
where it together with Mrk\,231 required a hot and dense gas component
to fit the observed \hto-SLED (cf. our Table 5 and its
comments). Stepwise collisional CH excitation may also be contributing
processes, which then would require the high gas densities and
temperatures resulting from the shocks which are observed to be in
action in the many hot spots of the outer circumnuclear torus of
NGC\,1365 (Galliano et al. \cite{gal12}; Fazeli et al. \cite{faz19})
and here were crucial for the \hto\ and CO excitation (see
Sect. 4.2.2). Extensive shocks and star formation-induced outflows also
are likely to be present in Arp 220, as well as in the IR bright
($L_{\rm FIR} \approx 2 \times 10^{11}$\lsol) Seyfert 2 
galaxy NGC\,1068 (Spinoglio et al. \cite{spi12}; Liu et
al. \cite{liu17}), which  will be our forthcoming CH, \CHp\ comparison galaxy . 

The CH assignment of the striking 532.7/536.8 GHz line doublet
requires some carefulness, since unfortunately the laboratory
frequencies of the HCN($6-5$) and \hcop($6-5$) lines at 531.7 and 535.1 GHz
are almost overlapping, considering the poor SPIRE spectral resolution
and the broad galaxy spectral lines. However, there are no visible
signs of the HCN($7-6$) and \hcop($7-6$) lines in our NGC\,1365 SPIRE
spectrum, which limits their contribution to our CH doublet to be at
most about 20 \%. This is consistent with the comparatively small
contributions from HCN and \hcop\ found by Rangwala et al. (\cite{ran14}) in
their complementary {\it Herschel} HIFI observations of a number of galaxies,
including Arp\,220.    

From the {\it Herschel} SPIRE line intensities listed in Table\,1, we
estimate the upper state column densities of the 533/537 GHz and
1471/1477 GHz CH doublets to be $23 \times 10^{12}$ cm$^{-2}$ and $15
\times 10^{12}$ cm$^{-2}$, for an assumed, accumulated source size of
14\as. We can then determine an excitation (population 
distribution) temperature for these states to be $T_{\rm ex}\approx
75$ K. This allows us to calculate the total column density in the
four lowest rotational states to be $\approx 8 \times 10^{13}$
cm$^{-2}$, under the assumption of a common excitation
temperature. The estimated excitation temperature is in fact a lower
limit since the 533/537 GHz doublet may contain some HCN/HCO$^+$ emission
and is at least as high as the kinetic temperature of the quiescent
molecular clouds, which suggests influences of FIR pumping and/or
efficient collisional excitation in the observed shock regions. 

In the same way, we estimate the upper ($J=1$) state column density of \CHp\ to
be $1.6 \times 10^{12}$ cm$^{-2}$, for an assumed source size of
14\as\ (Fig. 4). The minimum total \CHp\ column density is
estimated to be $\approx 3.5 \times 10^{12}$ cm$^{-2}$ for excitation
temperatures in the range 40 to 100 K (cf. Falgarone et
al. \cite{fal10b}). However, in view of the intense UV-illumination
from newly formed stars, the \CHp\ column density may well be an order of
magnitude larger, approaching that of CH. The reason is formation
by the reaction between C$^+$ and vibrot-excited \htwo, which also causes a
large enhancement of the rotationally excited \CHp\
states, as has been observed in the Orion Bar (Nagy et
al. \cite{nag13}, \cite{nag17}; see also Appendix C). In the
case of NGC\,1365 we have to rely on the observed existence of
multiple vibrot \htwo\ lines in the hot spots of the outer torus region,
and the existence of strong UV radiation in these regions of intense
star formation (Galliano et al. \cite{gal12}; Fazeli et
al. \cite{faz19}). For these reasons we may estimate the CH/\CHp\
abundance ratio to be $\lid 23$.

From the simultaneously observed SPIRE maps (Fig. 4), it appears
that the \hto\ and \CHp\ emissions peak in the NE torus region while the CO
and CH emissions are more evenly distributed across the circumnuclear
torus. The higher energy CO-SLED is nicely modeled by a low velocity
(10 \kms) shock, which also may satisfactorily explain the required CH 
excitation  as well as its high abundance in denser gas (see Appendix
C). The higher velocity (40 \kms) shock, possibly required to
model the \hto -SLED in the NE torus region, paired with the intense
UV radiation from the observed massive young stellar superclusters
(Galliano et al. \cite{gal12}) may explain the previously discussed
possible overabundance of \CHp\ (Gerin et al. \cite{ger16}; 
Godard et al. \cite{god19}; see also Appendix C). 

With these uncertainties in mind, we now proceed to estimate the CH and
\CHp\ abundances. Here a comparison \htwo\ column density is available from
our multi-transition SPIRE-CO modeling in Sect. 4.2.2. Correcting the
model value $N$(CO) $ \approx 4 \times 10^{15}$ cm$^{-2}$ for the beam
filling of a 14\as\ source 
size in a 40\as\ antenna beam, and assuming a CO/\htwo\ abundance ratio of
$10^{-4}$, we arrive at an average of $N$(\htwo) $ \approx 4 \times
10^{20}$ cm$^{-2}$ for the shocked gas regions of the outer 
torus. The CH and \CHp\ abundances then become $X$(CH) $ \approx 2
\times 10^{-7}$ and $X$(\CHp) $\approx 10^{-8}$. In the unlikely case
that the emissions were originating in the quiescent molecular clouds,
these abundances would be lower by a factor of about 30.

Only in NGC\,1068, out of the four prototypical AGN or star burst
dominated galaxies (NGC\,1068, Arp\,220, M82 and NGC\,253) selected for
{\it Herschel} CH observations by Rangwala et al. (\cite{ran14}), did
both the CH and the \CHp\ lines appear in emission, similar to our
case of NGC\,1365. In the other galaxies, \CHp\ was observed in
absorption, presumably from lower excitation foreground gas, while CH
was observed in emission (from higher density regions). Our IR
luminous comparison galaxy, NGC\,1068 ($L_{\rm FIR} \approx 2 \times
10^{11}$ \lsol), is exhibiting a number of similarities 
with NGC\,1365 ($L_{\rm FIR} \approx 5 \times 10^{10}$ \lsol), that is, a
circumnuclear torus – a prominent “star forming molecular
cloud ring” of radius $1-1.5$ kpc versus $\approx 0.8$ kpc in NGC\,1365 – 
and a smaller circumnuclear disk – CND of radius $100-150$ pc versus
$\approx 26$ pc in NGC\,1365 – as discussed in
more detail by Spinoglio et al. (\cite{spi12}) and Liu et al. (\cite{liu17}). 

The CH abundance estimated in NGC\,1068 by Rangwala et
al. (\cite{ran14}), from the 533/537 GHz doublet alone, is $ \approx
10^{-8}$ versus \htwo. The CH/\CHp\ abundance ratio 
is found to be $\approx 23$, just as in NGC\,1365, however in both
cases somewhat uncertain values 
because of lacking observations of the molecular excitation. The
NGC\,1068 abundances were nicely accommodated by XDR physics, chemistry
in the CND, driven by the X-ray emission from the AGN. A similar inner
torus generated XDR emission of CH and \CHp\ is likely to exist also in
NGC\,1365. However, here the dominant parts of the CH and \CHp\ emissions are
located to the UV-illuminated shock regions of the outer torus, as is
strongly indicated by our observed emission maxima at 
the NE torus FIR and molecular gas mass peak. In the absence of
{\it Herschel} Space Observatory, improved knowledge of the CH and
  \CHp\ molecular population distributions in NGC\,1365, as well as in
  NGC\,1068, could here be achieved by spectrally resolved {\it Sofia}
limited mapping observations of the CH doublet at 2 THz and the
\CHp\ ($J= 2-1$) transition at 1669 GHz.

In this context the results from {\it Herschel} PACS observations
  of multiple \ohp, \htop, and \htreop\ absorption lines in the ULIRGs,
  NGC\,4418 and Arp\,220, by Gonzalez-Alfonso et al. (\cite{gon13}) are
  useful to contemplate, since the importance of an observed molecular 
population distribution is emphasized. It turned out that in the
single case of Arp\,220 the pure inversion, metastable state lines of
\htreop\ revealed a very high population distribution temperature of
$\approx 500$ K, indicating hot gas or “formation pumping”. There,
excess energy from the \htreop\ formation process leaves the resulting
molecule in a high rotational temperature molecular state
distribution). The population distributions of \ohp\ and \htop\ did
not show such an anomaly. These results are very important for the
understanding of the molecular formation processes in
question. However, no \ohp, \htop\ or \htreop\ signals can be
identified in our NGC\,1365 SPIRE spectrum (Fig. 3 and Table 1).

In the light of our current discussion of CH and \CHp\ observations and
their modeling interpretations, and including the results from the very
recent modeling of UV irradiated molecular shocks by Godard et
al. (\cite{god19}), inspired by the ALMA discovery of \CHp($1-0$) emission
and absorption lines from a number of lensed star burst galaxies at a
redshift of $\approx 2.5$ (Falgarone et al. \cite{fal17}) – we are
inclined to believe that the CH and \CHp\ emission lines observed from
the circumnuclear torus region of NGC\,1365 are signposts of shock
action in this region, presumably penetrated by intense UV light
and molecular outflows caused by the observed extensive star formation.

\section{Summary and conclusions}

We have used the {\it Odin} satellite to observe the central region of the
barred spiral galaxy NGC\,1365 in the 557 GHz \hto\ line. After a total
of 81 hours of ON-source integration time, we have obtained a marginal
detection of this water vapor line at a velocity resolution of 5
\kms. We have combined these observations with {\it Herschel} PACS and SPIRE
observations of two positions in the galaxy’s nucleus, obtained from
the {\it Herschel} Science Archive, and produced simultaneously observed
SPIRE images of the distribution of o-\hto, p-\hto, CO, CH, \CHp, and
\NII. 

The water vapor emission is predominantly located in the shocked 15\as\
(1.3 kpc) region of the northeastern (NE) component of the molecular
torus surrounding the nucleus. Here, several compact radio
sources, hot-spot \HII\ regions, as well as co-located very massive
molcular cloud complexes, and young stellar superclusters have been
observed – all of which triggered by the rapid bar-driven inflow into
the circumnuclear torus, causing cloud-cloud collisions and shocks. The
two ground state \hto\ emission components, detected in the {\it Odin}
spectrum, have approximate velocities of –150 and –40 \kms\ with respect to the
systemic velocity 
of the galaxy, while an intersecting absorption appears at a velocity
of  –100 \kms, which agrees with kinematic studies of the bar-driven
circumnuclear torus gas flow in the vicinity of the NE component. For
the absorbing gas region, we find an \hto\ fractional abundance of 
$\gid 5 \times 10^{-8}$, which is well accomodated by PDR
chemistry. In the presumably denser emission regions, the \hto\
abundance appears to increase but may still be understood in terms of
PDR chemistry taking place in this region of strong UV radiation from
the young super star clusters. However, part of the ground state  \hto\ emission
may also be caused by shock excitation and chemistry, as will be
discussed next.  

 Model studies of the multi-transition SPIRE CO and \hto\ observations
 toward the NE component reveal the presence of a slow-velocity
 C-type shock with a shock speed of 10 to 40 \kms\ and an \htwo\ pre-shock
 density of  10$^4$ cm$^{-3}$, where the lower velocities are
 sufficient to model the observed CO spectral line energy distribution
 (SLED). The observed \hto\ SLED requires the higher shock velocities,
 which are also sufficient to drive the vibrational excitation of
 \htwo. In addition, FIR pumping may be required to satisfactorily explain
 the \hto\ SLED. A model abundance versus \htwo\ of 10$^{-4}$ was reached for
 CO as well as for \hto. If the \hto\ excitation were dominated by
   FIR pumping, the required \hto\ abundance is lower by a factor of $10
   -100$ and is most likely caused by slow velocity shock chemistry (as
   discussed in Sect. 4.2.2 and Appendix B ).   

While the \hto\ and \CHp\ emissions seem to be concentrated near the NE
torus component, the CO and CH emissions appear to be more evenly
distributed between the NE and the SW torus component, as revealed by
the SPIRE images. These relative distributions may reflect the fact
that the eastern bar-driven gas inflow into the NE torus region is
much more massive than the corresponding gas inflow from the western
bar into the SW torus region. In the case of CH we estimate an abundance
of $2 \times 10^{-7}$. The estimated \CHp\ abundance of $\approx
10^{-8}$ in the NE torus region is likely to be a lower limit because
of (here unconfirmed) formation via the abundant vibrationally and
rotationally excited  \htwo\ in this shock and UV irradiated
region. Such a \CHp\ abundance enhancement has indeed been observed in
the Orion Bar (See Appendix C).   

We have made a statistical image deconvolution (SID) of our single
dish SEST observations of the CO($3-2$) emission in the nuclear region
of NGC\,1365, yielding an angular resoltuion of 5\as, complementing
recent ALMA observations. These SID observations yield dynamical
masses of $1.2 \times 10^{10}$ \msol\ and $4.4 \times 10^{10}$ \msol\
inside radii of 1 and 2 kpc, respectively, where the gas mass amounts
to $\approx20$\%\ in the inner region and $\approx6$\%\ in the region
inside 2 kpc.    

We discuss a co-located collimated CO jet, a relativistic radio jet
and a wide-angle inhomogeneous ionized \OIII\ gas cone outflow from the
nuclear engine. Our VLA \HI\ absorption analysis reveals an \HI\ ridge
extending from the nucleus in a south-southeast direction, coinciding
with the southern edge of the \OIII\ cone. With an \HI\ column density
of a few $\times 10^{21}$ cm$^{-2}$, this ridge may have been created
by the action of outflow-driving X-ray photons colliding with dust
grains, causing water ice from the dust grains to be desorbed and
dissociated into 2H + O. A precessing nuclear engine appears to be
required to accomodate the various outflow components, and is also
consistent with the ALMA observations by Combes et al. (\cite{com19})
of a tilted massive inner gas torus or accretion disk.

\begin{acknowledgements}
  First of all we wish to thank Per Olof Lindblad for many valuable
  discussions and his numerous comments on the manuscript of this
  paper. We also wish to thank the anonymous referee for the
    careful reading of the manuscript and the many constructive
    comments, which much improved our paper. 
  For valuable contributions in the late-stage preparation
  of the {\it Odin} satellite observing program and its successful
  operation, we wish to thank H.-G. Floren at Stockholm Observatory,
  M. Battelino and B. Jakobsson at OHB Sweden, G. Persson at Onsala
  Space Observatory and S.-O. Silverlind at the Swedish Space
  Corporation in Esrange. 
\end{acknowledgements}

{}

\appendix

\section{VLA \HI\ and SEST SID CO ($3-2$) position-velocity maps}

Position-Velocity (P-V) maps of our VLA \HI\ and SEST SID CO ($3-2$)
observations, oriented parallel to the major axis of NGC\,1365, are
presented in Fig. A.1. 

\begin{figure*}[ht]
\includegraphics[angle=0, width=.24\textwidth]{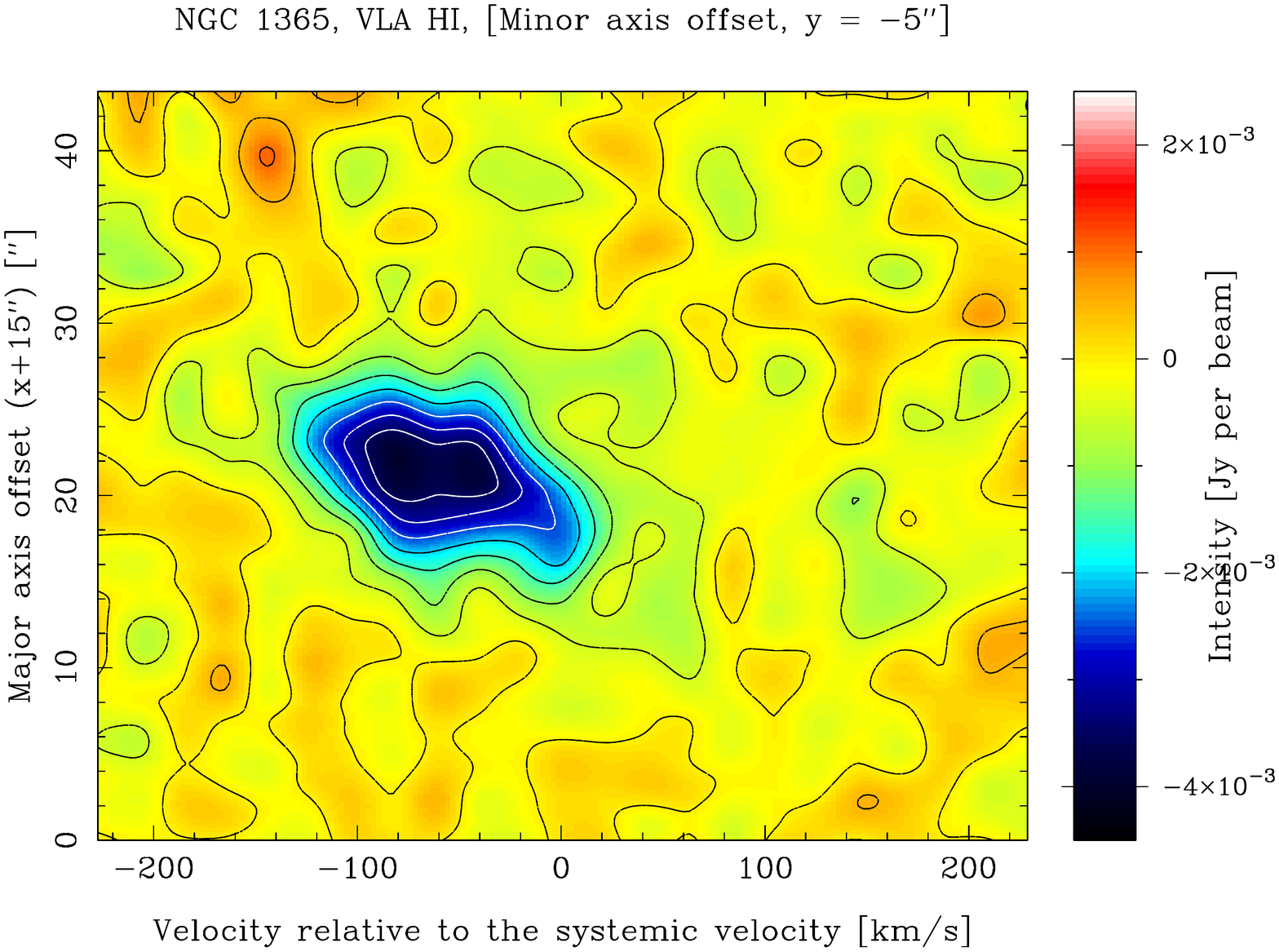}
\includegraphics[angle=0, width=.24\textwidth]{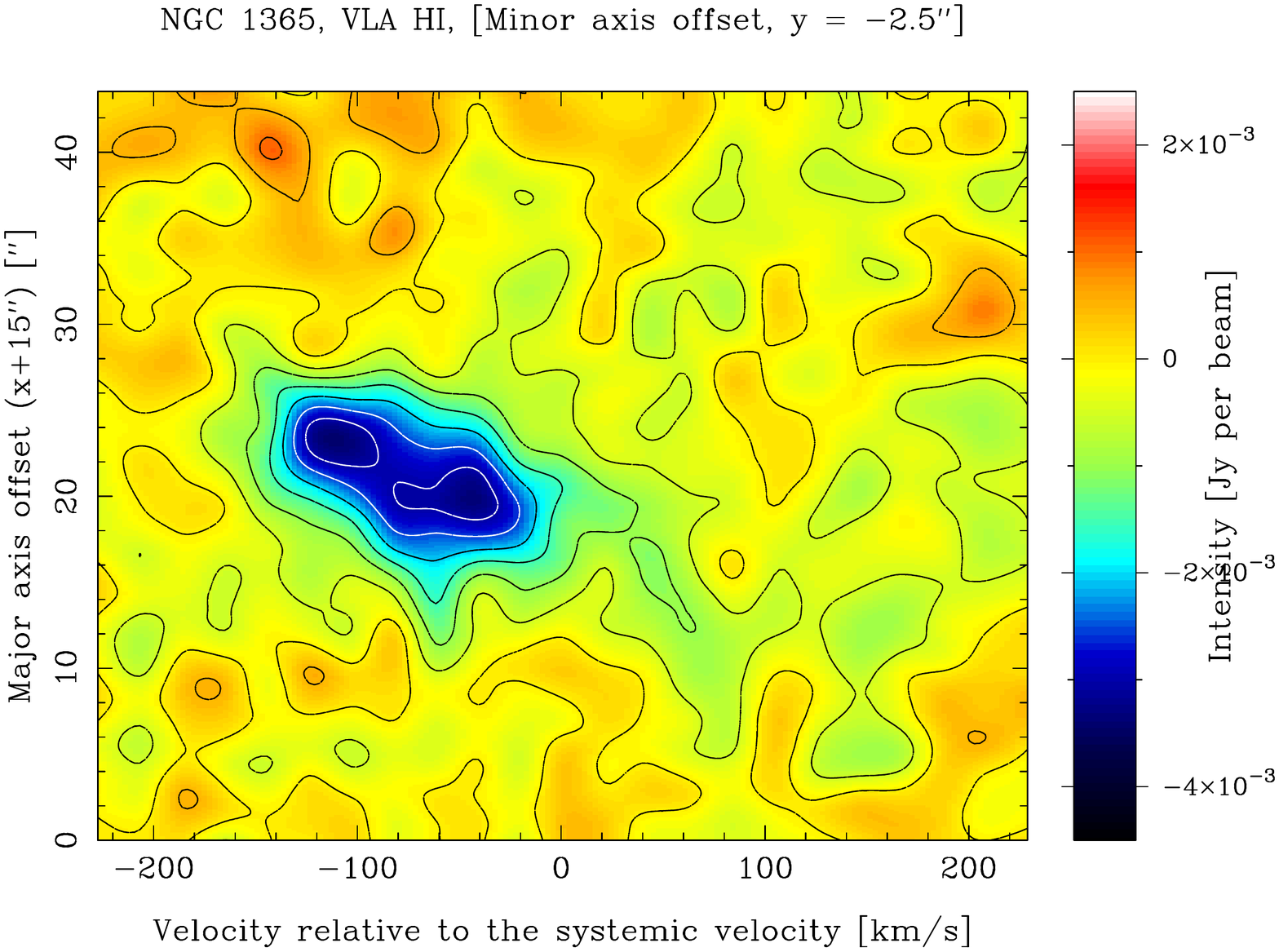}
\includegraphics[angle=0, width=.24\textwidth]{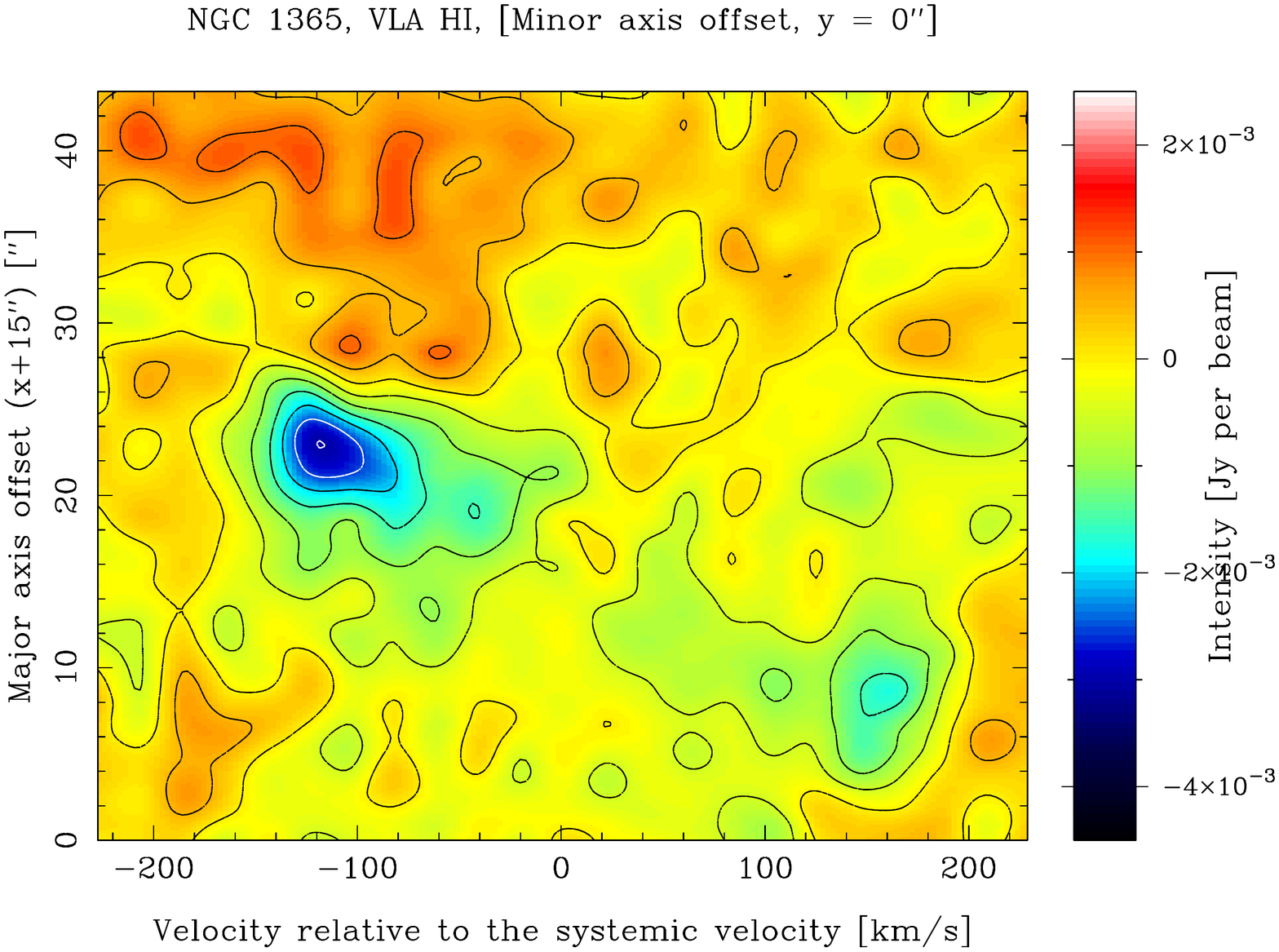}
\includegraphics[angle=0, width=.24\textwidth]{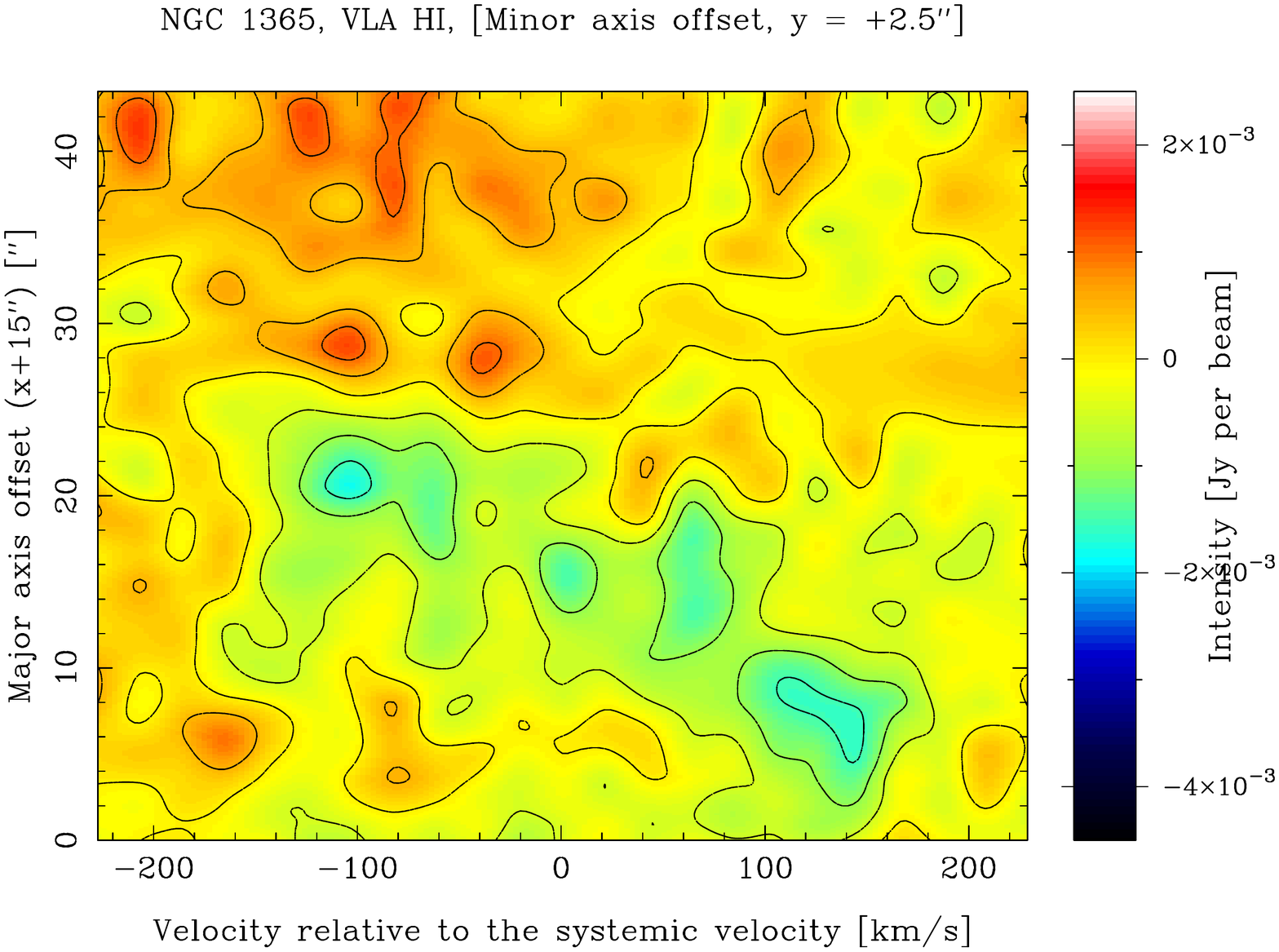} 
\includegraphics[angle=0, width=.24\textwidth]{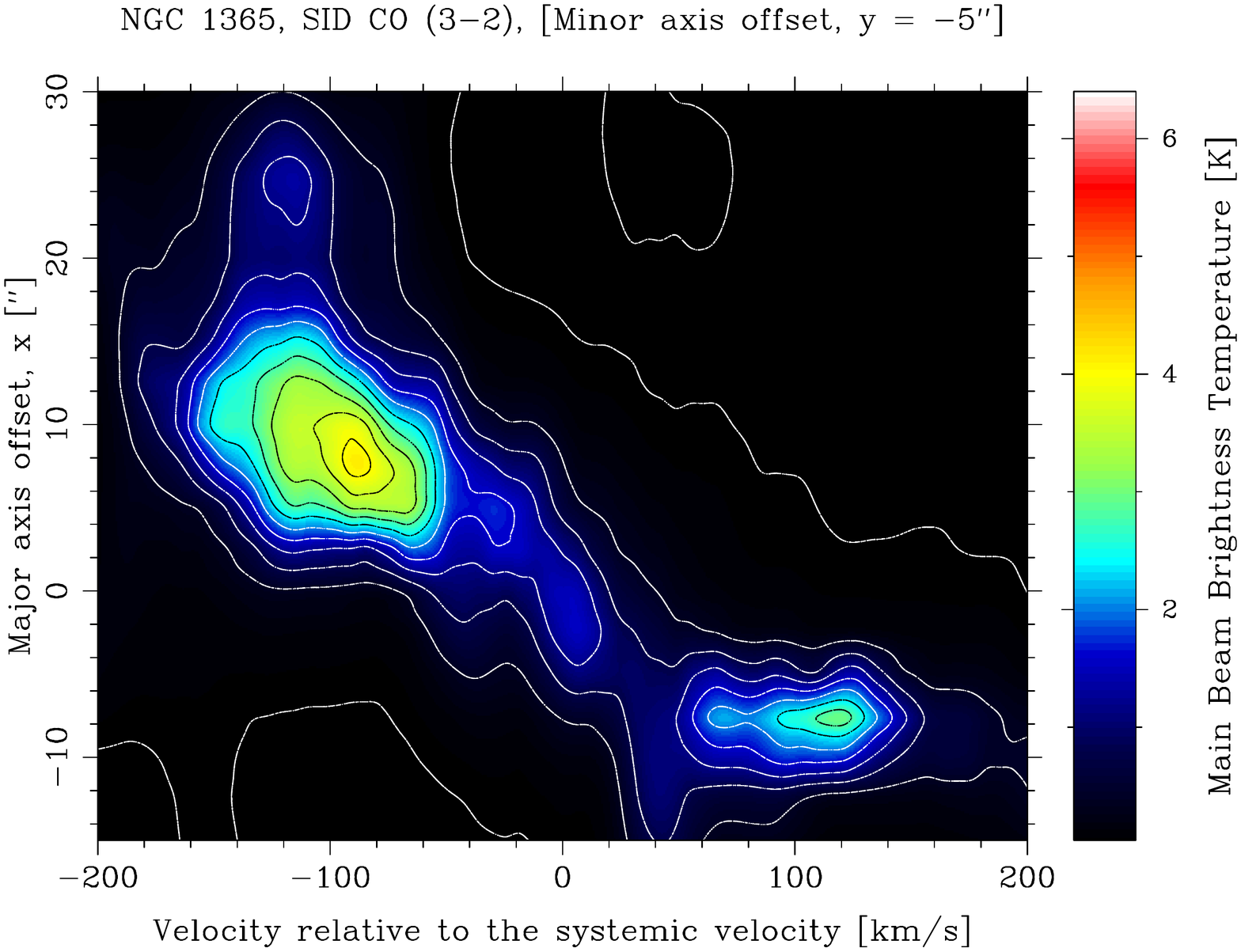}
\includegraphics[angle=0, width=.24\textwidth]{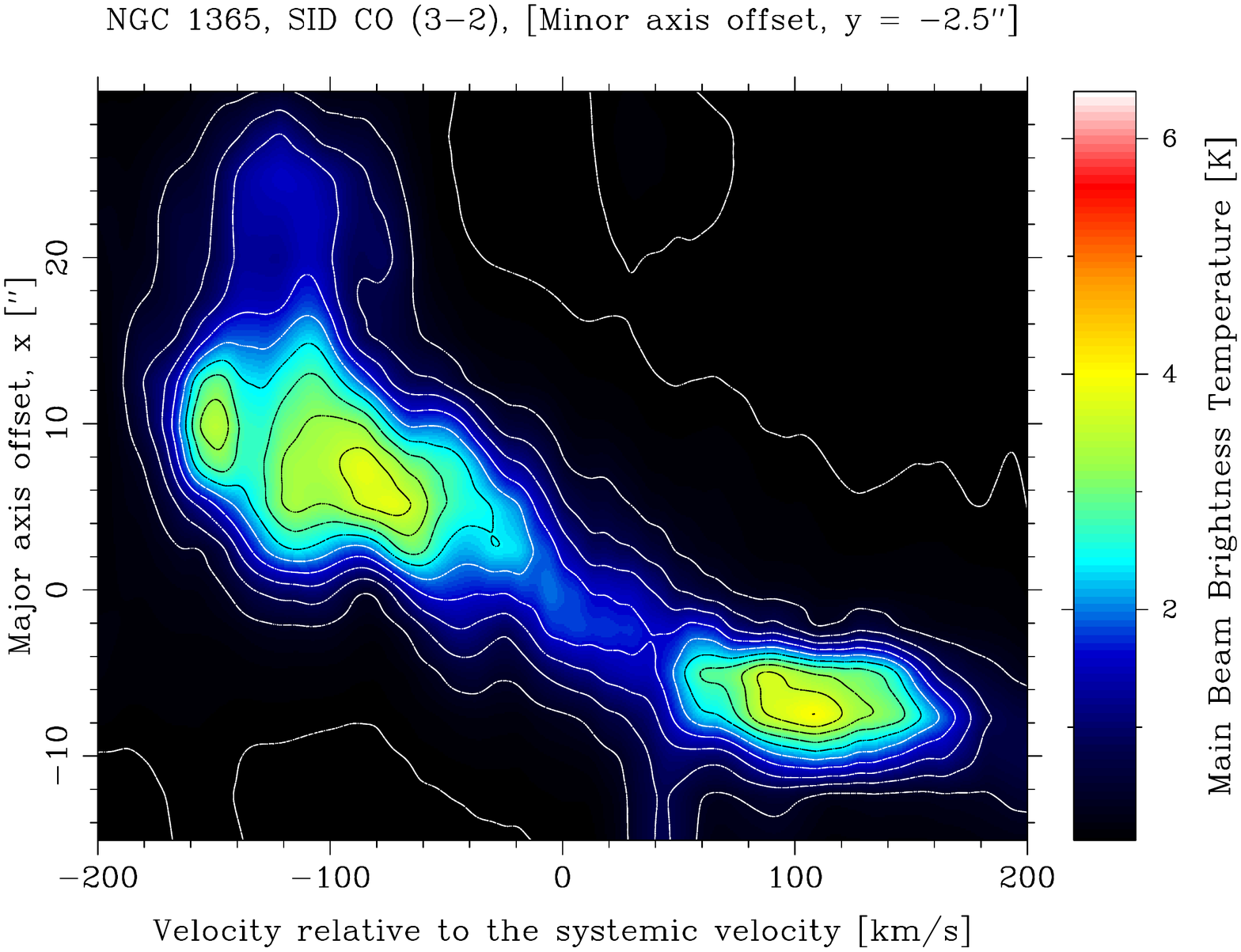}
\includegraphics[angle=0, width=.24\textwidth]{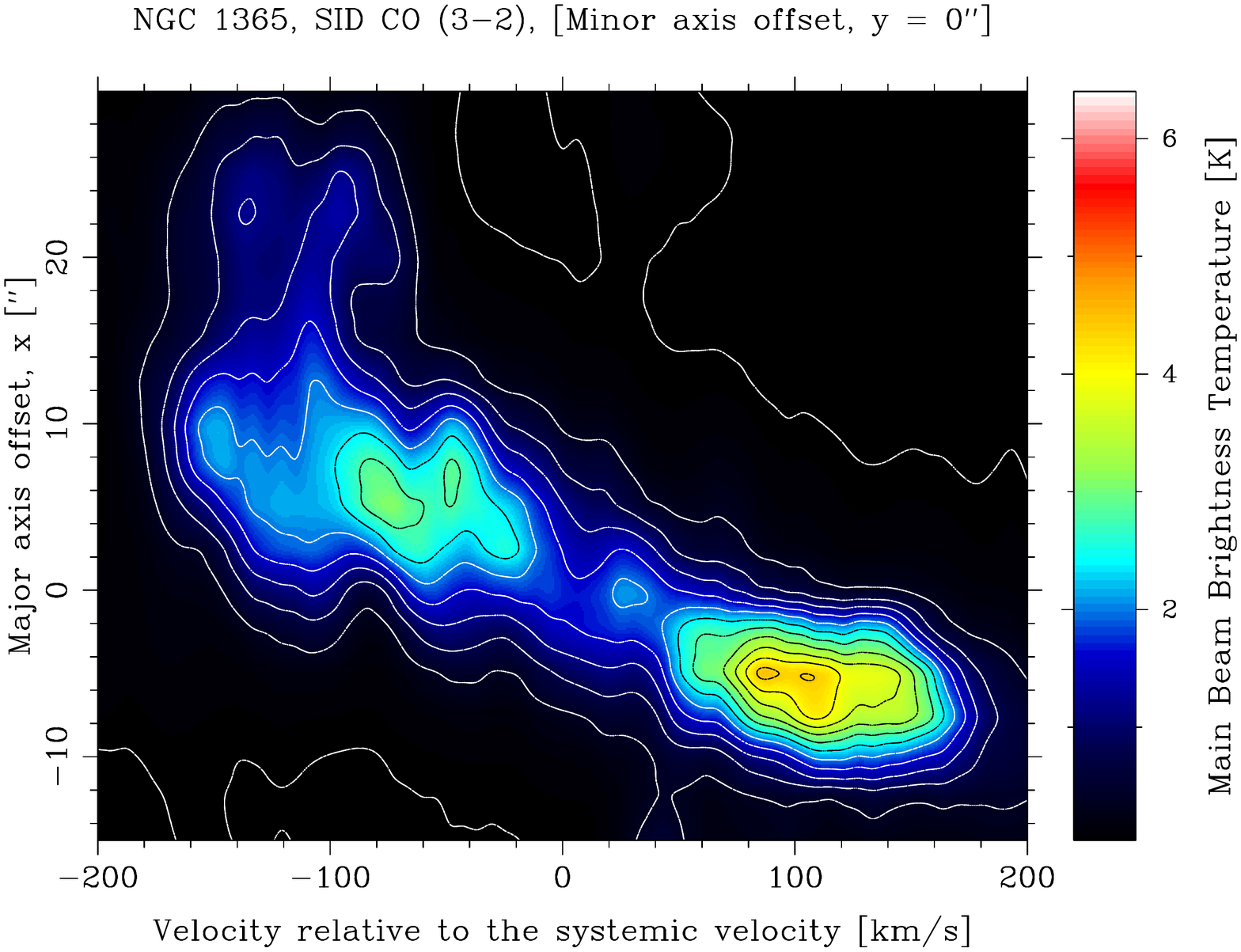}
\includegraphics[angle=0, width=.24\textwidth]{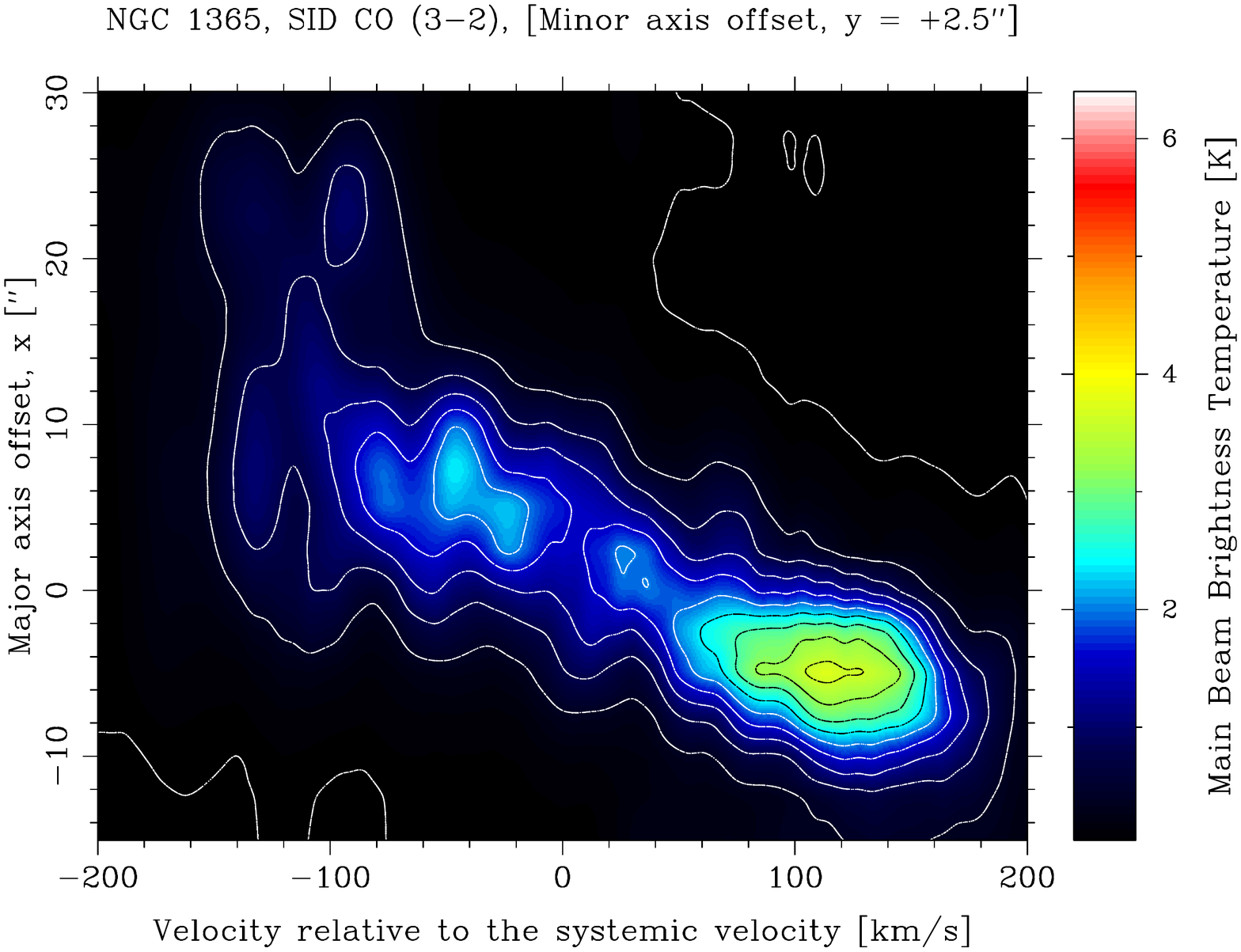} 
\caption{[{\it Top}]: VLA \HI\  P-V maps oriented
  parallel to the major axis of NGC\,1365 for different minor axis offsets
    ($y = -5\as, -2\farcs5, 0\as, +2\farcs5$). The major axis offsets
    should be subtracted by 15\as\ to obtain the $x$-offsets from the
    optical nucleus, with positive values in the NE direction. [{\it
      Bottom}]: The SEST SID CO($3-2$) P-V maps oriented
    parallel to the major axis of NGC\,1365 for the same minor axis
    offsets as in the \HI\ maps, with offsets ($x,y$) being from the
    optical nucleus. The velocity offsets are with respect to 1613 \kms.} 
  \label{A.1}
\end{figure*}

\section{On interstellar water chemistry}
\subsection{Water abundances in NGC\,1365 vs Galactic prototype
  regions} 

In Table B.1 we compare
our current NGC\,1365 \hto\ abundance estimates with recent {\it Odin}
results for the Galactic Center Sgr\,A region (Karlsson et
al. \cite{kar13}), and with {\it Odin} and {\it Herschel} HIFI results
for the local Orion molecular   cloud (to demonstrate observationally
determined \hto\ abundances caused by the PDR, warm cloud core, Hot
Core, bipolar outflow and shock chemistries which are likely to be
present also in the central regions of NGC\,1365). For comparison we
also list {\it Herschel} and {\it Odin}/SWAS results for the cold
dense cores of the Galactic molecular clouds DR\,21 and W\,3 and for
the S\,140 PDR interface region. 

\begin{table*}
\caption{Comparison between observed \hto\ abundances in NGC\,1365,
  our Galactic Center region (Sgr\,A) and some nearer Galactic sources.}
\begin{flushleft}
\begin{tabular}{llll}
\hline\hline\noalign{\smallskip}
Source & Region & Abundance vs \htwo\ [$\times 10^{-8}$] &   \\ 
(distance) & (kinetic temperature)  &   {\it Odin}  \tablefootmark{*}
  &  {\it Herschel}  \\   
\hline\noalign{\smallskip}
  NGC\,1365 \tablefootmark{a} & absorption ($\approx$20 K)  & $5-8$  &  \\
  (18.6 Mpc) & warm emission ($40-60$ K)  & $6-600$\tablefootmark{s}  &   \\
    & shocks (350 K)  &   & $200\tablefootmark{u} -
                            10,000\tablefootmark{t}$ \\ \\ 
  Sgr\,A  \tablefootmark{d} & +50 \kms\ Cloud (60 K)  & 5 &  \\
  (8.7 kpc) & Red wing of +50 \kms\ Cloud \tablefootmark{e} & 130 & \\
       & CND (150 K) \tablefootmark{f} & 9 & \\
       & Red wing of CND \tablefootmark{e} & 760 &
                                                   $100-1300$\tablefootmark{b}
  \\ 
       & +20 \kms\ Cloud (60 K) & 3 &  4\tablefootmark{c}  \\
       & Red wing of +20 \kms\ Cloud \tablefootmark{e} & 110 & \\
       & $-$30 \kms\ arm \tablefootmark{g} & 4 & \\  \\
  Orion KL \tablefootmark{m} & PDR interface region (70 K) & 9
                                           \tablefootmark{h,i} & \\
  (450 pc) & 2\am\ south core (75 K) & 11
                                       \tablefootmark{h,i} & \\
    & Compact Ridge (115 K) & 280 \tablefootmark{j} & 260
                                                      \tablefootmark{k}
  \\
    & Hot Core (200 K) & 1200 \tablefootmark{j} & 1000
                                                  \tablefootmark{l,k} \\
    & Low velocity outflow & 290 \tablefootmark{j} & 480
                                                     \tablefootmark{k}
  \\
    & High velocity outflow (vs all \htwo) & 2200 \tablefootmark{j} &
  \\
       &   &   & 7000 \tablefootmark{l} \\  
       & High velocity outflow (vs shocked \htwo] & 10000
                                                    \tablefootmark{h}  & \\
       &   &  28000 \tablefootmark{j}  & \\ \\
  DR\,21 & Cool dense core (23 K) &  & 0.1
                                       \tablefootmark{n,o}  \\
  (1.7 kpc) & Foreground cloud (10 K) &  & 2
                                           \tablefootmark{n,o} \\
    & Low velocity outflow &   & 280
                                 \tablefootmark{n,o}  \\  \\
  W3 IRS 5 & Cool dense core (40 K) & 0.2
                                           \tablefootmark{p} \\
  (2.3 kpc) &   &   &  \\ \\
  S\,140 & Clumpy PDR interface with dark cloud (55 K) & 5/0.1
                                                         \tablefootmark{q,r}
                                                         &  \\
      & Low velocity outflow & 40 \tablefootmark{q}  &  \\
  
\noalign{\smallskip}\hline\end{tabular}
\end{flushleft}
$^{(*)}$ based upon ortho-\hto, \htio\ and H$_2^{17}$O ($1_{10}-1_{01}$) observations; 
$^{(a)}$ this paper;
$^{(b)}$ from {\it Herschel} mapping of the CND/Sgr\,A* by
Armijos-Abendano et al. (\cite{arm19});
$^{(c)}$ from {\it Herschel} mapping of the +20 \kms\ Cloud by
Armijos-Abendano et al. (\cite{arm19});
$^{(d)}$ from Karlsson et al. (\cite{kar13}), assuming an
ortho-to-para ratio of 3;
$^{(e)}$ outflow/shock;
$^{(f)}$ Circumnuclear Disk (torus) surrounding the Sgr\,A* black
hole;
$^{(g)}$ foreground spiral arm;
$^{(h)}$ Olofsson et al. (\cite{olo03});
$^{(i)}$ Wirstr\"om et al. (\cite{wir06});
$^{(j)}$ {\it Odin} spectral scan, Persson et al. (\cite{per07});
$^{(k)}$ {\it Herschel}/HIFI spectral scan, Neill et
al. (\cite{nei13});
$^{(l)}$ Melnick et al. (\cite{mel10});
$^{(m)}$ mapped in detail by {\it Odin}, Hjalmarson et al. (\cite{hja05});
$^{(n)}$ based upon para-\hto\ ($1_{11}-0_{00})$ observations, van der
Tak et al. (\cite{tak10});
$^{(o)}$ assuming an ortho-to-para ratio of 3;
$^{(p)}$ from {\it Odin} mapping by Wilson et al. (\cite{wil03}) in
close agreement with SWAS results by Snell et al. (\cite{sne00});
$^{(q)}$ clump/interclump values;
$^{(r)}$ from {\it Odin} mapping by Persson et al. (\cite{per09}).
$^{(s)}$ the lower abundance estimates appear at higher densities and
can be accommodated by PDR models, while the high abundances (at a
density of $10^4$ cm$^{-3}$) would indicate that the existing shocks also
contribute to the observed ground state \hto\ emission. 
$^{(t)}$ shock model abundance (Flower \&\ Pineau des For\^ets
\cite{flo10}) in an effective (accumulated) 10\as\ size area of the
circumnuclear torus (see Sect. 4.4).
$^{(u)}$ caused by slow velocity shock chemistry in the case of (here
unconfirmed) dominant FIR excitation (Gonzalez-Alfonso et
al. \cite{gon10}, \cite{gon12}, \cite{gon14}).

\end{table*}

\subsection{Water chemistry in cold and warm regions}

To make our discussion of the interstellar \hto\ chemistry going on in
the central regions of NGC\,1365 more to the point, we here provide a
very simplified summary of the main formation and destruction routes,
for cases where the theoretical models are well established (for
details we refer to the extensive review paper by van Dishoeck et
al. \cite{dis13}) and where a sample of relevant observed \hto\ abundances have
been listed in Table B.1. 

\subsubsection{Low temperature gas-phase water chemistry}

The oxygen chemistry is initiated by cosmic-ray (CR) ionization of
\htwo\ which rapidly leads to the formation of the trihydrogen ion
(\htrep). Next step is \\

  \htrep\ + O   => \ohp\ + \htwo\ ; \htop\ + H	\ \ \ \ \ \ \ \ \ \ \
  \ 	(1) \\

followed by \\

  \ohp\  + \htwo\ = \htop\ + H \ \ \ \ \ \ \ \ \ \ \ \ \ \ \ \
  (2) \\

and, subsequently, \\ 

  \htop\ + \htwo\ = \htreop\ + H   \ \ \ \ \ \ \ \ \ \ \ \ \ \ \ \                          (3) \\

followed by dissociative recombination, \\

  \htreop\ + e$^-$ =>  \hto\  + H    ($\approx$17\%), \ \ \ \ \ \ \ \
  \ \ \ \ \ \ \ \    (4) \\ 

  but also
        ($\approx$83\%)	 	
        OH + \htwo\ ;  OH + 2H ; O + \htwo\ + H. \\

At low temperatures the main loss of gaseous \hto\ is via accretion
(adsorption, “sticking”) onto cold dust grain surfaces (where water
ice has been observed to be a main constituent). 
Only at grain temperatures of 100 K, or higher, the \hto\ molecules are
efficiently desorbed (evaporated) from the icy grain surfaces. While
the just (crudely) described ion-molecule reaction scheme, for a
standard Galactic CR flux, is capable of producing a gas-phase \hto\ 
abundance $X$(\hto) $\approx  10^{-6} - 10^{-7}$ versus \htwo, the
“freeze out” on cold grain surfaces may reduce this abundance by
several orders of magnitude, as is illustrated in Table B.1 by the
observed values of $(1-2) \times 10^{-9}$ in the DR\,21 and W\,3
molecular cloud cores.  

The expected gas-phase \hto\ abundance is closely related to the
\htrep\ production. An increase of the CR flux leads to a faster chemistry,
and also results in higher \htrep\ and \hto\ abundances (cf. Farquhar et
al. \cite{far94}; Nilsson et al. \cite{nil00}, on-line Figs. 16 and
21). In NGC\,1365 we have already envisioned a probable CR
focusing by the observed magnetic field aligned along the
bar. Alternatively, X-ray ionization of \htwo, rapidly leading to
enhanced \htrep\ abundance, also would do the same job. X-ray emission has
indeed been observed in NGC\,1365, especially in the nuclear region
(from the central AGN, black hole region; see Lindblad
\cite{lin99}). Theory for XDR (X-ray Dominated Region) physics, chemistry has
been developed by Meijerink \& Spaans (\cite{mei05}) and has been
applied to galaxy nuclei by Meijerink et al. (\cite{mei07}). Also
an increased UV flux from one or more newly formed massive stars,
influencing the surface layers of their mother molecular clouds
and a developing \HII\ region also is bound to produce an enhanced
\hto\ abundance (theory by e.g, Hollenbach et al. \cite{hol09},
\cite{hol12}, who also study the influences of increased CR
ionization). In Table B.1 we have listed observed \hto\ abundances of
$(5-10) \times 10^{-8}$ in the Orion PDR (Photon Dominated Region)
interface layer between the molecular cloud and the M\,42 \HII\
region, and in the S\,140 PDR. The water abundances of  $(2-3) \times
10^{-8}$, observationally estimated for the DR\,21 foreground cloud and
the Sgr\,A $-$30 \kms\ arm, both seen in absorption, also are likely 
results of PDR chemistry, but now in lower density regions of low
visual extinction where the general Galactic background UV radiation
can penetrate.

\subsubsection{Ice chemistry}

In parallel with the low temperature gas-phase chemistry, there are
also ongoing grain surface reactions. In addition to the formation of
\htwo, which requires a third body for its efficient formation, there
are also a number of  reactions (collisions) between atoms and
molecules sticking on the grain surfaces, and jumping between the
potential wells on the surface, hence forming ices containing \hto,
CH$_3$OH, NH$_3$, CO, CO$_2$, and CH$_4$. To contribute to the
gas-phase chemistry, these surface-sticking species need to be desorbed
(evaporated, i.e., converted from solid state to gas-phase
constituents). This desorption will require a grain temperature
increase, which may be caused by absorption of the strong UV light
from nearby newly formed stars, which is the case in PDR chemistry
(Hollenbach et al. \cite{hol09}) as
well as in Hot Core chemistry (Rodgers \&\ Charnley
  \cite{rod01}). According to Table B.1 the observed \hto\ abundances
are as high as $10^{-5}$ and $3 \times 10^{-6}$ in the high density
Orion Hot Core (where $T_{\rm kin} \approx 200$ K) and Compact Ridge
(where  $T_{\rm kin} \approx 115$ K) regions, respectively.

\subsubsection{High temperature water chemistry}

Detachment of ice layers of grain-sticking molecules may also be
caused by shocks, as may be the case in low velocity bi-polar outflows
from young stars, supernova shocks and Galactic density wave shocks
(cf. Flower \&\ Pineau des For\^ets \cite{flo03}; Melnick et
al. \cite{mel08}). In Table B.1 we list observed gas-phase \hto\ 
abundances of a few $\times 10^{-6}$ in the Orion and DR\,21 low velocity
outflows, and similar values for the red wing emissions from the Sgr\,A
molecular cloud cores. 

In addition to the evaporation of ices we also have to consider some
key high temperature gas phase reactions, that is, \\ \\
O + \htwo\ = OH + H \ \ \ \ \ \ \ \ \ \  (5) \\ \\
and subsequently \\ \\
OH + \htwo\ = \hto\ + H.	\ \ \ \ \ \ \ \     (6) \\ \\
Shock heating and compression in this way may result in very high post
shock gas-phase \hto\ abundances (cf. e.g., Bergin et al. \cite{ber98}; Flower
et al. \cite{flo003}; Kristensen et al. \cite{kri07}; Flower \& Pineau
de For\^ets \cite{flo10}), as illustrated in Table B.1 by an
observationally determined $X$(\hto) versus shocked \htwo\ approaching $3
\times 10^{-4}$ in the Orion high velocity outflow, where (almost) all
elemental oxygen has been “locked-up” in gas-phase water. Here we
should again remember that outflow driven shock excitation of \htwo\
has been observed by Galliano et al. (\cite{gal12}) in some of the
circumnuclear torus hot spots of NGC\,1365 (i.e., massive stellar
clusters with their associated massive star forming gas clouds).

\section{Summary of the molecular physics and the interstellar
  chemistry of CH and \CHp}

An energy level diagram for CH is provided by Rangwala et al. (\cite{ran14})
and more detailed ones may be found in Rydbeck et al. (\cite{ryd76}) and
Turner (\cite{tur88}). The rotational levels of the light weight CH molecule
appear in the submm/FIR range and show a doublet spectral line pattern
caused by the splitting of the individual rotational state energies
into $\Lambda$-doublets, resulting from the relative
orientations of the orbital momentum axis of the unpaired $\pi$ electron
and the molecular rotation axis (being parallel or orthogonal, denoted
as + or – parity). Magnetic hyperfine interaction further splits the
$\Lambda$-doublet state energies, which was crucial for the
identification of the interstellar radio transitions of CH, but causes
overlapping lines and even difficult blends of emission and absorption
at submm and FIR wavelengths. The observed doublet pattern of the
rotational lines is the result of the quantum mechanical selection
rule that a state parity change is required for an electric dipole
transition to take place. 

CH, \CHp, and CN were the first interstellar molecules to be identified
(in 1937-1941), via their absorption lines of the visual light from
background stars (see the review by Gerin et al. \cite{ger16}). The interstellar
detection of the radio signals at $\approx 
3.3$ GHz from the three hyperfine transitions within the
$\Lambda$-doubled  $^2\Pi (N = 1, J = 1/2)$ CH ground state was a
matter of searches across a wide frequency range, since the
frequencies were only very crudely known from molecular quantum
mechanics applied to optical spectroscopy (Rydbeck et al. \cite{ryd73}; Turner
\&\ Zuckerman \cite{tur74}). Subsequently the three CH hyperfine transitions
were extensively observed and analyzed, and it was shown that these
lines were weak masers (amplifying any background emission) in
molecular cloud cores, cold dark clouds as well as in lower density
spiral arm clouds (Rydbeck et al. \cite{ryd76}; Hjalmarson et
al. \cite{hja77}). The CH abundance versus \htwo\ was observed to decrease
(from $\approx 3 \times 10^{-8}$ to  $\approx 6 \times 10^{-10}$)
with increasing cloud density, which may well be expected in regions where this
reactive radical is consumed in the formation of more complex
molecules, and also is consistent with more recent interstellar
chemistry modeling (e.g., Herbst \&\ Leung 1986 a,b).  The lower
satellite line of CH at 3264 MHz displayed an interesting anomaly in
that its intensity rapidly increased with increasing FIR continuum
emission, which was explained as “FIR pumping” and resulted from the
parity structure of the $\Lambda$-doubled  rotational states (Rydbeck et
al. \cite{ryd76}). Subsequently CH detections in external galaxies could also
be done, using the very sensitive maser receiver from Onsala Space
Observatory installed on the Australian Parkes 64-m radio telescope
(Whiteoak et al. \cite{whi80}).  

The first astronomical observation of CH in the far-infrared was
performed by Stacey et al. (\cite{sta87}). Using NASA’s Kuiper Airborne
Observatory they were able to detect the $N = 1, J = 1/2$ ground state
$\Lambda$-doublet to the $N = 2, J = 3/2$ rotationally excited
$\Lambda$-doublet state transitions of CH as an absorption
line doublet at $\approx 149$ $\mu$m (2006.8/2010.8 GHz) against the FIR
thermal dust emission of Sgr\,B2. These transitions are “always”
expected to be observed in absorption since their critical densities
for collisional excitation are > 10$^{10}$ cm$^{-3}$. Stacey et al. estimate
the CH abundance versus \htwo\ to be $\approx 10^{-9}$ for the Sgr\,B2
(denser) molecular cloud and $\approx 10^{-7}$ for the (lower density)
absorbing clouds in the 3 kpc arm and the expanding molecular
ring. They also conclude that the enhanced 3264 MHz CH emission from
the Sgr\,B2 molecular cloud must be primarily a FIR excitation
effect. Recently much more detailed, higher spectral and spatial
resolution, 2 THz CH absorption doublet studies toward a number of
Galactic molecular clouds have been performed from SOFIA (the
Stratospheric Observatory for Infrared Astronomy; Wiesemeyer et
al. \cite{wie18}).    

The lowest rotationally excited CH ($N = 1; J = 3/2$) state, only 25.7 K
above the ground state, was detected by means of the very large 1000
foot (300 m) Arecibo and Green Bank 300 foot (91 m) telescopes in a
number of molecular cloud cores, in terms of four hyperfine
transitions within the $\Lambda$-doublet at $\approx 700$ MHz, all
seen in absorption (Ziurys \&\ Turner \cite{ziu85}; Turner \cite{tur88}). An
interesting conclusion was that the CH
abundance versus \htwo\ was as high as $2 \times 10^{-8}$ in regions of
density $2 \times 10^4$ cm$^{-3}$ – an abundance which was too high to
be consistent with chemical models for dense quiescent clouds, but
instead pointed at the action of shock chemistry. In fact, molecular
outflows (causing shocks) have been observed in the regions
considered. The results from the (early) modeling of low velocity
shocks ($5 - 20$ \kms) in dense (10$^4$ cm$^{-3}$) gas clouds by
Mitchell (\cite{mit84}) is a useful guideline here – and may be so also in our
case of NGC\,1365.   

The detection of the CH 533/537 GHz transition doublet in the
directions of several massive star-formation regions relied on
{\it Herschel} HIFI observations, and revealed intriguing blends of
absorptions from lower density (diffuse) spiral arm foreground clouds
and emission from the molecular cloud cores (Gerin et
al. \cite{ger10}). Thanks to the hyperfine splitting, nowadays known from
laboratory spectroscopy, paired with molecular quantum mechanics the
emission/absorption line pattern could be deconvolved. The CH
abundance vs \htwo\ in the foreground clouds was determined to be
$\approx 3.5 \times 10^{-8}$ in clouds having densities in the range
$100-1000$ cm$^{-3}$ – an abundance consistent with the results from
decades of optical CH absorption line observations, and precisely
modeled by PDR chemistry (originating from Black \&\ Dalgarno
\cite{bla73}; see Gerin et al. \cite{ger10} and  Gerin et
al. \cite{ger16} for a discussion and references).  From their SOFIA
observations and a careful analysis of the CH absorption line doublet
at $\approx 149$ $ \mu$m (from the ground state, just as the 533/537 GHz
transitions) Wiesemeyer et al. (\cite{wie18}) confirm this conclusion,
that is, that CH may be used as a reliable tracer of \htwo\ 
columns and not only in lower density (diffuse) clouds but also in
denser high mass star formation regions. This is contrary to the
previously mentioned expectations from chemical models for dense
quiescent clouds (e.g., Herbst \&\ Leung 1986 a,b) and suggests
influences of shock chemistry (caused by bipolar outflows) in the
denser star forming clouds (as was concluded already by Turner \cite{tur88},
based upon his detailed analysis of the $\approx 700$ MHz CH absorption line
data). 

The {\it Herschel} HIFI detections by Falgarone et al. (\cite{fal10a})
of the \CHp($1-$0), and $^{13}$\CHp($1-0$), absorption lines at 835
and 830 GHz in the directions of the same star-formation regions also
showing CH absorptions, confirmed the long standing problem that the
observed \CHp\ abundance versus \htwo\ $ \approx  5 \times 10^{-8}$,
i.e. similar to that of CH and several orders of magnitude higher
than the predictions by UV-driven steady state PDR chemistry
models. So-called turbulent dissipation models, where the activation
energy ($\approx 4600$ K), 
needed for the endothermic reaction C$^+$ + \htwo\ = \CHp\  + H, is drawn from
dissipation of turbulence, appear to provide the solution of this
dilemma (TDR;  Falgarone et al. \cite{fal10a}; see Gerin et
al. \cite{ger16} for a discussion and further references).  In their
{\it Herschel} HIFI observations of \CHp\ toward the massive
star-formation region DR\,21 molecular cloud ridge – one of the most
powerful molecular 
outflows seen in vibrationally excited \htwo\ – another \CHp\ surprise
appeared, in terms of a strong broad \CHp($1-0$) emission line from the
cloud core, together with the expected deep absorption features from
the DR\,21 molecular ridge and foreground gas (Falgarone et
al. \cite{fal10b}). The broad emission was satisfactorily interpreted
by means of a state-of-the-art C-shock model in dense, UV-illuminated gas
(Falgarone et al. \cite{fal10b}), while the absorptions were consistent with
TDR modeling results (Falgarone et al. \cite{fal10a}).

The importance of intense UV-illumination for the \CHp\ formation in
dense molecular cloud PDR surfaces became very clear from the {\it Herschel}
observations of \CHp\ in the Orion Bar (Nagy et al. \cite{nag13},
\cite{nag17}) and the 
mapping of \CHp\ and CH in an extended region including Orion KL as well
as the Orion Bar (Morris et al. \cite{mor16}). In the Orion Bar the
\CHp\ ($J=1-0$ and $2-1$) emission lines turned out to be markedly
broader than other 
molecular lines ($\Delta v \approx 5$ \kms\ vs $\approx 3$ \kms, as
observed by HIFI), and the strong higher energy lines observed by PACS
(up to $J=6-5$ at an upper state energy of 838 K) and indicate a kinetic
temperature of $500-1000$ K. This very high rotational excitation of \CHp\
together with the line broadening are clear signs of “formation
pumping”, where the activation energy of 4600 K, required for the
endothermic reaction C$^+$ + \htwo\ = \CHp\ + H to take place, is
supplied by vibrationally (and rotationally) excited \htwo, caused by
the intense UV-illumination in the observed regions. The \htwo\
excited state excess energy 
here will cause the high rotational excitation of \CHp\ as well as
line broadening by additional translational \CHp\ motion (as discussed
in detail by Nagy et al. \cite{nag13} and theoretically modeled by Godard and
Cernicharo \cite{god13}; see also Gerin et al. \cite{ger16} for discussions and
references). In the Orion Bar the observed \CHp\ population
distribution leads to a column density 40 times larger than the
“minimum” column density calculated from the \CHp ($J=1-0$) emission
line, assuming a “reasonable” excitation temperature in the range
$20-150$ K, and hence an estimated \CHp\ abundance larger than that of
CH (Table 2 of Nagy et al. \cite{nag17}). This \CHp\ formation scenario is
favored also for the extended Orion molecular cloud surface where
(some) CH also may result from dielectronic recombination of \CHp\ (Morris
et al. \cite{mor16}).

\end{document}